\documentclass[article,amsmath,amssymb,twocolumn,superscriptaddress,notitlepage]{revtex4-2}

\usepackage[utf8]{inputenc}
\usepackage{physics}
\DeclareMathOperator{\sgn}{sgn}
\usepackage{mathtools}
\usepackage{multirow}

\DeclarePairedDelimiter\floor{\lfloor}{\rfloor}
\usepackage{tabularx}
\newcolumntype{C}{>{\centering\arraybackslash}X}
\usepackage{here}
\usepackage{color}

\usepackage{bibunits}
\usepackage{soul}
\usepackage{textcomp}
\usepackage{bm}
\usepackage[linesnumbered, ruled,vlined]{algorithm2e}
\usepackage{ulem}

\begin{document}

\title{Quantum computing quantum Monte Carlo with hybrid tensor network for electronic structure calculations}

\author{Shu Kanno}
\email{shu.kanno@quantum.keio.ac.jp}
\affiliation{Mitsubishi Chemical Corporation, Science \& Innovation Center, Yokohama, 227-8502, Japan}
\affiliation{Quantum Computing Center, Keio University, Yokohama, 223-8522, Japan}

\author{Hajime Nakamura}
\affiliation{Quantum Computing Center, Keio University, Yokohama, 223-8522, Japan}
\affiliation{IBM Quantum – IBM Research Tokyo, Tokyo, 103-8510, Japan}

\author{Takao Kobayashi}
\affiliation{Mitsubishi Chemical Corporation, Science \& Innovation Center, Yokohama, 227-8502, Japan}
\affiliation{Quantum Computing Center, Keio University, Yokohama, 223-8522, Japan}

\author{Shigeki Gocho}
\affiliation{Quantum Computing Center, Keio University, Yokohama, 223-8522, Japan}
\affiliation{School of Fundamental Science and Technology, Faculty of Science and Technology, Keio University, Yokohama, 223-8522, Japan}

\author{Miho Hatanaka}
\affiliation{Quantum Computing Center, Keio University, Yokohama, 223-8522, Japan}
\affiliation{School of Fundamental Science and Technology, Faculty of Science and Technology, Keio University, Yokohama, 223-8522, Japan}

\author{Naoki Yamamoto}
\affiliation{Quantum Computing Center, Keio University, Yokohama, 223-8522, Japan}
\affiliation{Department of Applied Physics and Physico-Informatics, Keio University, 3-14-1 Hiyoshi, Kohoku-ku, Yokohama 223-8522, Japan}

\author{Qi Gao}
\affiliation{Mitsubishi Chemical Corporation, Science \& Innovation Center, Yokohama, 227-8502, Japan}
\affiliation{Quantum Computing Center, Keio University, Yokohama, 223-8522, Japan}

\begin{bibunit}[apsrev4-1]

\begin{abstract}
Quantum computers have a potential for solving quantum chemistry problems with higher accuracy than classical computers.
Quantum computing quantum Monte Carlo (QC-QMC) is a QMC with a trial state prepared in quantum circuit, which is employed to obtain the ground state with higher accuracy than QMC alone. 
We propose an algorithm combining QC-QMC with a hybrid tensor network to extend the applicability of QC-QMC beyond a single quantum device size.
In a two-layer quantum-quantum tree tensor, our algorithm for the larger trial wave function can be executed than preparable wave function in a device.
Our algorithm is evaluated on the Heisenberg chain model, graphite-based Hubbard model, hydrogen plane model, and MonoArylBiImidazole using full configuration interaction QMC.
Our algorithm can achieve energy accuracy (specifically, variance) several orders of magnitude higher than QMC, and the hybrid tensor version of QMC gives the same energy accuracy as QC-QMC when the system is appropriately decomposed. 
Moreover, we develop a pseudo-Hadamard test technique that enables efficient overlap calculations between a trial wave function and an orthonormal basis state. 
In a real device experiment by using the technique, we obtained almost the same accuracy as the statevector simulator, indicating the noise robustness of our algorithm.
These results suggests that the present approach will pave the way to electronic structure calculation for large systems with high accuracy on current quantum devices.
\end{abstract}

\maketitle
\section{Introduction}
\label{introduction}

Computationally accurate prediction of physical properties can accelerate the development of functional materials such as batteries~\cite{Ceder1998-re, Gao2021-hr}, catalysis~\cite{Norskov2009-fr}, and photochemical materials~\cite{Turro1991-qw, Michl1990-pn}.
The physical properties are mainly governed by the electrons in the materials, and the computational cost of calculating electronic structures increases exponentially with system size in general, which often prevents classical computers from achieving the required accuracy to predict the properties.

Quantum computers are expected to solve such classically intractable problems in quantum chemistry and materials science~\cite{Bauer2020-ug}.
Current quantum computers, called noisy intermediate scale-quantum (NISQ) devices~\cite{Preskill2018-sc}, have limitations on the numbers of qubits and quantum gates due to physical noise, and various approaches for the NISQ devices have been proposed~\cite{Bharti2022-ov, Cerezo2021-dy}.
One of the most popular algorithms is the variational quantum eigensolver (VQE)~\cite{Peruzzo2014-kp}, which is used to obtain the ground state energy by minimizing energy cost function using a variational quantum circuit, where its parameters are updated by a classical computer.
Contrary to the traditional quantum algorithms such as the quantum phase estimation~\cite{Yu_Kitaev1995-cf}, VQE requires reduced hardware resources. 
However, VQE suffers from many issues such as insufficient accuracy~\cite{Stilck_Franca2021-qx} and vanishing parameter gradients, so-called barren plateau~\cite{McClean2018-pn}.

In addition to studies to avoid those issues~\cite{Grant2019-ul, Skolik2021-ju, Cerezo2021-kj, Kanno2023-ip}, there are multiple variants of quantum Monte Carlo (QMC), which were proposed for further relaxing the hardware requirements~\cite{Huggins2022-ly, Yang2021-ie, Tan2022-fh, Xu2023-sn, Zhang2022-lf, Layden2023-ke, Lee2022-lq}.
QMC is a computational method using stochastic sampling techniques to solve large quantum many-body problems such as molecular systems containing hundreds of electrons~\cite{Austin2012-qi, Al-Hamdani2021-ns}.
To date, various types of QMC approaches have been proposed, including variational Monte Carlo \cite{McMillan1965-dd, Ceperley1977-yz}, auxiliary field quantum Monte Carlo \cite{Blankenbecler1981-is, Sugiyama1986-xm}, and full configuration interaction quantum Monte Carlo (FCIQMC)~\cite{Booth2009-lv}. 
In this study, we adopt FCIQMC, which is useful for quantum chemistry, a stochastic imaginary-time evolution is executed in the space of all the orthonormal basis states (such as the Slater determinants) that can be constructed from a given spatial orbital basis.
As in Refs.~\cite{Huggins2022-ly, Xu2023-sn, Zhang2022-lf}, a type of QMC called quantum computing QMC (QC-QMC) was introduced.
Combined with a quantum algorithm such as VQE, QC-QMC can improve the energy evaluation accuracy in ground-state calculations. 
Specifically, the wave function distribution, i.e., the walker distribution, is generated on a classical~\cite{Xu2023-sn} (or quantum~\cite{Huggins2022-ly, Zhang2022-lf}) device, and energy is evaluated by using a trial wave function prepared by VQE.
Contrary to VQE where the accuracy depends on parameter optimization, QC-QMC has no such optimization and the accuracy relies on the quality of the trial wave functions, the algorithms, and calculation settings in the energy evaluation. 
Avoiding  strict VQE optimizations lowers the hardware requirements; for example, there was a demonstration on a diamond model constructed in 16 qubits, the largest hardware experiment run nearly within the chemical accuracy (chemical precision)~\cite{Huggins2022-ly}.

While QC-QMC can provide highly accurate ground-state calculations, applying this method to large-scale systems is challenging.
Although the electronic correlations outside the active space can be obtained by classical post-processing~\cite{Huggins2022-ly}, the available active space size~\cite{Takeshita2020-ec, Roos2007-cm} is limited by the size of the trial wave function i.e., the number of qubits.
The proposals for constructing wave functions larger than the quantum device size include the divide-and-conquer algorithm~\cite{Yamazaki2018-pu, Fujii2022-bq}, embedding theory~\cite{Kawashima2021-os, Greene-Diniz2022-xo, Cao2022-hn}, circuit cutting~\cite{Peng2020-xc, Harada2023-ym}, perturbation~\cite{Sun2022-bh}, and tensor network~\cite{Huggins2019-qa, Yuan2021-ih}.
Hybrid tensor network (HTN)~\cite{Yuan2021-ih} is a general tensor network framework that can be implemented on a reduced number of qubits and gates by decomposing the wave function in the original system into smaller-sized tensors. These tensors are processed by quantum or classical computation, i.e., quantum/classical hybrid computation.
Such a decomposition can reduce the effective width and depth of the circuit, making it more robust to noise than the original circuit.
There are several quantum algorithms or ansatze depending on the tensor network structure, including the matrix product states~\cite{Schollwock2011-im}, projected entangled pair states~\cite{Verstraete2004-lj}, and tree tensor networks (TN)~\cite{Shi2006-he, Huggins2019-qa}.
In the two-layer TN adopted in this study, the quantum states of the subsystem in the first layer are integrated with the second layer by using a quantum or classical method, referred to as the quantum-quantum TN (QQTN) and quantum-classical TN, respectively. The deep VQE~\cite{Fujii2022-bq} and entanglement forging~\cite{Eddins2022-ny, Motta2023-mg} can be broadly classified into the quantum-quantum and quantum-classical ones, respectively.
If gaining the quantum advantage (rather than noise robustness) is a priority, QQTN is preferred.

\begin{figure}[h!]
 \centering
 \includegraphics[width=1\columnwidth]{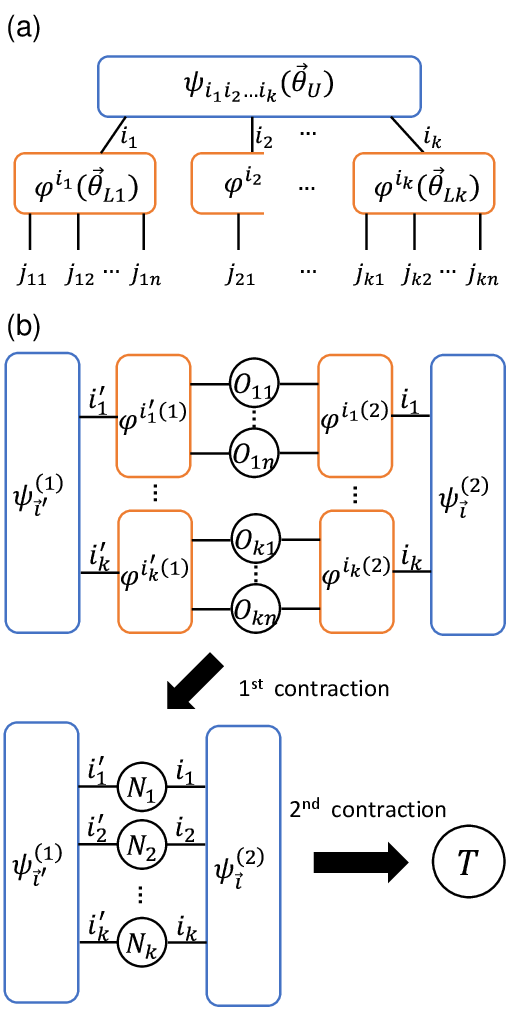}
\caption{
Representation and calculation of hybrid tensor network in the two-layer QQTN. 
(a) Hybrid tensor network representation of the two-layered QQTN state.
(b) Calculation of $T$ in Eq.~\eqref{Eq: ExpectationNonHermitian1}, where we omit the parameters. See Section~\ref{sec: HTN} for details.}
 \label{fig: HTNQMC}
\end{figure}

\begin{figure*}[]
 \centering
 \includegraphics[width=0.9\textwidth]{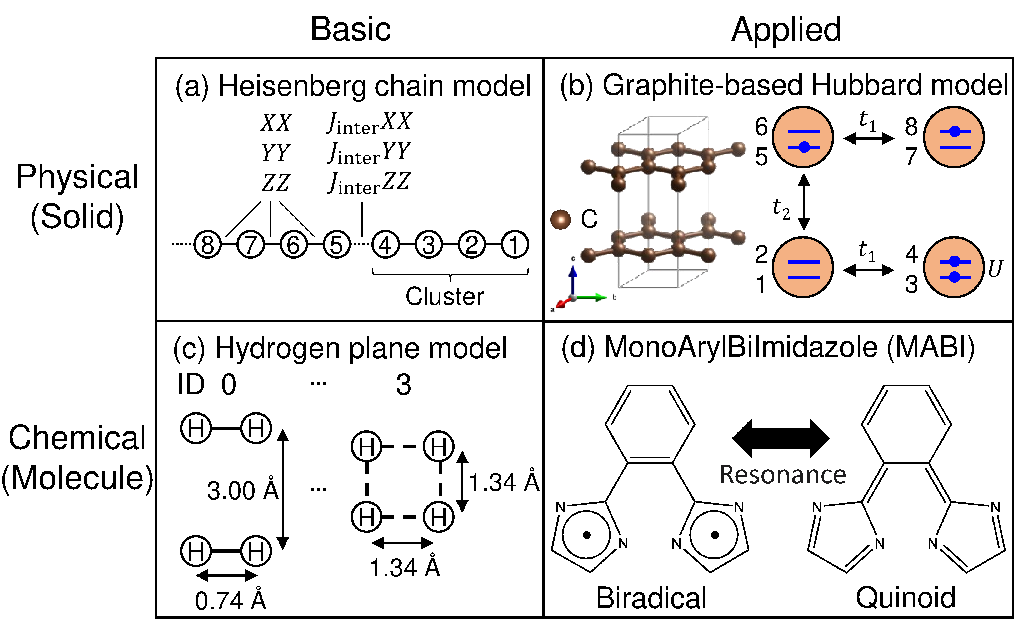}
\caption{Models for benchmarking HTN+QMC. The structure of graphite is drawn by VESTA~\cite{Momma2011-ck}. The qubit indices are labeled in (a) and (b)(see Section~\ref{sec: models and calculation condition} for details). (a) Heisenberg chain model. (b) Graphite-based Hubbard model. (c) Hydrogen plane model. (d) MonoArylBilmidazole (MABI).}
 \label{fig: Models}
\end{figure*}

In this study, we propose an algorithm of QC-QMC in combination with HTN of a two-layered QQTN. 
In particular, we consider the following QQTN:
\begin{equation}
 \begin{aligned}
&\ket{\psi_\mathrm{HTN}(\Vec{\theta}_\mathrm{HTN})} 
\\&=\sum_{\Vec{i}} \psi_{\Vec{i}}(\Vec{\theta}_{U}) \left(\prod_{m=1}^{k} \sum_{\Vec{j}_m}  \varphi^{i_m}_{\Vec{j}_m}(\Vec{\theta}_{Lm})\right) \left(\bigotimes_{m=1}^{k} \ket{\Vec{j}_m}\right)\\
&=\sum_{\Vec{i}} \psi_{\Vec{i}}(\Vec{\theta}_{U}) \bigotimes_{m=1}^{k} \ket{\varphi^{i_m}(\Vec{\theta}_{Lm})}.
\label{Eq: tensornetwork0}
\end{aligned} 
\end{equation}
The tensor network representation of this state is depicted in Fig.~\ref{fig: HTNQMC}(a).
The tensors $\psi_{\Vec{i}}(\Vec{\theta}_{U})$ ($\varphi^{i_m}_{\Vec{j}_m}(\Vec{\theta}_{Lm})$) are defined using wave functions $\ket*{\psi(\Vec{\theta}_{U})}$ ($\ket*{\varphi^{i_m}(\Vec{\theta}_{Lm})}$) as $\psi_{\Vec{i}}(\Vec{\theta}_{U}) = \braket*{\Vec{i}}{\psi(\Vec{\theta}_{U})}$ ($\varphi^{i_m}_{\Vec{j}_m}(\Vec{\theta}_{Lm}) = \braket*{\Vec{j}_m}{\varphi^{i_m}(\Vec{\theta}_{Lm})}$), where $\Vec{i} = i_1 i_2 \dots i_k$ ($\Vec{j}_m = j_{m1} j_{m2} \dots j_{mn}$) represent a $k$ ($n$)-qubit binary string, and the subscript $\Vec{j}_m$ of $\varphi^{i_m}_{\Vec{j}_m}(\Vec{\theta}_{Lm})$ is omitted in Fig.~\ref{fig: HTNQMC}. 
$\psi_{\Vec{i}}(\Vec{\theta}_{U})$ and $\varphi^{i_m}_{\Vec{j}_m}(\Vec{\theta}_{Lm})$ are assumed to be constructed by quantum circuits, where $\Vec{\theta}_{U}$ and $\Vec{\theta}_{Lm}$ are variational parameters of the upper tensor (blue) and the $m$-th lower tensor (orange), respectively. 
The subscripts $U$ and $L$ designate the upper and lower tensors, respectively. $k$ and $nk$ represent the number of subsystems and system qubits, respectively.
The algorithm details are given in Section~\ref{sec: method} (and Appendices~\ref{sec: fciqmc},~\ref{sec: statistical analysis of Emix}, and~\ref{sec: tensor contraction procedure}), where calculating the transition amplitude between a pair of the wave functions $T(\Vec{\theta}_\mathrm{HTN}^{(1)}, \Vec{\theta}_\mathrm{HTN}^{(2)})$ is a major part:
\begin{equation}
\begin{aligned}
T(\Vec{\theta}_\mathrm{HTN}^{(1)}, \Vec{\theta}_\mathrm{HTN}^{(2)}) = \mel{\psi_\mathrm{HTN}^{(1)}(\Vec{\theta}_\mathrm{HTN}^{(1)})}{O}{\psi_\mathrm{HTN}^{(2)}(\Vec{\theta}_\mathrm{HTN}^{(2)})},
\label{Eq: ExpectationNonHermitian1}
\end{aligned}
\end{equation}
where the observable $O$ is defined using tensor products $O = \bigotimes_{m, r} O_{mr}$ and $O_{mr}$ is an observable attached on the $r$-th qubit of the $m$-th subsystem $(r = 1,2,\dots,n)$. 
As shown in Fig.~\ref{fig: HTNQMC}(b), the transition amplitude $T(\Vec{\theta}_\mathrm{HTN}^{(1)}, \Vec{\theta}_\mathrm{HTN}^{(2)})$ is calculated by performing a quantum computation on the lower tensor, classically processing the obtained results to generate the contracted operators $N_m$, and then contracting with a quantum computation on the upper tensor.
Substituting $\ket{\psi_\mathrm{HTN}^{(1)}(\Vec{\theta}_\mathrm{HTN}^{(1)})} = \ket{\psi_\mathrm{HTN}^{(2)}(\Vec{\theta}_\mathrm{HTN}^{(2)})} = \ket{\psi_\mathrm{HTN}(\Vec{\theta}_\mathrm{HTN})}$ in Eq.~\eqref{Eq: ExpectationNonHermitian1} yields the estimation of the observable
\begin{equation}
\begin{aligned}
\mel{\psi_\mathrm{HTN}(\Vec{\theta}_\mathrm{HTN})}{O}{\psi_\mathrm{HTN}(\Vec{\theta}_\mathrm{HTN})},
\label{Eq: target function1}
\end{aligned}
\end{equation}
and by substituting $\ket{\psi_\mathrm{HTN}^{(1)}(\Vec{\theta}_\mathrm{HTN}^{(1)})}=\ket{\psi_\mathrm{HTN}(\Vec{\theta}_\mathrm{HTN})}$,  $\ket{\psi_\mathrm{HTN}^{(2)}(\Vec{\theta}_\mathrm{HTN}^{(2)})}=\ket{\phi_{h}}$, and $O=I^{\otimes nk}$ in Eq.~\eqref{Eq: ExpectationNonHermitian1}, we obtain the overlap with the ${h}$-th orthonormal basis state $\ket{\phi_{h}}$ (e.g., the Slater determinant)
\begin{equation}
\begin{aligned}
\braket{\psi_\mathrm{HTN}(\Vec{\theta}_\mathrm{HTN})}{\phi_{h}}.
\label{Eq: target function2}
\end{aligned}
\end{equation}

The HTN+QMC algorithm consists of two steps: first, we optimize $\Vec{\theta}_\mathrm{HTN}=(\Vec{\theta}_{U}, \Vec{\theta}_{Lm})$ to prepare the trial wave function, by calculating Eq.~\eqref{Eq: target function1}.
Second, we execute QMC by using the trial wave function $\ket{\psi_\mathrm{HTN}(\Vec{\theta}_\mathrm{HTN})}$, which was optimized in the first step, and using Eq.~\eqref{Eq: target function2}.
We will denote HTN+VQE when referring to the first step only and omit the parameter in $\ket{\psi_\mathrm{HTN}}$ hereafter. 
In this formalism, by using the contraction technique shown in Fig.~\ref{fig: HTNQMC}(b), we can construct a $nk$ qubit tree-type trial wave function in QMC by using a $n$ qubit device with a linear measurement overhead.
Note that the dimension of the subsystem represented by each lower tensor is determined by the number of the legs connected to the upper tensor.  Under such restrictions, therefore, how to decompose the target system into the subsystems is a crucial factor in achieving a high fidelity of the tensor product state.

We benchmarked the performance of our algorithm for the Heisenberg chain model, hydrogen ($\mathrm{H_4}$) plane model, graphite-based Hubbard model, and MonoArylBiImidazole (MABI), which are classified into four material categories: physical or chemical (solid or molecule) and basic or applied as in Fig.~\ref{fig: Models}. 
The first two models are commonly used as the benchmark model~\cite{Motta2019-sf, Huggins2022-ly}. Graphite is a two-dimensional layered material and is used as an anode in lithium-ion batteries~\cite{Thinius2014-xz}. MABI is a model system of the photochromic radical dimer PentaArylBiImidazole (PABI)~\cite{Kobayashi2017-ty}.
We also execute on a real device $ibmq\_kolkata$ by developing a technique to calculate the overlap between the trial wave function and orthonormal basis state required in HTN+QMC.  
We call this technique a pseudo-Hadamard test because the Hadamard test type circuit is used in the technique.
We mention that the proposed algorithm can be applied to other HTN structures and QMC types.
For example, by changing the overlap calculation in the original QC-AFQMC~\cite{Huggins2022-ly} to the calculation introduced in this paper, it could be extended to the HTN version.

Finally, while our main objective is to large-scale QC-QMC by using HTN, we should comment on the treatment of the sign problem in this study. 
In previous research on QC-QMC~\cite{Huggins2022-ly, Zhang2022-lf, Xu2023-sn}, there are two steps where quantum computation is utilized: one related to walker control, and the other (projected) energy evaluation. 
Both steps employ quantum computation for overlap or transition amplitude calculations. 
In this study, we adopted a method that applies quantum computation only to the energy calculation ~\cite{Xu2023-sn}. 
While this method reduces the variance of energy evaluation, it does not essentially address the sign problem of FCIQMC, where the required number of walkers exponentially increases with system size. To tackle the problem, we should address the method related to a walker control, for example, the construction of an orthonormal basis using a unitary transformation by VQE~\cite{Zhang2022-lf}, in which the QMC wave function is represented more sparsely than in the Slater determinant basis.
In Appendix~\ref{sec: HTN+FCIQMC with sparse basis construction}, we provide an overview of the HTN+FCIQMC algorithm including the sparse basis construction. 
However, this algorithm was not tested in this study due to its high computational cost for verifications and the fact that it is far from the main purpose of this study, which is to extend QC-QMC to HTN.

\section{Results}
\label{Results}
\subsection{Benchmarking models}
\label{sec: models and calculation condition}
The performance of HTN+QMC is benchmarked with the Heisenberg chain model, graphite-based Hubbard model, hydrogen plane model, and MABI.
The Heisenberg chain model, as shown in Fig.~\ref{fig: Models}(a), is defined as a chain of $k$ clusters consisting of four sites 
\begin{equation}
\begin{aligned}
H &= \sum_{p=1}^k H_p + J_\mathrm{inter} \sum_{p'=1}^{k-1} H_{p'},
\label{Eq: Heisenberg chain1}
\end{aligned}
\end{equation}
where 
\begin{equation}
\begin{aligned}
H_p &= \sum_{f=1}^{3} X_{4(p-1)+f}X_{4(p-1)+f+1}\\
&+Y_{4(p-1)+f}Y_{4(p-1)+f+1}\\
&+Z_{4(p-1)+f}Z_{4(p-1)+f+1},
\label{Eq: Heisenberg chain2}
\end{aligned}
\end{equation}
\begin{equation}
\begin{aligned}
H_{p'} &= X_{4p'}X_{4p'+1}+Y_{4p'}Y_{4p'+1}+Z_{4p'}Z_{4p'+1}.
\label{Eq: Heisenberg chain3}
\end{aligned}
\end{equation}
The intra- and inter-cluster interactions are 1 and $J_\mathrm{inter}$, respectively.
In the benchmark, we consider $k =2$ and $3$, i.e., 8- and 12-qubit models and $J_\mathrm{inter} =0.2, 0.4, \dots, 2.0$. We assume that the $J_\mathrm{inter}$ is described by atomic unit (i.e., Hartree), but except when necessary, the unit is omitted according to convention.

Figure~\ref{fig: Models}(b) shows our graphite-based Hubbard model (the graphite model hereafter), where two layers of graphene sheets are modeled with periodically aligned unit cells, in which two carbon atoms reside in each layer. Using two qubits to represent the up and down spin orbitals ($p_z$ orbitals) in each carbon atom, we define the Hamiltonian with 8 qubits as 
\begin{equation}
\begin{aligned}
H &=3 t_1 \sum_{q=1,2,5,6}(a_{q+2}^{\dagger} a_q+a_q^{\dagger} a_{q+2})\\
&+2 t_2 \sum_{q=1,2}(a_{q+4}^{\dagger} a_q+a_q^{\dagger} a_{q+4})\\
&+U \sum_{q=1,3,5,7} n_{q+1} n_q,
\label{Eq: Hubbard model}
\end{aligned}
\end{equation}
where $q$ is the spin-orbital index for the $p_z$ orbital in carbon, and $q=1,2,3$, and $4$ ($5,6,7$, and $8$) corresponds to the first (second) layer. 
$a_q^{\dagger}$ ($a_q$) is the creation (annihilation) operators on the $q$-th site and $n_q$ is the number operator $n_q = a_q^{\dag} a_q$. 
$t_1$ and $t_2$ are the hopping energy between the first and second nearest neighbor sites corresponding to the intra- and inter-layer interaction energy, respectively, and $U$ is the on-site Coulomb energy.
The prefactors for $t_1$ and $t_2$ arise from periodic boundary conditions, e.g., the prefactor for $t_2$ is 2 because two inter-layer interactions exist per carbon (one inside and the other outside the unit cell).
The reason for $t_2$ being only on two indices, $q=1$ and 2, is because graphite is AB stacking.
We determine the value of $t_1$, $t_2$, and $U$ using the electronic structure calculation. See Appendix~\ref{sec: methods and conditions} for details.

The Hamiltonians for the hydrogen plane model (8-qubit model) in Fig.~\ref{fig: Models}(c) and MABI (12-qubit model) in Fig.~\ref{fig: Models}(d) were constructed using their restricted Hartree-Fock orbitals of their ground states as molecular orbital bases.
In the hydrogen plane model, the two hydrogen molecules are arranged vertically as in the index (ID) 0 of Fig.~\ref{fig: Models}(c). 
Then, the intermolecular and intramolecular distances are shortened and lengthened, respectively, until the hydrogen plane becomes a square as in ID 3.
The two structures are interpolated between ID 0 and 3, benchmarking four structures.

\begin{figure*}[]
 \includegraphics[width=0.9\textwidth]{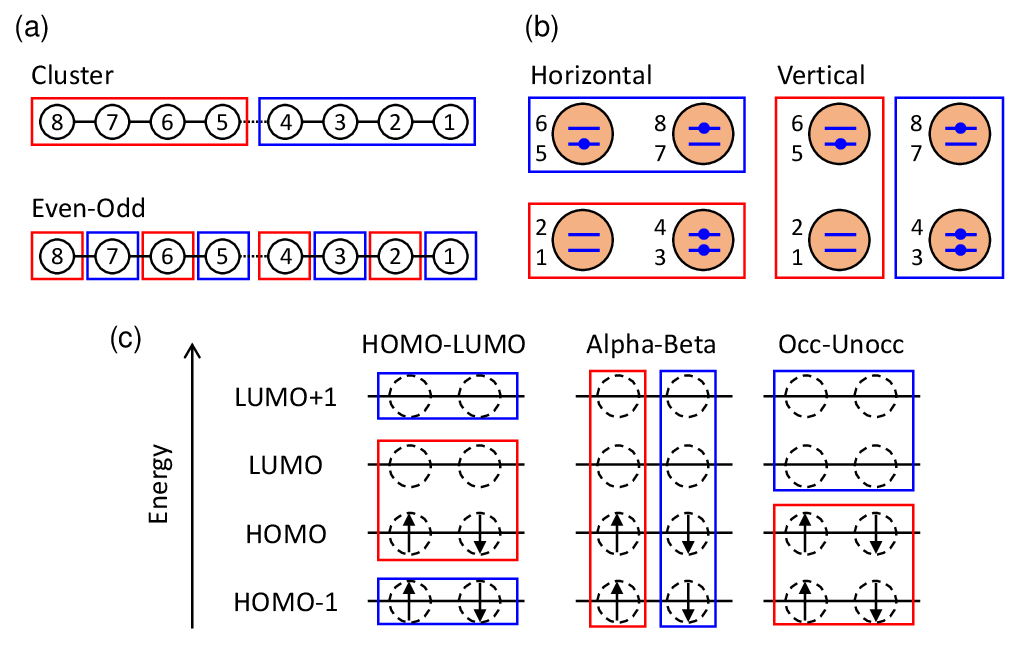}
 \caption{Decomposition settings. $n=4$ and $k=2$ are assumed for all the models in this figure. (a) Heisenberg chain model. (b) Graphite-based Hubbard model. (c) Chemical models in the case of the eight-qubit model. }
 \label{fig: Decomposition_settings}
\end{figure*}

Next, we present the specific conditions for constructing the benchmarking models.
The decomposition settings used for individual models are as follows: for the Heisenberg chain model, the cluster and even-odd settings refer to decomposing the model into subsystems by the cluster and even-odd indices, respectively, as illustrated in Fig.~\ref{fig: Decomposition_settings}(a);
for the graphite model, the horizontal and vertical settings refer to decomposing the model horizontally and vertically with respect to the sheet (in the ab-axis plane in Fig.~\ref{fig: Models}(b)), respectively, as in Fig.~\ref{fig: Decomposition_settings}(b);
for the chemical models (hydrogen plane model and MABI), the HOMO-LUMO, alpha-beta, and occ-unocc settings refer to decomposing the model using the pair of HOMO$-x$ and LUMO$+x$ as units ($x=0$ and $1$ in the hydrogen plane model and  $x=0,1$, and $2$ in MABI), based on alpha and beta spin-orbitals, as well as occupied and unoccupied orbitals, respectively, as in Fig.~\ref{fig: Decomposition_settings}(c), where the orbitals of the chemical models are shown in Appendix~\ref{sec: methods and conditions};
the no-decomposition setting refers to the calculation without decomposing the original system, i.e., not using HTN.
The cluster setting for the Heisenberg model, the horizontal setting for the graphite model, and the HOMO-LUMO setting for chemical models are adopted as a default.

\begin{figure}[]
 \includegraphics[width=0.9\columnwidth]{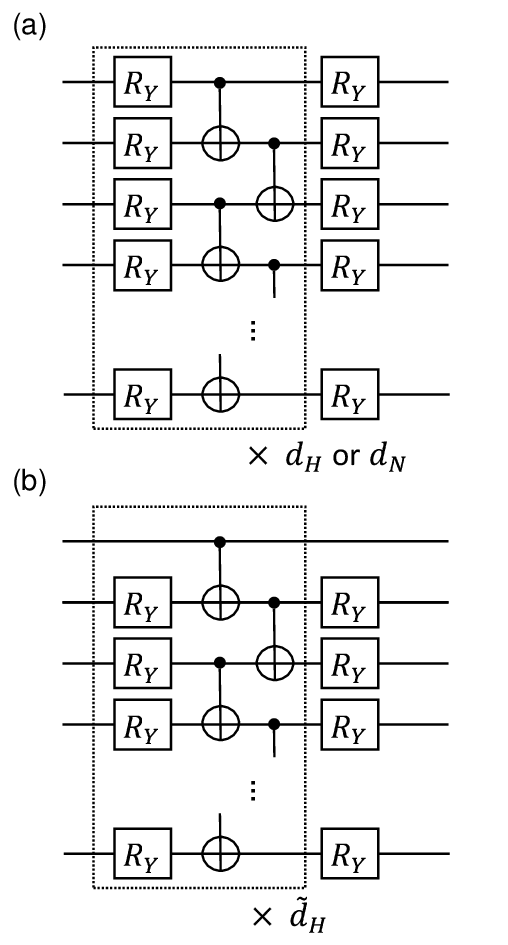}
\caption{Circuits of real amplitude ansatz in this study. (a) Circuit in statevector procedure. All the lines represent system qubits. (b) Circuit in real device procedure. The topmost line represents an ancilla qubit and the other lines represent system qubits.}
 \label{fig: RAansatz_1column}
\end{figure}

Figure~\ref{fig: RAansatz_1column} shows the quantum circuit called real amplitude ansatz, a popular hardware efficient ansatz employed for devices with linear connectivity layout, where $d_H$, $d_N$, and $\Tilde{d}_H$ denote the depth in HTN, the no-decomposition setting, and HTN used for real device, i.e., the block surrounded by the dotted line is repeated $d_H$ or $d_N$, and $\Tilde{d}_H$ times, respectively.
Figure~\ref{fig: RAansatz_1column} (a) shows the original circuit, which is used for the statevector simulation, whereas, in order to run on real devices, a variant with ancilla qubit was considered, as in Fig.~\ref{fig: RAansatz_1column} (b); details on the operation are given in Section~\ref{sec: real device execution}. 
In both ansatze, the rotation angle of the RY gate (i.e., the parameter) is updated at VQE and HTN+VQE iterations, while in the second step of QC-QMC and HTN+QMC, it is fixed.
Because HTN+VQE demands quite a high computational cost due to its iterative process, the real device ($ibmq\_kolkata$) was used only for the overlap calculation in the second step, i.e., not for HTN+VQE computations.

The depth represents a value for the real amplitude ansatze, which are $k$-qubit and $n$-qubit circuits for upper and lower tensors, respectively, in HTN, whereas $nk$-qubit circuits in the no-decomposition setting (in which ancilla qubit is not counted). 
Thus, the numbers of parameters in the HTN and no-decomposition settings are different even with $d_H = d_N$, whereas the numbers in the statevector and real device procedure of HTN are the same when $d_H = \Tilde{d}_H$. 
The numbers of parameters in the HTN settings with $d_H$ and the no-decomposition setting with $d_N$ are $nk(d_H+1)+k(d_H+1)$ and $nk(d_N+1)$, respectively.
We set $d_H=4$ as a default by referring to the benchmark results in Appendix~\ref{sec: benchmark results}.
The initial parameters for HTN+VQE are chosen randomly from $[0, 1)$ and $[0, 2\pi)$ in the statevector and the real device procedure, respectively, and those for VQE are $[-2\pi, 2\pi]$.
Note that only the results specifically relevant to the discussion are described in the main text, while the rest are presented in the appendices.

In all the QMC methods (QMC, QC-QMC, and HTN+QMC), we evaluate the mean and standard deviation of the (projected) energy over 5,000 to 10,000 iterations excluding the Heisenberg chain model, whereas 50,000 to 100,000 iterations in the Heisenberg chain model. Hereafter absolute deviation from the exact ground state is referred to by energy difference. 
In QMC, a leading single orthonormal basis state in the exact ground state (single reference state hereafter) is chosen as a trial wave function.
See Appendix~\ref{sec: conditions for VQE and QMC} for the details of the VQE and QMC conditions.

\subsection{Numerical results}
\label{sec: numerical results}
First, we demonstrate the performance in HTN+QMC for the Heisenberg chain model, followed by the analysis of all the benchmarking models. Next, we describe the operational principle of the pseudo-Hadamard test technique and present the corresponding results of the real device experiments for the hydrogen plane model and MABI.

\subsubsection{Statevector simulation}
\label{sec: execution example}
\begin{figure}[]
 \includegraphics[width=1\columnwidth]{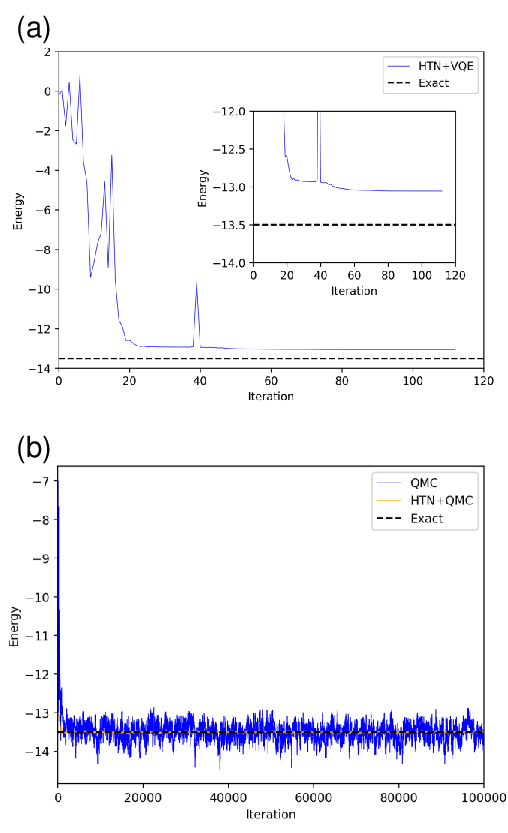}
\caption{Results of the energy in the Heisenberg chain models of the cluster setting with $n=4$, $k=2$, $J_\mathrm{inter} = 1.0$, and $d_H = 4$. The black dashed line represents the energy of the exact ground state. (a)  Result for HTN+VQE (blue line). The inset shows the enlarged view along the y-axis. (b) Results for QMC (blue line) and HTN+QMC (orange line).}
 \label{fig: execution example}
\end{figure}

Figure~\ref{fig: execution example}(a) shows the result of executing HTN+VQE for the Heisenberg chain model in the cluster setting with $d_H=4$, $k = 2$, and $J_\mathrm{inter} = 1.0$.
As in the inset, the energy difference from the exact ground state energy is $4.4\times10^{-1}$ at the end of the optimization.
Figure~\ref{fig: execution example}(b) shows the result for HTN+QMC,
and the energy difference with standard deviation from the exact value is $1.1\times10^{-4}\pm{1.4\times10^{-2}}$, which is more accurate than QMC. Specifically, that for QMC is $2.5\times10^{-2}\pm{2.1\times10^{-1}}$.
We discuss the details of this improvement in Sec.~\ref{sec: coclusion}, and we note the measurement cost scale in the QMC step of HTN+QMC in Sec.~\ref{sec: HTNQCQMC}. 

\begin{table*}[]
    \centering
    \caption{Energy differences with a standard deviation in VQE, QMC, and HTN+QMC, and fidelity for HTN+VQE and the single reference state for the Heisenberg chain models of the cluster setting with $k=2$ and $3$ and $J_\mathrm{inter}=1.0$, the graphite model of the horizontal setting with $k=2$, the hydrogen plane model of HOMO-LUMO setting with $k=2$ and ID 3, and MABI of HOMO-LUMO setting with $k=3$; $n=4$ and $d_H=4$ for all the models. The energy unit is Hartree.}
\label{tab: 4models benchmark results}
    \begin{tabular}{c c c c c c}
    \hline
        \multirow{2}{*}{Models} & \multirow{2}{*}{HTN+VQE} & \multirow{2}{*}{QMC} & \multirow{2}{*}{HTN+QMC} & Fidelity & Fidelity  \\ 
        &&&&(HTN+VQE)&(single reference state) \\ \hline
        Heisenberg chain ($k=2$) & $4.4\times10^{-1}$ & $2.5\times10^{-2}\pm{2.1\times10^{-1}}$ & $1.1\times10^{-4}\pm{1.4\times10^{-2}}$ & 0.92 & 0.13\\
        Heisenberg chain ($k=3$) & $1.0\times10^{0}$  & $3.0\times10^{-2}\pm5.5\times10^{-1}$ & $3.0\times10^{-3}\pm3.0\times10^{-2}$ & 0.80 & 0.06\\
        Graphite & $6.4\times10^{-4}$ & $4.6\times10^{-4}\pm6.4\times10^{-2}$ & $8.1\times10^{-6}\pm8.6\times10^{-5}$ & 1.00  & 0.09\\ 
        Hydrogen plane & $3.5\times10^{-2}$ & $9.1\times10^{-4}\pm6.7\times10^{-3}$ & $3.0\times10^{-5}\pm1.5\times10^{-3}$ & 0.44 & 0.45 \\ 
        MABI & $2.4\times10^{-2}$ & $6.6\times10^{-5}\pm2.6\times10^{-3}$ & $9.2\times10^{-6}\pm1.5\times10^{-3}$ & 0.94 & 0.86 \\  \hline
    \end{tabular}
\end{table*}

Here, we briefly comment on the other models; the computational results are described in detail in Appendix~\ref{sec: benchmark results}.
Table~\ref{tab: 4models benchmark results} shows the results summarized on HTN of the Heisenberg chain models with $k=2$ and $3$, the graphite model, the hydrogen plane model with ID 3, and MABI.
In all the models, the variance for HTN+QMC is smaller than that for HTN+VQE or QMC.  Except for the hydrogen plane model, the variance decreases as the fidelity increases. Here, the fidelity is defined as the square of the overlap between the target and exact ground states. See Appendix~\ref{sec: statistical analysis of Emix} for the statistical analysis for the fidelity and the variance in simple cases. 
For the hydrogen plane model, the variance for HTN+QMC is smaller than that for QMC, even though the fidelity for HTN+VQE (0.44) is almost the same as that for the single reference state (0.45).
The fidelity is not the only factor that determines the quality of the trial wave function~\cite{Amsler2023-vh}.
In the hydrogen plane model, the trial state generated by the HTN+VQE shares multiple determinants with the exact ground state, and it can make energy estimation robust to fluctuations in the QMC wave function.
Thus, the performance of QMC may be improved even if the trial wave function does not have high fidelity, and it is important to evaluate the trial wave function quality through QMC computation.

\subsubsection{Real device experiments}
\label{sec: real device execution}

\begin{figure}[]
 \centering
 \includegraphics[width=0.9\columnwidth]{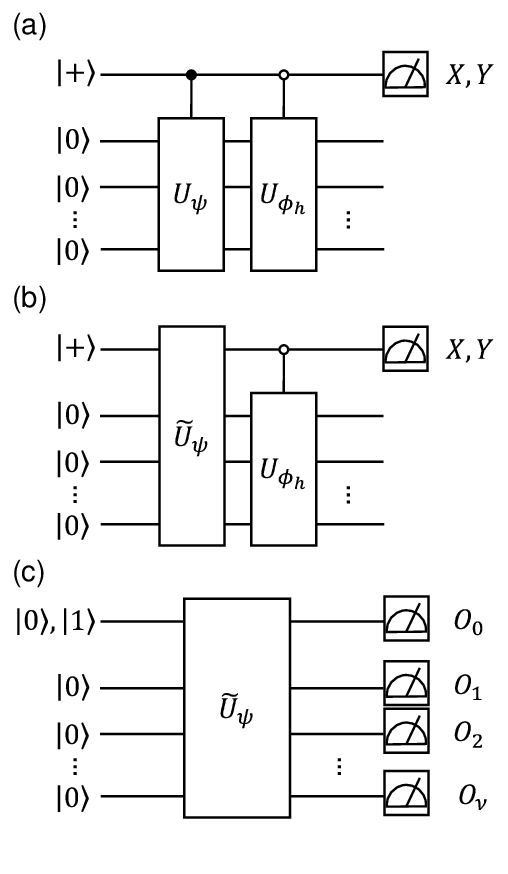}
\caption{Quantum circuits for calculating the overlap. 
The topmost line represents the ancilla qubit and the other lines represent the system qubits. 
(a) Hadamard test. (b) Pseudo-Hadamard test. (c) VQE circuit with constraint to determine $\Tilde{U}_{\psi}$ used for the pseudo-Hadamard test. 
The ancilla qubit is set to $\ket{0}$ when calculating the constraints.}
\label{fig: Circuit_RD_1column.eps}
\end{figure}

We consider performing the proposed HTN+QMC algorithm on a real device. 
For this purpose, we need to reduce the overlap computational cost between the trial wave function and the orthonormal basis state (as in Eq.~\eqref{Eq: target function2}). 
Conventionally, the overlap $\braket{\psi}{\phi_h}$ between a wave function $\ket{\psi}=U_{\psi}\ket{0}^{\otimes \nu}$ and the orthonormal basis state $\ket{\phi_h}=U_{\phi_h}\ket{0}^{\otimes \nu}$, can be calculated using the Hadamard test, as in Fig.~\ref{fig: Circuit_RD_1column.eps}(a), where $U_{\psi}$ typically consists of a collection of one- and two-qubit gates depending on the ansatz and $\nu$ is the number of system qubits.
The controlled-$U_{\psi}$ contains a much larger number of multi-qubit gates compared to $U_{\psi}$ because all one- and two-qubit gates in $U_{\psi}$ are modified to the controlled gates.
Several techniques to calculate the overlaps in fermionic systems have been proposed that can avoid deep circuits, for example, by using the particle preserving ansatz~\cite{Huggins2020-fk, Xu2023-sn}.
However, the techniques are inadequate for HTN+QMC due to no electron number preservation in the subsystems, and we explain the details in Appendix~\ref{sec: studies on overlap calculations}.
Note that $U_{\phi_h}$ requires at most $\nu$ two-qubit gates because it applies the CNOT gates from the ancilla to target qubits, the latter of which is set to $\ket{1}$ in $\ket{\phi_h}$.

We develop a gate-efficient technique to calculate the overlap executable on an arbitrary ansatz called the pseudo-Hadamard test. 
See Appendix~\ref{sec: details of real device execution} for extending the technique to HTN. 
Figure~\ref{fig: Circuit_RD_1column.eps}(b) shows the circuit of the pseudo-Hadamard test, in which $U_{\psi}$ is replaced with $\Tilde{U}_{\psi}$ involving the ancilla qubit.
This circuit calculates the overlap $\braket{\Tilde{\psi}}{\phi_h}$ if $\Tilde{U}_{\psi}$ satisfies the following conditions:
\begin{equation}
\begin{aligned}
\label{eq: overlap requirement}
    &\Tilde{U}_{\psi}\ket{1}\ket{0}^{\otimes \nu} = \ket{1}\ket*{\Tilde{\psi}},\\
    &\Tilde{U}_{\psi}\ket{0}\ket{0}^{\otimes \nu} = \ket{0}\ket{0}^{\otimes \nu}.
\end{aligned}
\end{equation}
The gate $\Tilde{U}_{\psi}$ can be determined through the VQE under constraints that the ancilla qubit is measured to be one in the first condition and all the qubits are measured to be zero in the second condition in Eq.~\eqref{eq: overlap requirement}, which is formulated as
\begin{equation}
\begin{aligned}
\label{Eq: general formalism for pseudo-Hadamard test}
& \underset{\Tilde{U}_{\psi}}{\min}
& & \bra{1}\bra{0}^{\otimes \nu} \Tilde{U}_{\psi}^{\dag} (I \otimes H) \Tilde{U}_{\psi} \ket{1}\ket{0}^{\otimes \nu}\\
& \text{s.t.}
& & \bra{1}\bra{0}^{\otimes \nu} \Tilde{U}_{\psi}^{\dag} (n_0 \otimes I^{\otimes \nu}) \Tilde{U}_{\psi} \ket{1}\ket{0}^{\otimes \nu} = 1,\\
& & & \bra{0}\bra{0}^{\otimes \nu} \Tilde{U}_{\psi}^{\dag} (\sum_{\kappa=0}^{\nu} I^{\otimes \kappa} \otimes n_{\kappa} \\
& & & \quad \otimes I^{\otimes \nu-\kappa}) \Tilde{U}_{\psi} \ket{0}\ket{0}^{\otimes \nu}=0,
\end{aligned}
\end{equation}
where the index of the ancilla qubit is set to 0.
$n_{\kappa}$ is the number operator of $\kappa$-th qubit in Eq.~\eqref{Eq: Hubbard model}.
Figure~\ref{fig: Circuit_RD_1column.eps}(c) shows the quantum circuit used to run such VQE, where $O_{\iota}$ ($\iota = 0,1,\dots,\nu$) is an observable of the $\iota$-th qubit. 
The constraints can be relaxed through the appropriate selection of ansatz, such as the real amplitude ansatz without rotation gates on the ancilla qubit, as shown in Fig.~\ref{fig: RAansatz_1column}(b). 
In this case, VQE can be simplified as
\begin{equation}
\begin{aligned}
\label{Eq: simple formalism for pseudo-Hadamard test}
& \underset{\Tilde{U}_{\psi}}{\min}
& & \bra{1}\bra{0}^{\otimes \nu} \Tilde{U}_{\psi}^{\dag} (I \otimes H) \Tilde{U}_{\psi} \ket{1}\ket{0}^{\otimes \nu}\\
& \text{s.t.}
& & \bra{0}\bra{0}^{\otimes \nu} \Tilde{U}_{\psi}^{\dag} (\sum_{\kappa=1}^{\nu} I^{\otimes \kappa} \otimes n_{\kappa} \\
& & &\quad\otimes I^{\otimes \nu-\kappa}) \Tilde{U}_{\psi} \ket{0}\ket{0}^{\otimes \nu}=0.
\end{aligned}
\end{equation}
We adopt this simplified formalism hereafter.
Note that $\Tilde{U}_{\psi} \ket{0}\ket{0}^{\otimes \nu}$ takes either $\ket{0}\ket{0}^{\otimes \nu}$ or $-\ket{0}\ket{0}^{\otimes \nu}$ in the constraint, but the signs cancel out in a numerator and denominator of the projected energy in QC-QMC and HTN+QMC (see Eq.~\eqref{Eq: projected energy}), allowing one to ignore the sign.
In cases where the technique is applied outside of QMC, the sign could be determined based on the overlap between a trial wave function and a specific orthonormal basis state, e.g., the Hartree-Fock state~\cite{Xu2023-sn}.
We also mention the extensions of the technique; a transition amplitude of $\mel{\Tilde{\psi}}{\Tilde{O}}{\phi_h}$ $(\Tilde{O}=I \bigotimes_{\iota=1}^{\nu} O_{\iota})$ can be calculated by measuring each system qubit in the basis corresponding to $O_{\iota}$ for the pseudo-Hadamard test circuit;
that of $\mel{\Tilde{\psi}}{\Tilde{O}}{\Tilde{\psi}^{'}}$ can be also calculated by using the additional constraint $\bra{1}\bra{0}^{\otimes \nu} \Tilde{U}_{\psi}^{\dagger} \Tilde{U}_{\psi^{'}}^{\dagger} \Tilde{U}_{\psi} (\sum_{\kappa=1}^{\nu} I^{\otimes \kappa} \otimes n_{\kappa} \otimes I^{\otimes \nu-\kappa}) \Tilde{U}_{\psi}^{\dagger} \Tilde{U}_{\psi^{'}} \Tilde{U}_{\psi} \ket{1}\ket{0}^{\otimes \nu}=0$, where $\ket{\Tilde{\psi}^{'}}$ is a trial wave function defined by $\Tilde{U}_{\psi^{'}}$ with a property of $\Tilde{U}_{\psi^{'}}\ket{0}\ket{0}^{\otimes \nu} = \ket{0}\ket*{\Tilde{\psi}^{'}}$, e.g., the ansatz changing the control qubit condition of CNOT gates in Fig.~\ref{fig: RAansatz_1column}(b) from one to zero ; 
a transition probability $\abs{\mel{\Tilde{\psi}}{\Tilde{O}}{\Tilde{\psi}^{'}}}^2$~\cite{Ibe2022-jp, Sawaya2023-wx} can be trivially obtained by the square of the transition amplitude.

\begin{figure}[!h]
 \centering
 \includegraphics[width=1\columnwidth]{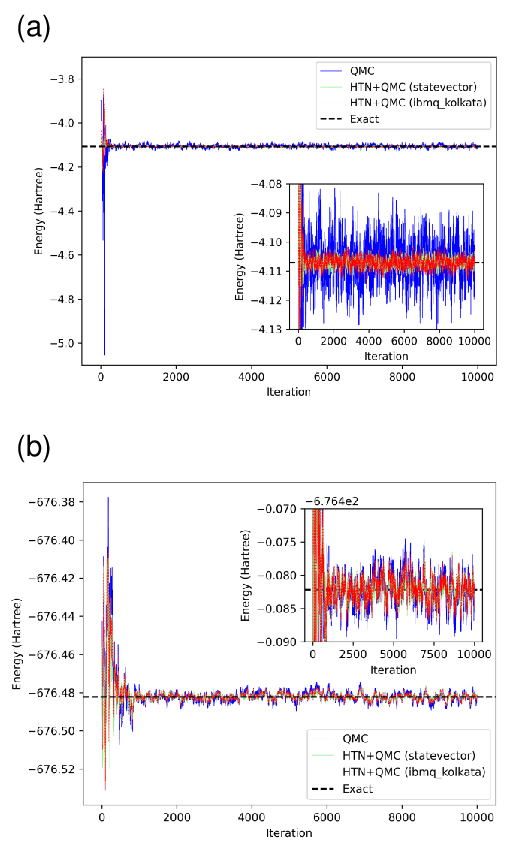}
\caption{Results of the energy on the real device execution for the hydrogen plane model and MABI. HOMO-LUMO setting and $\Tilde{d}_H=4$ are adopted for the models. The blue line represents the QMC result, the light green and red lines represent the HTN+QMC results of the statevector and real device procedures, respectively, and the black dashed line represents the exact ground state energy.
The inset in each figure presents an enlarged view along the y-axis.
(a) Result of the hydrogen plane model with $n=4$ and $k=2$. (b) Result of MABI with $n=4$ and $k=3$. 
}
\label{fig: QMC_RD_1column.eps}
\end{figure}

We applied the pseudo-Hadamard test technique to the HTN+QMC calculations on a real device $ibmq\_kolkata$ for the hydrogen plane model with ID 3 and MABI. 
The best trial wave function was selected from 100 seeds of HTN+VQE runs under the constraint ($\Tilde{d}_H = 4$ and statevector execution), where at most five qubits including an ancilla qubit, are used in the overlap calculation.
The energy accuracy decreased due to newly introduced constraints in the HTN+VQE, and thus we obtained the trial wave function of comparable accuracy by running 100 times, whereas the results without constraints in Table~\ref{tab: 4models benchmark results} were obtained with only one run.
See Appendix~\ref{sec: extension of the overlap technique to HTN} for details on the extension to HTN and the calculation conditions.
Figure~\ref{fig: QMC_RD_1column.eps}(a) shows the result for the hydrogen plane model.
A smaller fluctuation around the exact ground state energy was found in HTN+QMC than in QMC. 
The energy differences with a standard deviation obtained by the QMC, HTN+QMC (statevector) and HTN+QMC ($ibmq\_kolkata$) are $9.1\times10^{-4}\pm6.7\times10^{-3}$, $2.3\times10^{-5}\pm1.5\times10^{-3}$, and $1.8\times10^{-5}\pm1.8\times10^{-3}$ Hartree, respectively.
Importantly, there is only a slight difference in the fluctuation between the real device and statevector, indicating that there would be some robustness of HTN+QMC to noises. 
A previous study~\cite{Huggins2022-ly} showed that QC-QMC is robust to the depolarization noise. 
We also show the analytical and numerical result of the noise robustness in Appendix~\ref{sec: analysis of noise resilience}, and the results show that such robustness also holds in the current settings of HTN+QMC, even though it might be significant in HTN+VQE.
The MABI results in Fig.~\ref{fig: QMC_RD_1column.eps}(b) also showed small variance in the real device experiment, where the results are $6.6\times10^{-5}\pm2.6\times10^{-3}$, $2.3\times10^{-5}\pm1.6\times10^{-3}$, and $4.2\times10^{-5}\pm1.8\times10^{-3}$ Hartree in QMC, HTN+QMC (statevector) and HTN+QMC ($ibmq\_kolkata$), respectively.
These numerical and theoretical results suggest that HTN+QMC can be a promising candidate for calculating large systems beyond the scale of quantum devices while maintaining accuracy.
Note that although we utilized the readout error mitigation and the dynamic decoupling in the overlap calculation, we confirmed that a comparable level of accuracy was achieved even without those options in HTN+QMC for the hydrogen plane model, where the energy difference with a standard deviation excluding the options is $2.6\times10^{-5}\pm1.4\times10^{-3}$ Hartree.

\begin{figure*}[]
 \includegraphics[width=1\textwidth]{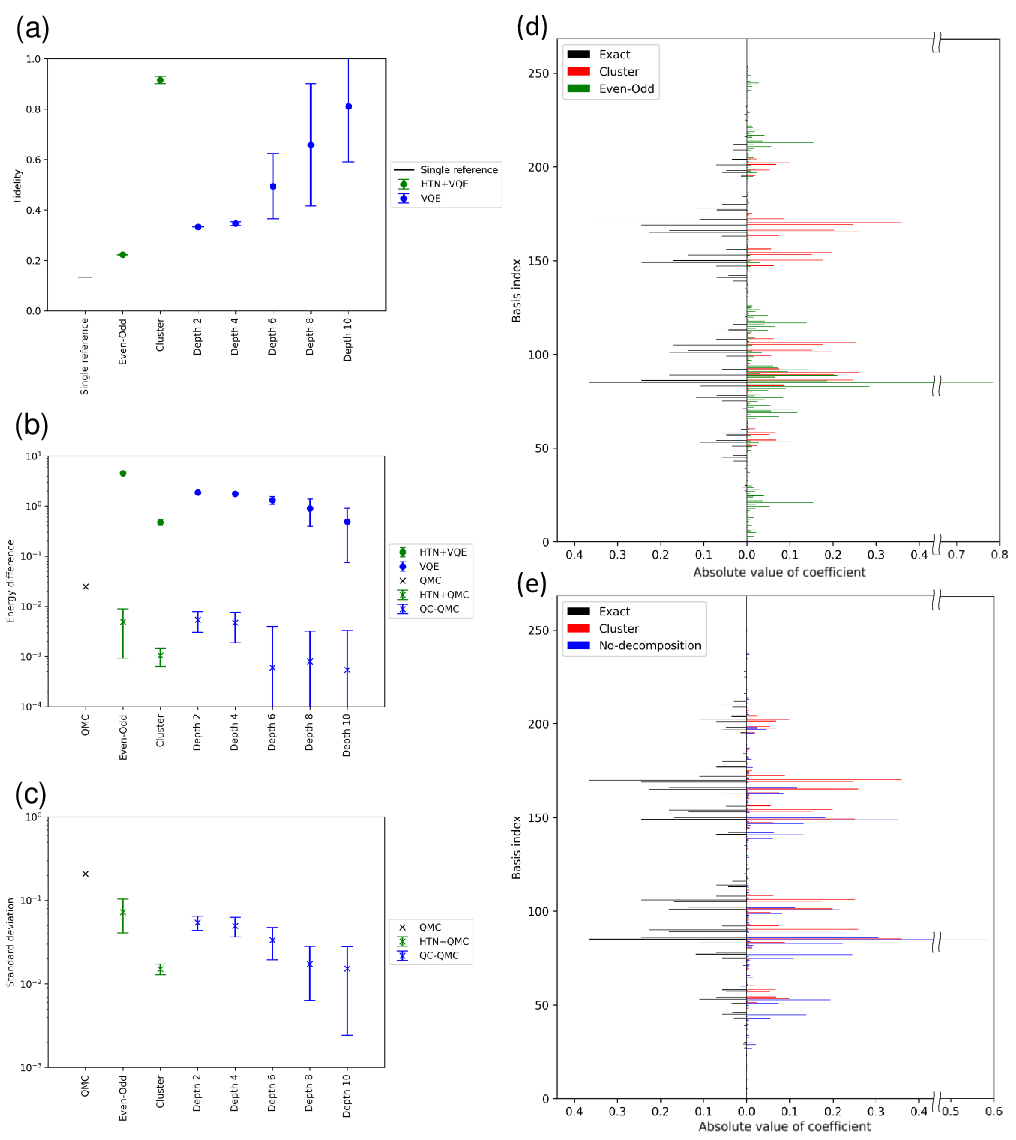}
\caption{Results of the analysis for the Heisenberg chain model with $J_\mathrm{inter} = 1.0$. (a), (b), and (c) show the decomposition dependencies. The average and error bar is obtained over 10 different random seeds used for the initial parameters in VQE or HTN+VQE. 
The fidelity of the single reference state is plotted by the black bar in (a), and the energy difference and standard deviation by the black cross markers in (b) and (c), respectively.
The green (blue) circle and cross markers denote the results for the decomposition (no-decomposition) setting in HTN+VQE and HTN+QMC (VQE and QC-QMC), respectively, with $d_H = 4$ ($d_N = 2,4,\dots,10$). 
In HTN+VQE and HTN+QMC, $n=4$ and $k=2$ for all the settings.
(d) and (e) show the wave function distributions for the Heisenberg chain model with $J_\mathrm{inter} = 1.0$. The basis index is the value when the orthonormal basis state is expressed in the decimal number, e.g., $\ket{10000010}$ corresponds to 130. 
The black, red, green, and blue bars represent the coefficients for the exact ground state, HTN+VQE results with cluster setting, HTN+VQE results with even-odd setting, and VQE results with the no-decomposition setting, respectively.  
(a) Fidelity. (b) Energy difference. (c) Standard deviation. (d) Comparison of the exact ground state, cluster setting, and even-odd setting. (e) Comparison of the exact ground state, cluster setting, and no-decomposition setting ($d_N=6$).}
\label{fig: Ana_Dist_Hei_2column.eps}
\end{figure*}

\section{Discussion}

\label{sec: coclusion}
We discuss dependencies of accuracy on the system decompositions in HTN.
We consider the Heisenberg chain model with $k=2$, $n=4$, and $J_\mathrm{inter}=1.0$, computed in the cluster, even-odd, and no-decomposition settings.
Figures~\ref{fig: Ana_Dist_Hei_2column.eps}(a), (b), and (c) show the decomposition dependencies of the fidelity, energy difference, and standard deviation for the Heisenberg chain model.
In cases of the decomposition settings, the fidelity (Fig.~\ref{fig: Ana_Dist_Hei_2column.eps}(a)) in the cluster setting (0.92) is four times higher than that in the even-odd setting (0.22).
We also calculated the bipartite entanglement entropy of the exact ground state when decomposing the cluster and even-odd settings to be 0.66 and 3.46, respectively.
Thus, an appropriate choice of the decomposition setting is crucial in the performance of the trial wave function. 
The fidelity of the cluster setting is equal to or slightly higher than that for the no-decomposition setting with $d_N = 10$, whereas that for the even-odd setting is lower than that for the no-decomposition setting with $d_N = 2$.
Here, the numbers of the parameters for the cluster setting with $d_H=4$ and the no-decomposition setting with $d_N=10$ are $nk(d_H+1)+k(d_H+1) = 50$ and $nk(d_N+1)=88$, respectively. 
That is, the cluster setting with fewer parameters shows comparable performance to the no-decomposition setting.
The increase in fidelity is related to the decrease in the standard deviation of HTN+QMC as in Fig.~\ref{fig: Ana_Dist_Hei_2column.eps}(c) (and the decrease in the energy difference in HTN+VQE and HTN+QMC as in Fig.~\ref{fig: Ana_Dist_Hei_2column.eps}(b)).
A similar tendency in fidelity, energy difference, and standard deviation can be found in the graphite model, the hydrogen plane model, and MABI (see Appendix~\ref{sec: other data of comparison of the decomposition data} for details).
These results suggest that if the system is appropriately decomposed, i.e., if the interaction between the subsystems is small, HTN can prepare a trial wave function that performs as well as or better than the wave function generated by the quantum circuit of the original system size.
For quantitative evaluations of the decomposition, we proposed measures in Appendix~\ref{sec: other data of comparison of the decomposition data}, specifically, interaction strength between subsystems and mutual information.
Note that one of the ways to increase the fidelity in VQE or HTN+VQE is to improve the initial wave function.  
As shown in Appendix~\ref{sec: other data of comparison of the decomposition data}, the fidelity could increase in some cases by using the initial state which is close to the Hartree-Fock state.  
In addition, while the parameters of all the tensors are sequentially optimized in this study, separate optimization of one tensor by another could be an alternative~\cite{Haghshenas2022-zn}.

The high performance in the cluster setting seems to arise when the wave function represented by QQTN in Fig.~\ref{fig: HTNQMC}(a) has a problem-inspired structure, whereas the ansatz used for each tensor is hardware efficient.
To analyze the wave function in detail, we first compare the distribution of the absolute coefficient of the orthonormal basis state, calculated using one of the 10 random seeds in Fig.~\ref{fig: Ana_Dist_Hei_2column.eps} (a), (b), and (c).
Fig.~\ref{fig: Ana_Dist_Hei_2column.eps}(d) shows the wave function distribution, where note that the value on the horizontal axis is the absolute value of the coefficient (not the square of the absolute value as in fidelity). The basis index is the value when the orthonormal basis state is expressed in the decimal number, e.g., $\ket{10000010}$ corresponds to 130. The cluster setting exhibits a distribution very close to that of the exact ground state compared to the even-odd setting.
Note that for the even-odd setting, the distribution was calculated with the basis state encoding, where the qubit index was reordered from $8,7,6,5/4,3,2,1$ to $8,6,4,2/7,5,3,1$ as shown in Even-Odd of Fig.~\ref{fig: Decomposition_settings}(a).
This qubit index was then changed back to that of Cluster in the plotting, allowing for direct comparison with the even-odd setting.  

We next examine the wave function, represented as a linear combination of tensor products of the subsystems. In the cluster setting for the Heisenberg chain model, we approximately describe
the exact ground state $\ket{\psi_{g-C}}$ using the four dominant terms with respective coefficients of $c_1 = -0.37, c_2 = 0.24, c_3 = -0.23$, and $c_4 = 0.18$, as
\begin{equation}
\begin{aligned}
\ket{\psi_{g-C}} 
&\simeq c_1 (
\ket{\textbf{0101}} \otimes \ket{\textbf{0101}} +
\ket{1010} \otimes \ket{1010}) \\
&+ c_2 (
\ket{\textbf{0101}} \otimes \ket{\textbf{0110}} + 
\ket{1001} \otimes \ket{0101} \\
& \quad + 
\ket{1010} \otimes \ket{1001} + 
\ket{0110} \otimes \ket{1010} ) \\
&+ c_3 (
\ket{\textbf{1010}} \otimes \ket{\textbf{0101}} +
\ket{0101} \otimes \ket{1010} ) \\ 
&+ c_4 (
\ket{0101} \otimes \ket{1001} + 
\ket{0110} \otimes \ket{0101} \\
& \quad + 
\ket{1001} \otimes \ket{1010} + 
\ket{\textbf{1010}} \otimes \ket{\textbf{0110}}).
\label{Eq: cluster representation}
\end{aligned}
\end{equation}
The state is energetically favored because of the (sub)antiferromagnetic spin configurations, with at most one spin pair having the same parity (e.g., 00 and 11), considering that all terms in the Hamiltonian of Eqs.~\eqref{Eq: Heisenberg chain1}, \eqref{Eq: Heisenberg chain2}, and \eqref{Eq: Heisenberg chain3} are positive.
Each term consists of a \textit{primitive} basis state (in bold) and the spin and spatial inverted states, where the ground state redefined in this primitive basis $\{\ket{\textbf{0101}}, \ket{\textbf{1010}}\} \otimes \{\ket{\textbf{0101}}, \ket{\textbf{0110}}\}$ is regarded as a \textit{separable} state (with the same number of up/down spins in each subsystem), accurately prepared by the lower tensors with one leg connecting to the upper tensor. 
In case of the even-odd setting, where the qubit index is reordered and the number of up (or down) spins in the subsystem takes 0 to 4, the state in such a primitive basis, e.g., $\{ \ket{\textbf{0000}}\otimes\ket{\textbf{1111}}, \ket{\textbf{0001}} \otimes \ket{\textbf{1110}}, \ket{\textbf{1100}} \otimes \ket{\textbf{0011}}, \ket{\textbf{1101}} \otimes \ket{\textbf{0010}} \}$, is no longer separable, resulting in a lower fidelity than in the cluster setting.
Through the above mechanism, the decomposition setting affects the fidelity, where matching between the tensor network and target states is a crucial factor, being in analogy with the discussion in the area law with tensor networks.
Note that the number of legs limits the dimension of the basis for state preparation.

HTN+VQE results with the cluster setting in Fig.~\ref{fig: Ana_Dist_Hei_2column.eps}(d) show that the coefficients for the above 12 basis states are of the same magnitude as $c_1,c_2,\dots,c_4$.
In contrast, the even-odd setting gives a much less accurate wave function; half of the basis states in Eq.~(\ref{Eq: cluster representation}) was negligibly small in magnitude, and the large coefficient of $\ket{0101} \otimes \ket{0101}$ is 0.78, leading to a different distribution from the exact.

Next, Fig.~\ref{fig: Ana_Dist_Hei_2column.eps}(e) shows the distribution of the no-decomposition setting, where the distributions of the ground state and cluster settings are reproduced from Fig.~\ref{fig: Ana_Dist_Hei_2column.eps} (a) for comparison.
The depth is set to $d_N=6$ such that the number of parameters in the no-decomposition and cluster settings becomes comparable, that is, 50 and 48, respectively.
The distribution for the no-decomposition setting deviates from that for the exact wave function, and half of the basis states in Eq.~(\ref{Eq: cluster representation}) were negligibly small in magnitude.
The no-decomposition setting would achieve higher performance than HTN. However, when the number of ansatz parameters is restricted, HTN may perform better if we find an ansatz and decomposition suited to the structure of the system.
In fact, the coefficient of the basis state $\ket{00101101}$, which has the $10$-th largest magnitude in the no-decomposition setting, is 0.14; in contrast, the corresponding values are -0.057 in the ground state and -0.00018 in the cluster setting. 
Thus, the cluster setting can efficiently prepare a trial wave function that incorporates system correlations with fewer parameters by eliminating the basis states with a small contribution to the ground state of the system.
Note that these observations are verified by calculating over 10 random seeds.

As summary, we proposed an algorithm HTN+QMC that combines QC-QMC with HTN for calculating quantum chemistry problems beyond the size of a quantum device.
As demonstrated on the benchmark models, our algorithm exhibits energy with smaller variance  than QMC.
We developed the gate-efficient pseudo-Hadamard test technique involving an ancilla qubit for conducting HTN+QMC experiments on real devices; the at most five qubits experiment showed an accuracy comparable to the statevector simulation for the hydrogen plane models (8 qubit model) and MABI (12 qubit model). 

While this study assumed the target system size to be larger than that of the quantum device, there may be cases in which the proposed algorithm should be used even if the target system size equals that of the quantum device.
For example, a quantum computer with thousands of qubits has appeared~\cite{Matthews2021-pl}. However, an accurate solution may not be obtained when calculating a chemical model of that size owing to noise in the quantum device.
In such a case, decomposing the system into subsystems of about hundreds of qubits would lead to a more accurate result than that obtained by directly executing on a thousand qubits.
In addition, the tree tensor that we have assumed has the advantage that the computation of the lower tensor can be performed in parallel.

The research on QC-QMC has not yet been extensive, and applying techniques developed for NISQ devices to QC-QMC is a possible future research item, as in HTN in this study.
HTN is also an intriguing research area, where there are several classical decomposition approaches, which may inspire ideas for efficient quantum algorithms to deal with the correlations between subsystems  ~\cite{Jimenez-Hoyos2015-sh, Abraham2020-ih, Li2021-uk, Parker2013-di}.
Another interesting direction is applying QC-QMC to fields outside the electronic structure calculations such as machine learning for designing advanced materials.

\section{Methods}
\label{sec: method}

\subsection{Variational quantum eigensolver}
\label{sec: VQE}

Variational quantum eigensolver (VQE) takes a variational approach to obtain the ground state of a given Hamiltonian. Here, a trial wave function $\ket{\psi(\Vec{\theta})}$ is generated by a quantum circuit with variational parameters $\Vec{\theta}$, which is repeatedly updated using the classical computer until a termination condition is satisfied, e.g., the expectation value of the Hamiltonian $\bra{\psi(\Vec{\theta})} H \ket{\psi(\Vec{\theta})}$ takes a minimum. 
The expressibility of $\ket{\psi(\Vec{\theta})}$ depends on the quantum circuit, called ansatz.
There are several problem-inspired ansatze, such as the unitary coupled cluster ansatz~\cite{Taube2006-jj} and Hamiltonian variational ansatz~\cite{Wecker2015-fu}. However, despite the high performance, the ansatze requires deeper quantum gates. 
In real device experiments, compromised strategy with hardware-efficient ansatze~\cite{Kandala2017-lh} is often considered.

The Hamiltonian $H$ in the electronic structure problems can be represented as
\begin{equation}     
\begin{aligned}     
H = \sum_a c_a \bigotimes_b P_{ab},
\label{Eq: Hamiltonian}
\end{aligned} 
\end{equation} 
where $c_a$ is the $a$-th coefficient of $H$ and $P_{ab} \in \{X, Y, Z, I\}$ is the $a$-th Pauli or identity operator on the $b$-th site.
In this study, we assume $c_a \in \mathbb{R}$ in all models (although $c_a \in \mathbb{C}$ in general).
The number of terms is estimated to be $\order{N_{so}^4}$ with $N_{so}$ being the number of spin-orbitals~\cite{Tilly2022-pb}.
We choose the Jordan-Wigner mapping~\cite{Jordan1928-hf} for the fermion-qubit translation.

\subsection{Quantum computing quantum Monte Carlo}
\label{sec: QCQMC}

In QC-QMC~\cite{Huggins2022-ly}, a stochastic algorithm such as the imaginary-time evolution is used to iteratively update the discretized coefficients of the wave function, i.e., walkers.
As we explained in Sec.~\ref{introduction}, 
This study will validate a version of QC-QMC that uses quantum computation only for energy evaluation~\cite{Xu2023-sn}.
The following projected energy $E_\mathrm{proj}$ is used as a common energy estimator: 
\begin{equation}
\begin{aligned}
E_\mathrm{proj} &= \frac{\bra{\xi} H \ket{\psi_\mathrm{QMC}}}{\braket{\xi}{\psi_\mathrm{QMC}}},
\label{Eq: projected energy}
\end{aligned}
\end{equation}
where $\ket{\xi}$ is a trial wave function. $\ket{\psi_\mathrm{QMC}}$ denotes 
a wave function generated by QMC, defined as
\begin{equation}
\begin{aligned}
\ket{\psi_\mathrm{QMC}} &= \sum_{h} w_{h} \ket{\phi_{h}},
\label{Eq: fciqmc}
\end{aligned}
\end{equation}
where $w_{h}$ is the ${h}$-th coefficient and is expressed using discretized units (walkers).  $\ket{\phi_{h}}$ is the ${h}$-th orthonormal basis state as in a Slater determinant and can be represented by a binary string, i.e., a computational basis. 
The procedure of generating the wave function $\ket{\psi_\mathrm{QMC}}$ depends on the QMC method. 
In this work, we choose FCIQMC. See Appendix~\ref{sec: fciqmc} for process details. 
$E_\mathrm{proj}$ is not equivalent to the expectation value of the Hamiltonian $\mel{\psi_\mathrm{QMC}}{H}{\psi_\mathrm{QMC}}/\braket{\psi_\mathrm{QMC}}{\psi_\mathrm{QMC}}$ (that is, not the pure estimator); however, we can obtain the exact ground state energy $E_g$ when 
$\ket{\psi_\mathrm{QMC}}$ approaches the ground state $\ket{\psi_g}$, 
i.e., $\ket{\psi_\mathrm{QMC}} \sim \ket{\psi_g}$ as 
\begin{equation}
\begin{aligned}
E_\mathrm{proj} &\sim \frac{\bra{\xi} H \ket{\psi_g}}{\braket{\xi}{\psi_g}} = E_g,
\label{Eq: projected energy approximation}
\end{aligned}
\end{equation}
where $H \ket{\psi_g}=E_g \ket{\psi_g}$ is used.

The trial function $\ket{\xi}$ is fixed throughout the QMC, and a more sophisticated $\ket{\xi}$ can be used to lower the statistical error.
A trivial case is that, if the trial wave function coincides with the ground state, i.e., $\ket{\xi}=\ket{\psi_g}$, then we can obtain $E_g$ for any $\ket{\psi_\mathrm{QMC}}$ with zero variance~\cite{Apaja2018-fi}.  See Appendix~\ref{sec: statistical analysis of Emix} for derivation details. 
For example, the Hartree-Fock state, the linear combination of mean-field states, and the Jastrow-type states~\cite{Austin2012-qi} are used as the trial wave functions conveniently available in classical computations. 
Therefore, we can considerably minimize energy estimation errors by preparing a suboptimal trial wave function in the exponentially large Hilbert space.

In QC-QMC, trial wave function $\ket{\xi}$ is prepared by a quantum algorithm. 
Specifically, by decomposing $\ket{\psi_\mathrm{QMC}}$, Eq.~(\ref{Eq: projected energy}) is rewritten as
\begin{equation}
\begin{aligned}
E_\mathrm{proj} &= \frac{\sum_{h {l}} w_{h} \braket{\xi}{\phi_{l}} \bra{\phi_{l}} H \ket{\phi_{h}}}{\sum_{{h}} w_{h} \braket{\xi}{\phi_{h}}}.
\label{Eq: projected energy2}
\end{aligned}
\end{equation}
The matrix elements $\bra{\phi_{l}} H \ket{\phi_{h}}$ can be obtained by trivial classical calculations. 
In contrast, quantum computation is required to calculate the overlap ($\braket{\xi}{\phi_{l}}$ and $\braket{\xi}{\phi_{h}}$) for which the Hadamard test or classical shadow can be applied~\cite{Huang2020-ld, Zhao2021-kw}, henceforth, $\braket{\xi}{\phi_{l}}$ and $\braket{\xi}{\phi_{h}}$ will both be described as $\braket{\xi}{\phi_{h}}$.
We discuss in Appendix~\ref{sec: statistical analysis of Emix} that the variance of projected energy will decrease with the fidelity of the trial wave function in a simplified case. 
In this study, fidelity is defined as the square of the overlap between the target and ground states, assuming zero noise.

\subsection{Hybrid tensor network}
\label{sec: HTN}

In the two-layer QQTN, the original $nk$-qubit system is decomposed into $k$ subsystems of $n$ qubit states $\ket{\varphi^{i_m}}$, which are integrated with the $k$-qubit state $\ket{\psi}$. These states can be associated with tensors, defined as the lower tensor $\varphi_{\Vec{j}_m}^{i_m} = \braket{\Vec{j}_m}{\varphi^{i_m}}$ and upper tensor $\psi_{\Vec{i}} = \braket{\Vec{i}}{\psi}$ with the vector indices $\Vec{j}_m = j_{m1} j_{m2} \dots j_{mn}$ and $\Vec{i} = i_1 i_2 \dots i_k$ in $n$-qubit and $k$-qubit binary strings, respectively. 
The wave function $\ket{\psi_\mathrm{HTN}}$ is then decomposed into tensor products of $\ket{\varphi^{i_m}}$
\begin{equation}
 \begin{aligned}
\ket{\psi_\mathrm{HTN}} &=\sum_{\Vec{i}} \psi_{\Vec{i}} \bigotimes_{m=1}^{k} \ket{\varphi^{i_m}},
\label{Eq: tensornetwork0_re}
\end{aligned} 
\end{equation}
which is the same as Eq.~\eqref{Eq: tensornetwork0} although omitting the parameters,  where the number of vector coefficients $\{\psi_{\Vec{i}}\}$ is $2^{Lk}$ with $L$ denoting the number of bits (legs) used to construct each $i_m$. 
$L=1$ is adopted in this study. 
If $L=n$, $\ket{\psi_\mathrm{HTN}}$ becomes the general $nk$-qubits wave function.
When $L \ll n$, $\ket{\psi_\mathrm{HTN}}$ lives in a subspace much smaller than the entire $2^{nk}$ dimensional Hilbert space; however, it can be larger than the subspace consisting only of the classical tensor due to the exponentially large ranks of $\psi_{\Vec{i}}$ and $\varphi^{i_m}_{\Vec{j}_m}$. Moreover, the performance of QQTN depends on the decomposition setting of the system, which determines the entanglement between the subsystems.

Tensor-network quantum circuits can be implemented in various ways.
For example, we can define a lower tensor as $\ket{\varphi^{i_m}} = U_{Lm}\ket{i_m}\ket{0}^{\otimes n-1}$, $U_{Lm}\ket{i_m}^{\otimes n}$, $U_{Lm}^{i_m}\ket{0}^{\otimes n}$, etc., where the first two formulations give the index in initial states and the third a unitary matrix using unitary matrices $U_{Lm}$ and $U_{Lm}^{i_m}$. 
In the present study, we choose $\ket{\varphi^{i_m}} = U_{Lm}\ket{i_m}\ket{0}^{\otimes n-1}$ and $\ket{\psi} = U_U\ket{0}^{\otimes k}$ for lower and upper tensors, respectively.

In our QQTN formulation, an observable is defined as the tensor product $O = \bigotimes_{m, r} O_{mr}$, where $O_{mr}$ is observable on the $r$-th qubit of the $m$-th subsystem (lower tensor).
Transition amplitude is then defined as
\begin{equation}
\begin{aligned}
T = \mel{\psi_\mathrm{HTN}^{(1)}}{O}{\psi_\mathrm{HTN}^{(2)}},
\label{Eq: ExpectationNonHermitian1_re}
\end{aligned}
\end{equation}
which is the same as Eq.~\eqref{Eq: ExpectationNonHermitian1} although omitting the parameters, where $\ket{\psi_\mathrm{HTN}^{(l)}} =\sum_{\Vec{i}} \psi_{\Vec{i}}^{(l)} \bigotimes_{m} \ket{\varphi^{i_m (l)}} (l=1,2)$ are the two different of states for $\ket{\psi_\mathrm{HTN}}$.
As explained in the next section, $T$ is used in VQE and QMC to evaluate an observable and overlap, respectively.

We first calculate $N^{i_m'(1) i_m (2)} = \mel{\varphi^{i_m' (1)}}{\bigotimes_r O_{mr}}{\varphi^{i_m (2)}}$ for the lower tensors by quantum computations, construct $N_m = \mqty(N^{00} & N^{01} \\ N^{10} & N^{11})$, which is a trivial classical process, and then integrate the results as $T = \mel{\psi^{(1)}}{\bigotimes_m N_m}{\psi^{(2)}}$ on the upper tensor by quantum computations.
Appendix~\ref{sec: tensor contraction procedure} shows the details of this procedure, which is based on the Hadamard test circuit as in Ref.~\cite{Kanno2021-zn}.
Our algorithm requires only $\max(n, k)$ qubits except for ancilla qubits.
In this study, only $8k+2$ terms are measured to calculate $T$; i.e., the overhead for calculating the expectation value is a linear scale for the system size $nk$.
Note that the number of measurements is $2 \times 4^L k +2$ in general.

\subsection{Proposed algorithm: HTN+QMC}
\label{sec: HTNQCQMC}

The procedure of HTN+QMC consists of two steps.

\begin{enumerate}
    \item Perform VQE by minimizing $\bra{\psi_\mathrm{HTN}} H \ket{\psi_\mathrm{HTN}}$ to obtain trial wave function $\ket{\xi}=\ket{\psi_\mathrm{HTN}}$.     
    \item Perform QC-QMC by using the obtained trial wave function 
    $\ket{\psi_\mathrm{HTN}}$; the quantum computer is used to compute $E_\mathrm{proj}$ 
    to accurately estimate the ground state energy. 
\end{enumerate}
Here, we will denote HTN+VQE when only the first step is mentioned.
HTN+QMC can be performed for $nk$-qubit system by using only $\order{\mathrm{max}(n, k)}$ qubits except for ancilla qubits. 
The Hamiltonian and projected energy can be evaluated in both steps by calculating $T$ in Eq.~(\ref{Eq: ExpectationNonHermitian1_re}).

In the first step, by splitting index $b$ in Eq.~(\ref{Eq: Hamiltonian}) 
into indices $m$ and $r$, we can rewrite $H$ and its expectation value as
\begin{equation}
\begin{aligned}
H &= \sum_a c_a \bigotimes_{m,r} P_{amr},
\label{Eq: rewrite Hamiltonian1}
\end{aligned}
\end{equation}
and
\begin{equation}
\begin{aligned}
\bra{\psi_\mathrm{HTN}} H \ket{\psi_\mathrm{HTN}} &= \sum_a c_a 
\bra{\psi_\mathrm{HTN}} \bigotimes_{m,r} P_{amr} \ket{\psi_\mathrm{HTN}},
\label{Eq: rewrite Hamiltonian2}
\end{aligned}
\end{equation}
respectively. 
The expectation value is evaluated through Eq.~(\ref{Eq: ExpectationNonHermitian1_re}) by setting $\ket{\psi_\mathrm{HTN}^{(1)}} = \ket{\psi_\mathrm{HTN}^{(2)}} = \ket{\psi_\mathrm{HTN}}$ and replacing $O_{mr}$ with $P_{amr}$; coefficient index $a$ is implicitly included in the expression for $O_{mr}$.
The wave function $\ket{\psi_\mathrm{HTN}}$ is prepared by parameterized quantum circuits equivalent to the unitary matrices $U_{Lm}$ and $U_{U}$.

In the second step, the overlap $\braket{\xi}{\phi_{h}} = \braket{\psi_\mathrm{HTN}}{{\phi_{h}}}$ in Eq.~(\ref{Eq: projected energy2}) is calculated by substituting $\ket{\psi_\mathrm{HTN}^{(1)}}=\ket{\psi_\mathrm{HTN}}$,  $\ket{\psi_\mathrm{HTN}^{(2)}}=\ket{\phi_{h}}$, and $O=I^{\otimes nk}$ in $T$ in Eq.~(\ref{Eq: ExpectationNonHermitian1_re}), where the circuit parameters are fixed to the values obtained in the first step.
Specifically, we can prepare an arbitrary basis state $\ket{\phi_{h}} = \bigotimes_{m,r} \ket{j_{mr}(h)} = (\bigotimes_{m,r} X^{j_{mr}(h)}) \ket{0}^{\otimes nk}$ by setting $U_U^{(2)} = I^{\otimes k}$ and $U_{Lm}^{(2)} = \bigotimes_m X^{j_{mr}(h)}$, where $j_{mr}(h)$ is a function of $h$ and takes a value on $\{0,1\}$, and $X$ is the Pauli $X$ operator. 
Using the overlaps $\braket{\psi_\mathrm{HTN}}{\phi_{h}}$ of all $\ket{\phi_{h}}$, corresponding to the 
orthnormal basis states appearing through the QMC wave function and matrix elements $\bra{\phi_{l}} H \ket{\phi_{h}}$ during the QMC execution at hand, we can perform QMC through iterative evaluations of the projected energy (Eqs.~(\ref{Eq: projected energy}) and (\ref{Eq: projected energy2})) in principle.
The scale of the measurement cost in the QMC step is $\order{N_{so}^4 N_W^* (\tau_\mathrm{fin}/\Delta\tau)2^{2L}k/\varepsilon^2}$, where $N_{so}$ comes from $\bra{\phi_{h}} H \ket{\phi_{h'}}$, $N_W^*$ is the maximum number of walkers that can be taken in a QMC iteration, $\tau_\mathrm{fin}$ is a total time in QMC, $\Delta\tau$ is a time step in QMC, $2^{2L}k$ comes from the overhead of HTN contraction, $\varepsilon$ is an additive error~\cite{Kiser2023-hi, Yuan2021-ih}.
We mention that in our actual run, for simplicity, the overlaps of all orthonormal basis states were computed before QMC in the statevector simulation. For real device execution, the overlaps of the states for all target electron numbers were computed.

\begin{acknowledgements}
    This work was supported by MEXT Quantum Leap Flagship Program Grants No. JPMXS0118067285 and No. JPMXS0120319794, JSPS KAKENHI Grant No. JP20K05438, and COI-NEXT JST Grant No. JPMJPF2221, and partly supported by UTokyo Quantum Initiative.
    A part of this work was performed for Council for Science, Technology and Innovation (CSTI), Cross-ministerial Strategic Innovation Promotion Program (SIP), “Promoting the application of advanced quantum technology platforms to social issues”(Funding agency : QST).
    The part of calculations were performed on the Mitsubishi Chemical Corporation (MCC) high-performance computer (HPC) system “NAYUTA”, where “NAYUTA” is a nickname for MCC HPC and is not a product or service name of MCC.
    We acknowledge the use of IBM Quantum services for experiments in this paper. The views expressed are those of the authors, and do not reflect the official policy or position of IBM or the IBM Quantum team.
    S.K. thanks Kenji Sugisaki and Rei Sakuma for the technical discussion, and Hajime Sugiyama for the technical support on HPC.
    We would like to thank Editage (www.editage.jp) for English language editing.
\end{acknowledgements}

\putbib[bibHTNQMC]
\end{bibunit}
\clearpage
\appendix

\begin{bibunit}[apsrev4-1]
\begin{figure*}[]
 \centering
 \includegraphics[width=1\textwidth]{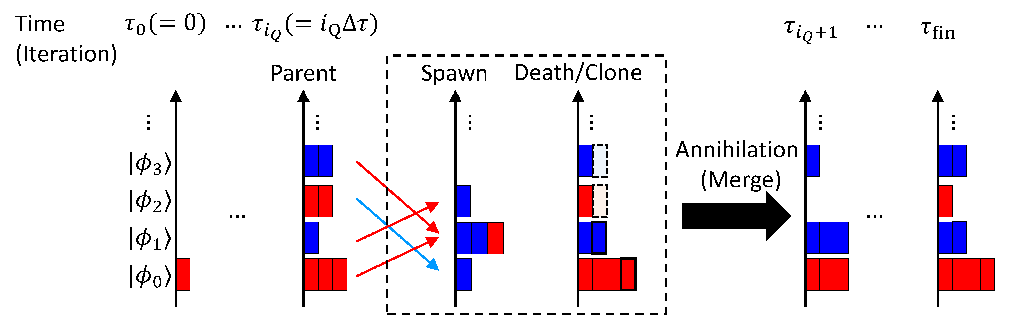}
\caption{Procedure of FCIQMC. The walker of $\ket{\phi_{h}}$ with a plus (minus) sign is represented as a red (blue) rectangle.}
 \label{fig: FCIQMC}

\end{figure*}

\section{Procedure of full configuration interaction quantum Monte Carlo}
\label{sec: fciqmc}
In FCIQMC, the wave function $\ket{\psi_\mathrm{QMC}}$ is defined as
\begin{equation}
\begin{aligned}
\ket{\psi_\mathrm{QMC}} &= \sum_{h} w_{h} \ket{\phi_{h}},
\label{Eq: fciqmc2}
\end{aligned}
\end{equation}
where $w_{h}$ is the real coefficient of $h$-th orthonormal basis state (such as the Slater determinant) $\ket{\phi_{h}}$.  
Those coefficients will be updated iteratively through the imaginary-time evolution according to
\begin{equation}
\begin{aligned}
\frac{dw_{h'}}{d\tau} &= - \sum_{h} (H_{h' h} - S \delta_{h' h}) w_{h},
\label{Eq: ite1}
\end{aligned}
\end{equation}
which can be written in a discrete form as 
\begin{equation}
\begin{aligned}
\Delta w_{h'} &= - \sum_{h \neq h'} H_{h' h} \Delta\tau \times w_{h} - (H_{hh} - S) \Delta\tau \times  w_{h},
\label{Eq: ite2}
\end{aligned}
\end{equation}
where $H_{h' {h}} =  \bra{\phi_{h'}} H  \ket{\phi_{h}}$ is a matrix element of the Hamiltonian, $\delta_{h' h}$ is the Kronecker delta, $\tau$ is an imaginary time, $\Delta\tau$ is a time increment, and $S$ is an energy shift.
The first and second terms are contributions from the off-diagonal and diagonal elements of the Hamiltonian, respectively.

The procedure of FCIQMC is illustrated in Fig.~\ref{fig: FCIQMC}.
The walker with the positive (negative) sign is depicted as red (blue) rectangular.
The imaginary time in the $i_Q$-th iteration is denoted by $\tau_{i_Q}$.
For the walkers in the $i_Q$-th iteration, called the parent walkers, the following operations are performed to obtain the $i_Q+1$-th walkers.
\begin{itemize}
    \item Spawning step [the off-diagonal terms in Eq.~(\ref{Eq: ite2})]:   
    For each parent walker with index ${h}$, a new walker with index ${h'}$ is generated with a probability of $\abs{H_{h' {h}}}\Delta\tau$, to which $\sgn(-H_{h' {h}})$ is assigned.
    If $\abs{H_{h' {h}}}\Delta\tau > 1$, however, $\floor{\abs{H_{h' {h}}}\Delta\tau}$ walkers and a walker are generated with a probability of 1 and $\abs{H_{h' {h}}}\Delta\tau - \floor{\abs{H_{h' {h}}}\Delta\tau}$, respectively.
    
    \item Death/cloning step [the diagonal term in Eq.~(\ref{Eq: ite2})]: 
    For each parent walker with an index ${h}$, remove (copy) a walker of the same index with probability $\abs{(H_{h' {h}} - S)} \Delta\tau$ if the sign of $(H_{h' {h}} - S) \Delta\tau$ is positive (negative).
    If $\abs{(H_{h' {h}} - S)} \Delta\tau > 1$, remove (copy) the walker with probability 1.

    \item Annihilation step:
    Merge the walkers obtained in the spawning and death/cloning steps.
    The walkers with opposite signs are canceled out.
\end{itemize}
In the death/cloning step, the initial value of $S$ is a constant (e.g., the value of the Hartree-Fock energy), and when the number of walkers exceeds the set value $N_\mathrm{shift}$, $S$ is updated every $A$ iteration as $S(\tau_{i_Q})=S(\tau_{i_Q}-A\Delta\tau)-\frac{\zeta}{A\Delta\tau}\ln\frac{N_W(\tau_{i_Q})}{N_W(\tau_{i_Q}-A\Delta\tau)}$ so that the number of the walkers remains constant (called variable shift mode), where $S(\tau_{i_Q})$ is $S$ on the time $\tau_{i_Q}$, $\zeta$ is a damping parameter, and $N_W(\tau_{i_Q})$ is the number of walkers on $\tau_{i_Q}$.
We note that although the computational cost, i.e., the number of walkers required by this algorithm, may increase exponentially with the size of the system in general, there are several proposals to make the algorithms executable for practical systems~\cite{Booth2009-lv, Cleland2010-au, Petruzielo2012-rk, Guther2020-um}.
For example, the number of walkers required for $\mathrm{N_2}$ molecule can be reduced by three orders of magnitude ($10^8$ to $10^5$) by restricting spawning only on the basis state for which the number of walkers exceeds a certain threshold~\cite{Cleland2010-au}.

\section{Statistical analysis of projected energy in simple cases}
\label{sec: statistical analysis of Emix}
Here, in the simple cases, we discuss the statistical error of the projected energy, which arises from preparing the trial wave function by a quantum algorithm.
Let $\epsilon$ and $(1-F)$ denote the deviations of the wave function generated by QMC $\ket{\psi_\mathrm{QMC}}$ and the trial wave function $\ket{\xi}$ from the ground state $\ket{\psi_g}$, the wave functions can be represented as
\begin{equation}
\begin{aligned}
\ket{\psi_\mathrm{QMC}(\epsilon)} &= (1-\epsilon)\ket{\psi_g} + \epsilon \sum_s u_s \ket{\psi_s^{\perp}},
\label{Eq: deviated wf1}
\end{aligned}
\end{equation}
and
\begin{equation}
\begin{aligned}
\ket{\xi(F)} &= F\ket{\psi_g}+(1-F) \sum_{s'} v_{s'} \ket{\psi_{s'}^{\perp}},
\label{Eq: deviated wf2}
\end{aligned}
\end{equation}
respectively. 
$F^2$ is a fidelity of $\ket{\xi(F)}$, $\ket{\psi_s^{\perp}}$ and $\ket{\psi_{s'}^{\perp}}$ are the orthogonal states for $\ket{\psi_g}$, $u_s$ and $v_{s'}$ are the coefficients, and $\epsilon, F \in [0,1]$.
The projected energy $E_\mathrm{proj}(\epsilon; F)$ can be described as
\begin{equation}
\begin{aligned}
E_\mathrm{proj}(\epsilon; F) &= \frac{\bra{\xi(F)} H \ket{\psi_\mathrm{QMC}(\epsilon)}}{\braket{\xi(F)}{\psi_\mathrm{QMC}(\epsilon)}}\\
&= \frac{(1-\epsilon)FE_g + \epsilon(1-F) \sum_s E_s u_s v_s}{(1-\epsilon)F+\epsilon(1-F)\sum_s u_s v_s},
\label{Eq: eval projected energy1}
\end{aligned}
\end{equation}
with the ground state energy $E_g$ and eigenvalue of $\ket{\psi_s^{\perp}}$ denoted by $E_s$.
Below, we assume $E_g = 0$ without loss of generality.
Then the mean and variance of $E_\mathrm{proj}(\epsilon; F)$, $\mathbb{E}(E_\mathrm{proj}(\epsilon; F))$ and $Var(E_\mathrm{proj}(\epsilon; F))$, are respectively represented as
\begin{equation}
\begin{aligned}
\mathbb{E}(E_\mathrm{proj}(\epsilon; F)) &= \mathbb{E}\left(\frac{\epsilon(1-F) \sum_s E_s u_s v_s}{(1-\epsilon)F+\epsilon(1-F)\sum_s u_s v_s}\right),
\label{Eq: eval projected energy1-2}
\end{aligned}
\end{equation}
and
\begin{equation}
\begin{aligned}
Var(E_\mathrm{proj}(\epsilon; F)) &= \mathbb{E}(E_\mathrm{proj}(\epsilon; F)^2) - \mathbb{E}(E_\mathrm{proj}(\epsilon; F))^2\\
&=\mathbb{E}\left(\frac{\{\epsilon(1-F)\sum_s E_s u_s v_s\}^2}{\{(1-\epsilon)F+\epsilon(1-F)\sum_s u_s v_s\}^2}\right)\\
&-\mathbb{E}\left(\frac{\epsilon(1-F) \sum_s E_s u_s v_s}{(1-\epsilon)F+\epsilon(1-F)\sum_s u_s v_s}\right)^2,
\label{Eq: eval projected energy1-3}
\end{aligned}
\end{equation}
where the integral variable of the expected value is the QMC time step, on which $\epsilon$ and $u_s$ depend.
Now consider $\mathbb{E}(E_\mathrm{proj}(\epsilon; F))$ and $Var(E_\mathrm{proj}(\epsilon; F))$ in the two cases.
In the first case of $F=1$, both $\mathbb{E}(E_\mathrm{proj}(\epsilon; F))$ and $Var(E_\mathrm{proj}(\epsilon; F))$ are trivially zero, that is, if the trial wave function is the ground state, $E_\mathrm{proj}$ is equal to the ground state energy regardless of $\ket{\psi_\mathrm{QMC}(\epsilon)}$ (excluding $\epsilon=1$).

In the second case, where $E_\mathrm{proj}(\epsilon; F)$ fluctuates around the ground state energy in a small amount, assuming $\mathbb{E}(E_\mathrm{proj}(\epsilon; F)) = 0$ and the expected value of the higher order terms of $\epsilon$ be negligible, the variance becomes
\begin{equation}
\begin{aligned}
Var(E_\mathrm{proj}(\epsilon; F)) &=\mathbb{E}\left(\frac{\{\epsilon(1-F)\sum_s E_s u_s v_s\}^2}{\{(1-\epsilon)F+\epsilon(1-F)\sum_s u_s v_s\}^2}\right)\\
&=\mathbb{E}\left(\frac{(1-F)^2}{F^2}(\sum_s E_s u_s v_s)^2 \epsilon^2 + \order{\epsilon^3}\right)\\
&=\order{\left( \frac{1-F}{F} \right)^2}.
\label{Eq: eval projected energy2}
\end{aligned}
\end{equation}
$\left( \frac{1-F}{F} \right)^2$ is monotonically decreasing, and thus when the fidelity of the trial wave function in the quantum algorithm is larger than that in the classical algorithm, we can expect a decrease in the variance of the projected energy.

\begin{figure}[]
 \includegraphics[width=1\columnwidth]{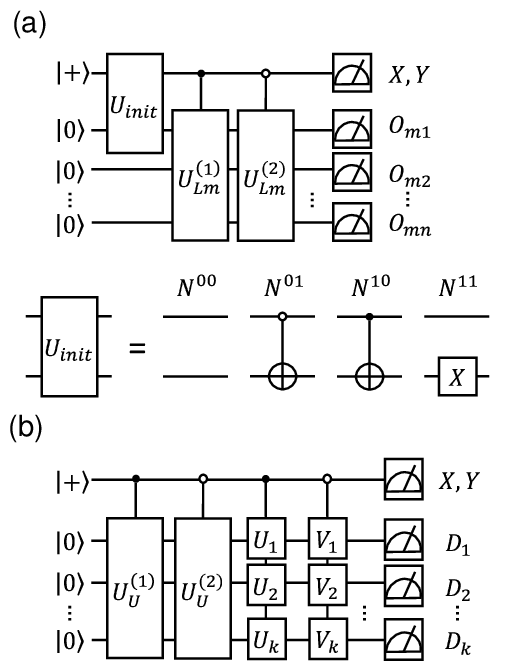}
\caption{Quantum circuits for calculating $T$. The topmost line represents an ancilla qubit and the other lines represent system qubits. (a) Calculation for the lower tensor. (b) Calculation for the upper tensor.}
 \label{fig: Circuit}
\end{figure}

\section{Calculation of the transition amplitude by using HTN}
\label{sec: tensor contraction procedure}
We explain the procedure for calculating $T$ in Eq.~(\ref{Eq: ExpectationNonHermitian1_re}) based on the Hadamard test~\cite{Kanno2021-zn}, which is reproduced as 
\begin{equation}
\begin{aligned}
T &= \sum_{\Vec{i}'~\Vec{i}} \psi_{\Vec{i}'}^{(1)*} \psi_{\Vec{i}}^{(2)} \prod_{m=1}^k \mel{\varphi^{i_m'(1)}}{\bigotimes_r O_{mr}}{\varphi^{i_m(2)}}\\
&= \sum_{\Vec{i}'~\Vec{i}} \psi_{\Vec{i}'}^{(1)*} \psi_{\Vec{i}}^{(2)} \prod_{m=1}^k N^{i_m' i_m}.
\label{Eq: ExpectationNonHermitian2}
\end{aligned}
\end{equation}
The first part includes the calculations of the matrices $N^{i_m' i_m}$ for the lower tensors.
Figure~\ref{fig: Circuit}(a) shows the quantum circuit for this purpose when the state in the lower tensor is defined by $\ket{\varphi^{i_m(l)}} = U_{Lm}^{(l)} \ket{i_m(l)}\ket{0}^{\otimes n-1}$. 
$U_{init}$ is one of the circuits in the lower panel, which is selected for each of the matrix elements $N^{00}, N^{01}, N^{10}$, or $N^{11}$, and creates the superposed state $\frac{1}{\sqrt{2}}( \ket{1}\ket{i_m'(1)}\ket{0}^{\otimes n-1}+\ket{0}\ket{i_m(2)}\ket{0}^{\otimes n-1})$, which is then transformed to $\frac{1}{\sqrt{2}}( \ket{1}U_{Lm}^{(1)}\ket{i_m'(1)}\ket{0}^{\otimes n-1}+\ket{0}U_{Lm}^{(2)}\ket{i_m(2)}\ket{0}^{\otimes n-1})$ through $U_{Lm}^{(1)}$ and $U_{Lm}^{(2)}$ operations, and measured in the measurement basis corresponding to $O_{mr}$.
We can obtain the real and imaginary parts of $N^{i_m' i_m}$ ($\Re(N^{i_m' i_m})$ and $\Im(N^{i_m' i_m})$) by specifying respective $X$ and $Y$ for the measurement basis on the ancilla qubit.
Calculation of $N^{i_m' i_m}$ for each lower tensor, therefore, requires $4\times2=8$ different measurements, where factor 4 comes from $N^{00}, N^{01}, N^{10}$, and $N^{11}$, and the factor 2 from $\Re(N^{i_m' i_m})$ and $\Im(N^{i_m' i_m})$.

The second part is the calculation of the upper tensor.  In terms of the upper tensor state, $T$ can be rewritten as
\begin{equation}
\begin{aligned}
T &= \sum_{\Vec{i}'~\Vec{i}} \psi_{\Vec{i}'}^{(1)*} \psi_{\Vec{i}}^{(2)} \prod_{m=1}^k N^{i_m' i_m} \\
&=  \bra{\psi^{(1)}} \bigotimes_{m=1}^k N_m \ket{\psi^{(2)}},
\label{Eq: ExpectationNonHermitian3}
\end{aligned}
\end{equation}
where the $2 \times 2$ non-Hermitian matrix $N_m$ is reformulated by the singular value decomposition (SVD), 
\begin{equation}
\begin{aligned}
T &= \bra{\psi^{(1)}} \bigotimes_{m=1}^k U_m^{\dag} D_m V_m \ket{\psi^{(2)}},
\label{Eq: ExpectationNonHermitian4}
\end{aligned}
\end{equation}
where $N_m = U_m^{\dag} D_m V_m$, $U_m$ and $V_m$ are unitary matrices, and $D_m$ is a diagonal matrix.
Since $N_m$ is a $2 \times 2$ matrix, the SVD can be executed classically.
The quantum circuit for calculating Eq.~(\ref{Eq: ExpectationNonHermitian4}) is shown in Figure~\ref{fig: Circuit}(b), where the superposed state $\frac{1}{\sqrt{2}}(\ket{1}U_{U}^{(1)} \ket{0}^{\otimes k}+\ket{0}U_{U}^{(2)} \ket{0}^{\otimes k})$ is transformed to $\frac{1}{\sqrt{2}}(\ket{1}\bigotimes_{m=1}^k U_m U_{U}^{(1)} \ket{0}^{\otimes k}+\ket{0} \bigotimes_{m=1}^k V_m U_{U}^{(2)} \ket{0}^{\otimes k})$ through $\bigotimes_{m=1}^k U_m$ and $\bigotimes_{m=1}^k V_m$, and measured in the computational basis. The measurement result, $\ket{0}$ or $\ket{1}$ gives the diagonal element of $D_m$, i.e., $(D_m)_{00}$ or $(D_m)_{11}$, respectively~\cite{Kanno2021-zn}.
The real and imaginary parts of $T$ can be obtained by specifying $X$ and $Y$ for the measurement basis on the ancilla qubit, respectively.
The calculation for the upper tensor thus requires measurements on two different bases. 
In total, the calculations of the lower and upper tensor require $8k+2$ sets of measurements, i.e., the overhead for calculating the expectation value is a linear scale for the system size $nk$.
Note that when the Hamiltonian and wave function are real as in this study, the calculations of the imaginary parts in the lower and upper tensors are unnecessary, reducing the overhead to $4k+1$.
We additionally note that the QQTN can be expressed as a $nk$-qubit quantum circuit $(\prod_m U_{Lm}^{(l)}) U_U^{(l)} (\bigotimes_m \ket{0}^{\otimes n})$ under the current assumption of $\ket{\varphi^{i_m (l)}}$ and $\ket{\psi^{(l)}}$, where $U_{Lm}^{(l)}$ operates qubits corresponding to $m$-th subsystem (i.e., $m$-th $\ket{0}^{\otimes n}$), and $U_U^{(l)}$ operates the qubits corresponding the first qubit of each subsystem~\cite{Yuan2021-ih}. As a final remark, the robustness of the barren plateau in this type of circuit has been reported~\cite{Martin2022-pj}. 

\begin{algorithm}[]
\caption{outline of HTN+FCIQMC with sparse basis construction}
\label{Alg: HTN+FCIQMC}
\SetKwInput{KwInput}{Input} 
\SetKwFunction{FMain}{Main}
\DontPrintSemicolon
  
\KwInput{Hamiltonian $H$, unitary gates from HTN+VQE $(U_U$, $U_{Lm})$,  total time $ \tau_\mathrm{fin}$, time step $\Delta\tau$, energy shift $S$, initial parent walker distribution}

\For{iter = $0:(\tau_\mathrm{fin}/\Delta\tau)$}{
    \tcp{Spawning step}
    \For{each walker indexed to $\Vec{I}$ in the parent walker distribution}{
    Generate $\abs{H_{\Vec{I}' \Vec{I}}}$ distribution by the Bayesian estimation\;
    Sample $\Vec{I}'$ ($\neq \Vec{I}$) from the distribution and calculate $H_{\Vec{I}' \Vec{I}}$\;
    \For{sampled $\Vec{I}'$}{
        \If{$\abs{H_{\Vec{I}' \Vec{I}}}\Delta\tau \leq 1$}{A new walker with index $\Vec{I}'$ is generated with a probability of $\abs{H_{\Vec{I}' \Vec{I}}}\Delta\tau$, to which $\sgn(-H_{\Vec{I}' \Vec{I}})$ is assigned. }
        \Else{$\floor{\abs{H_{\Vec{I}' \Vec{I}}}\Delta\tau}$ walkers and a walker are generated with a probability of 1 and $\abs{H_{\Vec{I}' \Vec{I}}}\Delta\tau - \floor{\abs{H_{\Vec{I}' \Vec{I}}}\Delta\tau}$, respectively, to which $\sgn(-H_{\Vec{I}' \Vec{I}})$ are assigned.
        }}
    }
    \tcp{Death/cloning step}
    \For{each walker indexed by $\Vec{I}$ in the parent walker distribution}{
    Calculate $H_{\Vec{I} \Vec{I}}$\;
    \If{the sign of $(H_{\Vec{I} \Vec{I}} - S) \Delta\tau \geq 0$ }{Remove a walker of the same index with probability $\mathrm{min}((H_{\Vec{I}' \Vec{I}} - S) \Delta\tau, 1)$.
    }
    \Else{Copy a walker of the same index with probability $\mathrm{min}(\abs{(H_{\Vec{I} \Vec{I}} - S)} \Delta\tau, 1$)}
    }
    \tcp{Annihilation step}
    Get a new parent walker distribution by merging the walkers obtained in the spawning and death/cloning steps.
    The walkers with opposite signs are canceled out.\;
    Calculate and print $E_\mathrm{proj}$\; 
    }
\end{algorithm}

\section{Outline of HTN+FCIQMC with sparse basis construction}
\label{sec: HTN+FCIQMC with sparse basis construction}
We show an outline of a sparse basis construction version of HTN+FCIQMC (hereafter we simply call HTN+FCIQMC).
In the original QC-FCIQMC~\cite{Zhang2022-lf}, a unitary gate $U$ is obtained through VQE, and then an orthonormal basis is constructed by operating $U$ on the initial state indexed by a binary string $\Vec{I}$ on the quantum circuit, i.e., $U\ket{\Vec{I}}$.
Using this basis, rather than a classically constructible orthonormal basis such as the Slater determinant, we would obtain a QMC wave function with a more localized (sparse) distribution, which could alleviate the sign problem~\cite{Huggins2022-ly, Lee2022-lq, Zhang2022-lf}.

Such an orthonormal basis can also be constructed for HTN as follows.
Assuming $\Vec{I} = \Vec{I}_U\Vec{I}_{L1}\Vec{I}_{L2}\dots\Vec{I}_{Lk}$ where $\Vec{I}_U$ and $\Vec{I}_{Lm}$ are $k$ and $n-1$ qubit binary strings, respectively.
The HTN state indexed by $\Vec{I}$ is defined as
\begin{equation}
 \begin{aligned}
&\ket{\psi_\mathrm{HTN}(\Vec{I})} 
\\&=\sum_{\Vec{i}} \psi_{\Vec{i}}(\Vec{I}_U) \left(\prod_{m=1}^{k} \sum_{\Vec{j}_m} \varphi^{i_m}_{\Vec{j}_m}(\Vec{I}_{Lm})\right) \left(\bigotimes_{m=1}^{k} \ket{\Vec{j}_m}\right)\\
&=\sum_{\Vec{i}} \psi_{\Vec{i}}(\Vec{I}_{U}) \bigotimes_{m=1}^{k} \ket{\varphi^{i_m}(\Vec{I}_{Lm})},
\label{Eq: tensornetwork_indexed}
\end{aligned} 
\end{equation}
where the tensors $\varphi^{i_m}_{\Vec{j}_m}(\Vec{I}_{Lm})$ and $\psi_{\Vec{i}}(\Vec{I}_{U})$ are defined using wave functions $\ket*{\varphi^{i_m}(\Vec{I}_{Lm})}$ and $\ket*{\psi(\Vec{I}_{U})}$ as $\varphi^{i_m}_{\Vec{j}_m}(\Vec{I}_{Lm}) = \braket*{\Vec{j}_m}{\varphi^{i_m}(\Vec{I}_{Lm})}$ and $\psi_{\Vec{i}}(\Vec{I}_{U}) = \braket*{\Vec{i}}{\psi(\Vec{I}_{U})}$, respectively, and other indices are the same as in Eq.~\eqref{Eq: tensornetwork0}. We assume that $\ket{\psi_\mathrm{HTN}(\Vec{I}=\Vec{0})}$ corresponds to the approximate ground state of HTN+VQE, i.e., the trial wave function.

The calculation procedure of a transition amplitude 
\begin{equation}
\begin{aligned}
H_{I' I} &= \mel{\psi_\mathrm{HTN}(\Vec{I}')}{H}{\psi_\mathrm{HTN}(\Vec{I})}
\end{aligned}
\end{equation}
is the same as that of $\mel{\psi_\mathrm{HTN}^{(1)}}{O}{\psi_\mathrm{HTN}^{(2)}}$ in Sec.~\ref{sec: HTN} (and Appendix~\ref{sec: tensor contraction procedure}) by substituting $\ket{\psi_\mathrm{HTN}^{(l)}}$ for $\ket{\psi_\mathrm{HTN}(\Vec{J})}$ ($l \in \{1,2\}$, $\Vec{J} \in \{\Vec{I}', \Vec{I}\}$, and decomposing $H$ into a sum of observables as in Eq.~\eqref{Eq: rewrite Hamiltonian1}); that is, 
\begin{equation}
 \begin{aligned}
\ket{\varphi^{i_m}(\Vec{J}_{Lm})} &= U_{Lm} (I \bigotimes_{r=2}^n X^{j_{mr}(\Vec{J}_{Lm})}) \ket{i_m}\ket{0}^{\otimes n-1} \\
&= U_{Lm} \ket{i_m}\ket*{\Vec{J}_{Lm}},
\end{aligned} 
\end{equation}
\begin{equation}
 \begin{aligned}
\ket{\psi_{\Vec{i}}(\Vec{J}_{U})} &= U_U (\bigotimes_{m=1}^k X^{j_{m}(\Vec{J}_U)}) \ket{0}^{\otimes k} \\
&= U_U \ket*{\Vec{J}_U},
\end{aligned} 
\end{equation}
where the Pauli $X$ operators and binary numbers $j_{mr}(\Vec{J}_{Lm}), j_{m}(\Vec{J}_U) \in \{0,1\}$ are used for mapping each of the binary string in an initial state.

Algorithm~\ref{Alg: HTN+FCIQMC} shows an outline of HTN+FCIQMC, which can be performed in the same manner as in Appendix~\ref{sec: fciqmc} by substituting the index $h$ for $\Vec{I}$, except for the spawning step.
In the original paper~\cite{Zhang2022-lf}, since it is hard to calculate all the $H_{\Vec{I}' \Vec{I}}$ for exponential number of indices $\Vec{I}'$ for the walker index $\Vec{I}$, $H_{\Vec{I}' \Vec{I}}$ to be evaluated are identified by calculating $\abs{H_{\Vec{I}' \Vec{I}}}^2$ from a projective measurement on the state $U^{\dag}HU\ket{\Vec{I}}$ ($H$ decomposes into a sum of Pauli terms).
However, since the identification procedure requires the implementation of state preparation and measurement in a single circuit, it is not executable with HTN. 
Thus, we suggest using the Bayesian estimation for obtaining $\abs{H_{\Vec{I}' \Vec{I}}}$, which has been studied in another QC-QMC algorithm~\cite{Xu2023-sn}. 
Specifically, we start with a suitable distribution and iteratively calculate $H_{\Vec{I}' \Vec{I}}$ for the sampled $\Vec{I}'$ and update the distribution until some termination condition is satisfied (e.g., maximum iteration). 
From the final distribution, we sample $\Vec{I}'$, calculate $H_{\Vec{I}' \Vec{I}}$, generate walkers with a certain probability, and assign a sign to the walker.
The death/cloning and annihilation steps in the algorithm are the same as in FCIQMC, but the spawning and death/cloning steps are executed by using a quantum computer in addition to the projected energy calculation.
Here, the projected energy $E_\mathrm{proj}$ in Eq.~\eqref{Eq: projected energy} is represented as 
\begin{equation}
\begin{aligned}
E_\mathrm{proj} &= \frac{\bra{\xi} H \ket{\psi_\mathrm{QMC}}}{\braket{\xi}{\psi_\mathrm{QMC}}}\\
&= \frac{\sum_{\Vec{I}} w_{\Vec{I}} \mel{\psi_\mathrm{HTN}(\Vec{0})}{H}{\psi_\mathrm{HTN}(\Vec{I})}}{\sum_{\Vec{I}} w_{\Vec{I}} \braket{\psi_\mathrm{HTN}(\Vec{0})}{\psi_\mathrm{HTN}(\Vec{I})}},
\label{Eq: projected energy with walker control}
\end{aligned}
\end{equation}
where $\ket{\xi} = \ket{\psi_\mathrm{HTN}(\Vec{0})}$, and $\ket{\psi_\mathrm{QMC}} = \sum_{\Vec{I}} w_{\Vec{I}} \ket{\psi_\mathrm{HTN}(\Vec{I})}$.
Note that the initial distribution may be uniform or an approximate distribution that is classically prepareable~\cite{Xu2023-sn}. 
In addition, $H_{\Vec{I}' \Vec{I}}$ can be reused once calculated, which reduces the quantum computational cost in the algorithm.

\section{Methods and conditions for constructing specific benchmarking models}
\label{sec: methods and conditions}
For the graphite, 
we first calculated the band structure using density functional theory in the Quantum ESPRESSO package~\cite{Giannozzi2009-zq, Giannozzi2017-jz, Giannozzi2020-yi}, in which we adopted the generalized gradient approximation by Perdew-Burke-Ernzerhof (PBE)~\cite{Perdew1996-qc} as the exchange-correlation functional and optimized norm-conserving Vanderbilt (ONCV) pseudopotential~\cite{Hamann1979-ti, Hamann2013-tx}.
The wave function cutoff, k-point grids, and the number of bands were 64 Rydberg, $8\times8\times3$, and 30, respectively.
Then we calculated $t_1$, $t_2$, and $U$ for target orbitals using the maximally localized Wannier function~\cite{Marzari1997-sz, Souza2001-un} and constrained random phase approximation~\cite{Aryasetiawan2004-sq} in the RESPACK package~\cite{Fujiwara2003-xg, Nakamura2008-hg, Nohara2009-rm, Nakamura2009-te, Nakamura2016-vj, Nakamura2021-sh}.
The target orbitals were the four $p_z$ orbitals in each carbon atom in the unit cell.
The polarization-function cutoff was 6.4 Rydberg.
We obtained $t_1 = -1.05 \times 10^{-1}$, $t_2 = 1.03 \times 10^{-2}$, and $U = 3.00 \times 10^{-1}$ Hartree.

\begin{figure}[]
 \centering
 \includegraphics[width=1\columnwidth]{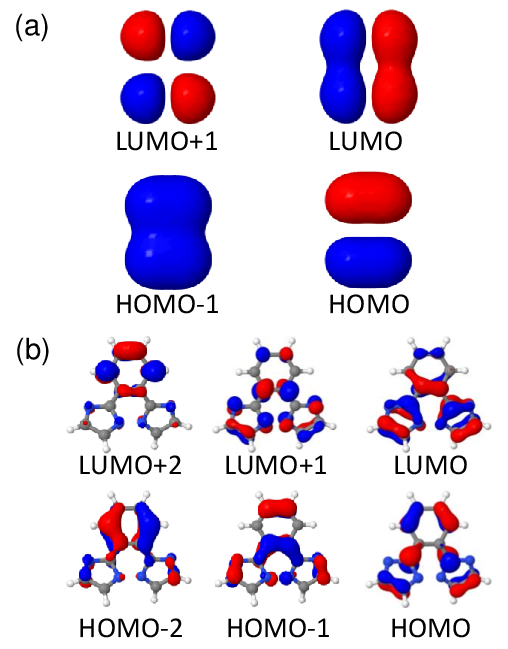}
\caption{Orbitals for the hydrogen plane model and MABI, where the structures are drawn by  Jmol~\cite{Jmol_development_team2016-ly}. (a) Orbitals of the hydrogen plane model with ID 2. (b) Orbitals of MABI.}
\label{fig: Orbitals_1column.eps}

\end{figure}

The MABI structure was determined by $\mathrm{S_0}$ state geometry optimization at the three-state-averaged CASSCF(14e,14o)/6-31G level using Molpro 2015~\cite{Werner2012-ky, Werner2020-ph}, where MABI geometry in Cartesian coordinates is shown in Table~\ref{tab: MABI geometry}.
The PySCF package~\cite{Sun2020-sa} was adopted for constructing the Hamiltonian for the hydrogen plane models and MABI with STO-3G and 6-31G being as the respective basis functions.
The orbitals of the hydrogen plane model with ID 2 and MABI are shown in Fig.~\ref{fig: Orbitals_1column.eps}.
The CASCI(6e,6o) problem was considered for preparing the 12-qubit Hamiltonian in MABI.

\begin{table}[!ht]
    \centering
    \caption{MABI geometry. The unit is Angstrom.}
    \label{tab: MABI geometry}
    \begin{tabular}{c c c c}
    \hline
    Atom & x & y & z \\ \hline
        N & 0.3978306615 & 2.8693899395 & -0.3692443149 \\ 
        N & -0.3978306615 & -2.8693899395 & -0.3692443149 \\ 
        C & -0.0272717639 & 1.5204757231 & -0.3278195176 \\ 
        C & 0.0272717639 & -1.5204757231 & -0.3278195176 \\ 
        N & -0.7339328460 & 1.1810526470 & -1.5025890590 \\ 
        N & 0.7339328460 & -1.1810526470 & -1.5025890590 \\ 
        C & -0.7152844974 & 2.2570066792 & -2.2475658131 \\ 
        C & 0.7152844974 & -2.2570066792 & -2.2475658131 \\ 
        C & 0.0141406967 & 3.3230854653 & -1.5354846933 \\ 
        C & -0.0141406967 & -3.3230854653 & -1.5354846933 \\ 
        C & 0.1345889910 & 0.7148615165 & 0.7894921301 \\ 
        C & -0.1345889910 & -0.7148615165 & 0.7894921301 \\ 
        C & 0.4536375163 & 1.3333761293 & 2.0531301366 \\ 
        C & -0.4536375163 & -1.3333761293 & 2.0531301366 \\ 
        C & 0.2527776487 & 0.6610600524 & 3.2195655062 \\ 
        C & -0.2527776487 & -0.6610600524 & 3.2195655062 \\ 
        H & -0.4501558264 & -1.1474113899 & 4.1551410403 \\ 
        H & 0.4501558264 & 1.1474113899 & 4.1551410403 \\ 
        H & 0.7765766894 & 2.3531013152 & 2.0441189047 \\ 
        H & -0.7765766894 & -2.3531013152 & 2.0441189047 \\ 
        H & -1.1937555617 & 2.3162253096 & -3.1985170250 \\ 
        H & 1.1937555617 & -2.3162253096 & -3.1985170250 \\ 
        H & 0.2078031320 & 4.3179022040 & -1.8673641207 \\ 
        H & -0.2078031320 & -4.3179022040 & -1.8673641207 \\ \hline
    \end{tabular}
\end{table}

\begin{figure*}[]
 \includegraphics[width=0.8\textwidth]{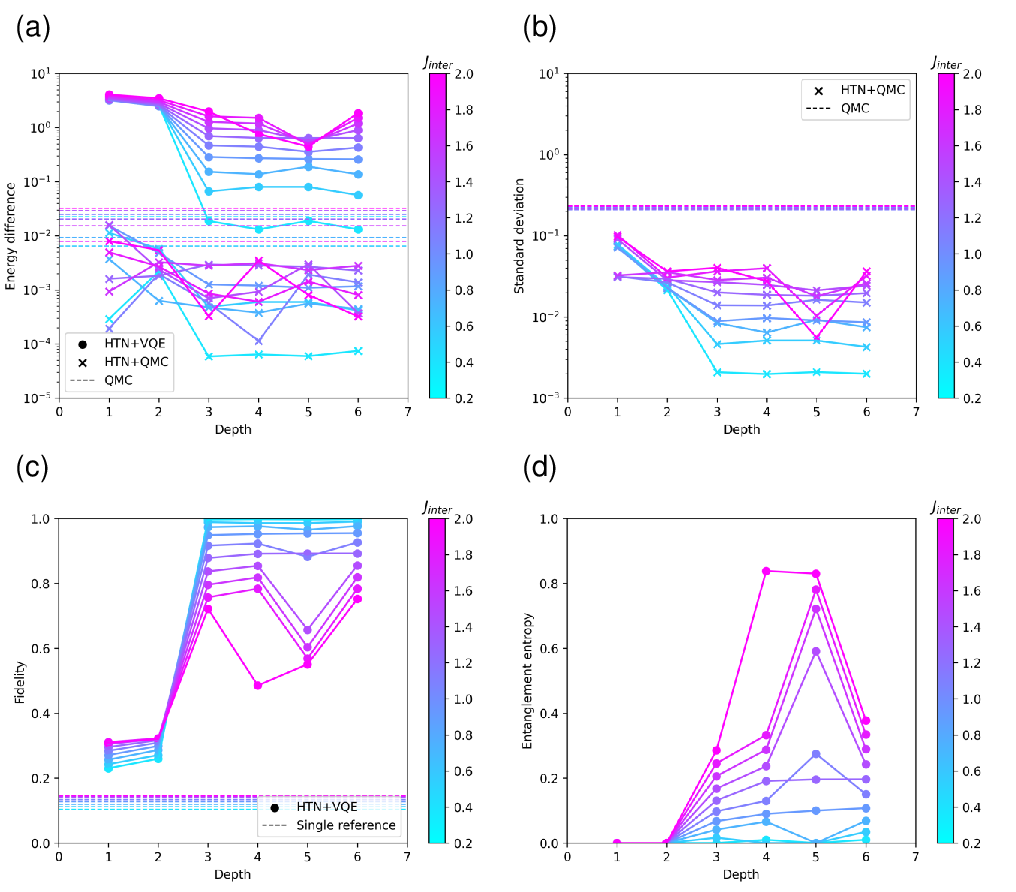}
\caption{Results of the Heisenberg chain model of the cluster setting with $n=4, k=2$, and  $d_H=1,2,\dots,6$. The values of $J_\mathrm{inter}$ are shown in color bars. The circle and cross marks denote the results of HTN+VQE and HTN+QMC, respectively. The dashed lines in (a) and (b) [(c)] denote the results of QMC [a single reference state]. (a) Energy difference. (b) Standard deviation. (c) Fidelity. (d) Bipartite entanglement entropy.}
 \label{fig: Benchmark_Heisenberg_2column}
\end{figure*}

\begin{figure*}[]
 \includegraphics[width=0.8\textwidth]{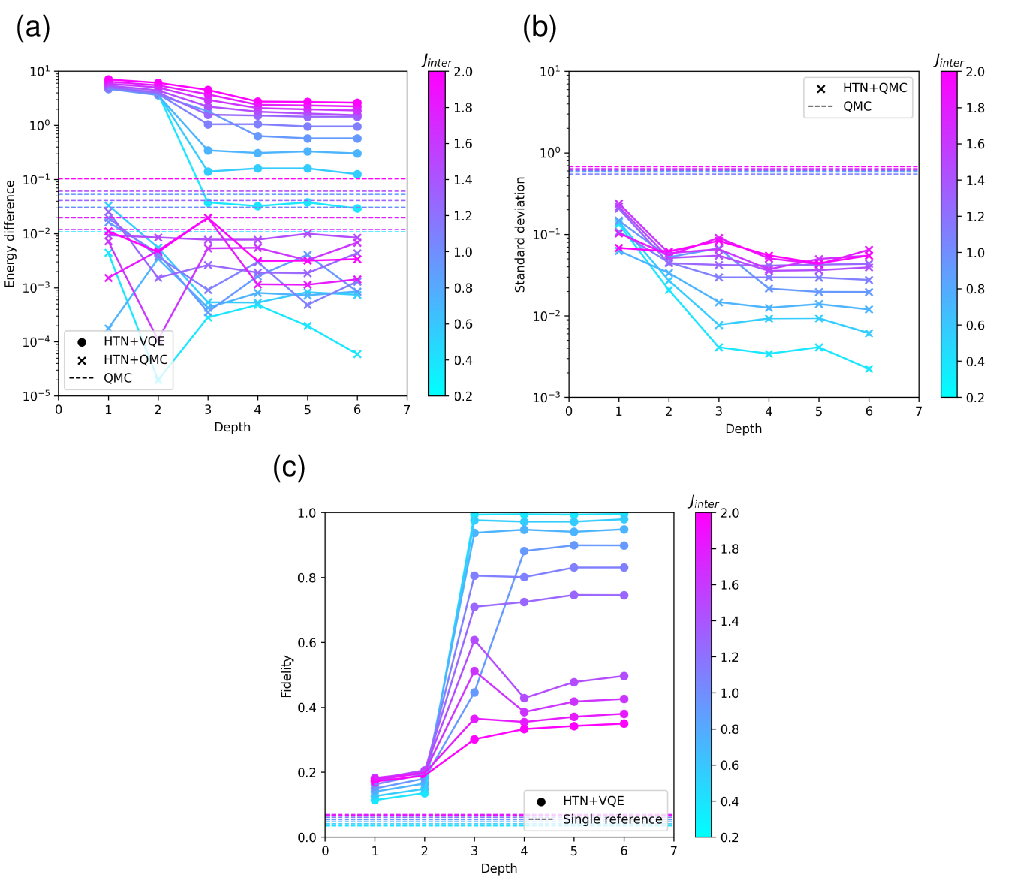}
\caption{Results of the Heisenberg chain model of the cluster setting with $n=4, k=3$, and $d_H=1,2,\dots,6$. The values of $J_\mathrm{inter}$ are shown in color bars. The circle and cross marks denote the results of HTN+VQE and HTN+QMC, respectively. The dashed lines in (a) and (b) [(c)] denote the results of QMC [a single reference state]. (a) Energy difference. (b) Standard deviation. (c) Fidelity. }
 \label{fig: Benchmark_Heisenbergk3_2column}
\end{figure*}

\begin{figure*}[]
 \includegraphics[width=0.8\textwidth]{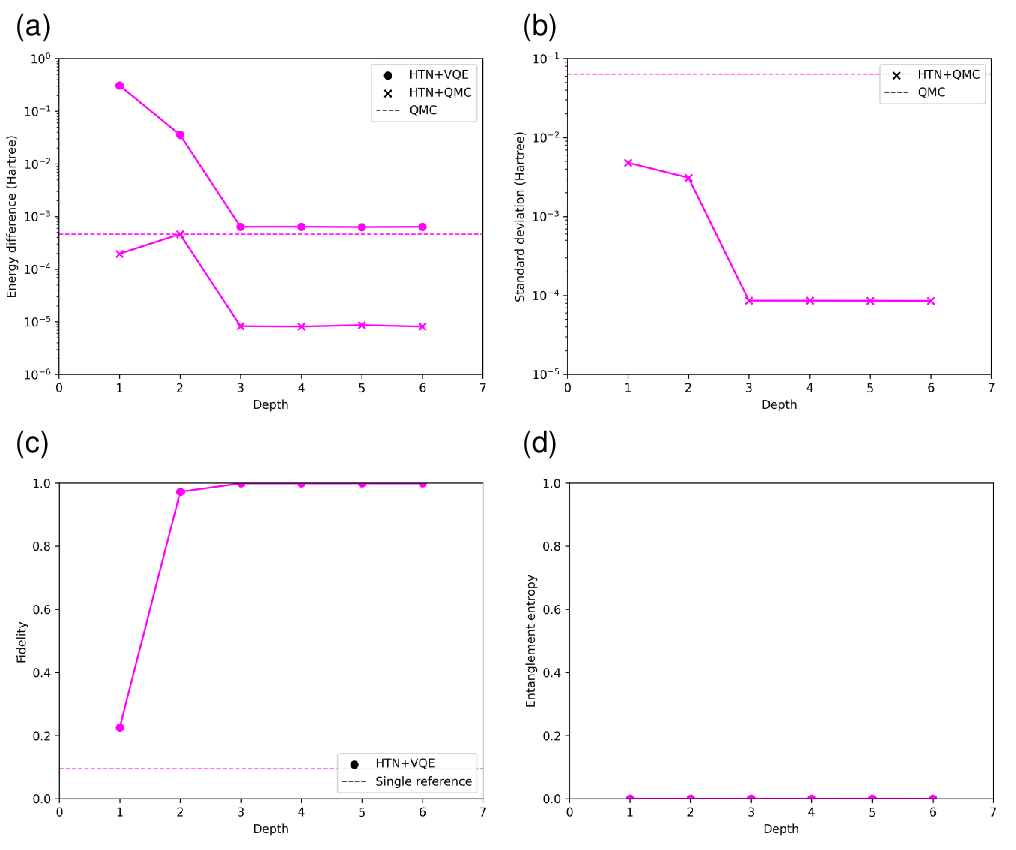}
\caption{Results of the graphite model of the horizontal setting with $n=4, k=2$, and $d_H=1,2,\dots,6$. The circle and cross marks denote the results of HTN+VQE and HTN+QMC, respectively. The dashed lines in (a) and (b) [(c)] denote the results of QMC [a single reference state]. (a) Energy difference. (b) Standard deviation. (c) Fidelity. (d) Bipartite entanglement entropy.}
 \label{fig: Benchmark_Graphite_2column}
\end{figure*}

\begin{figure*}[]
 \includegraphics[width=0.8\textwidth]{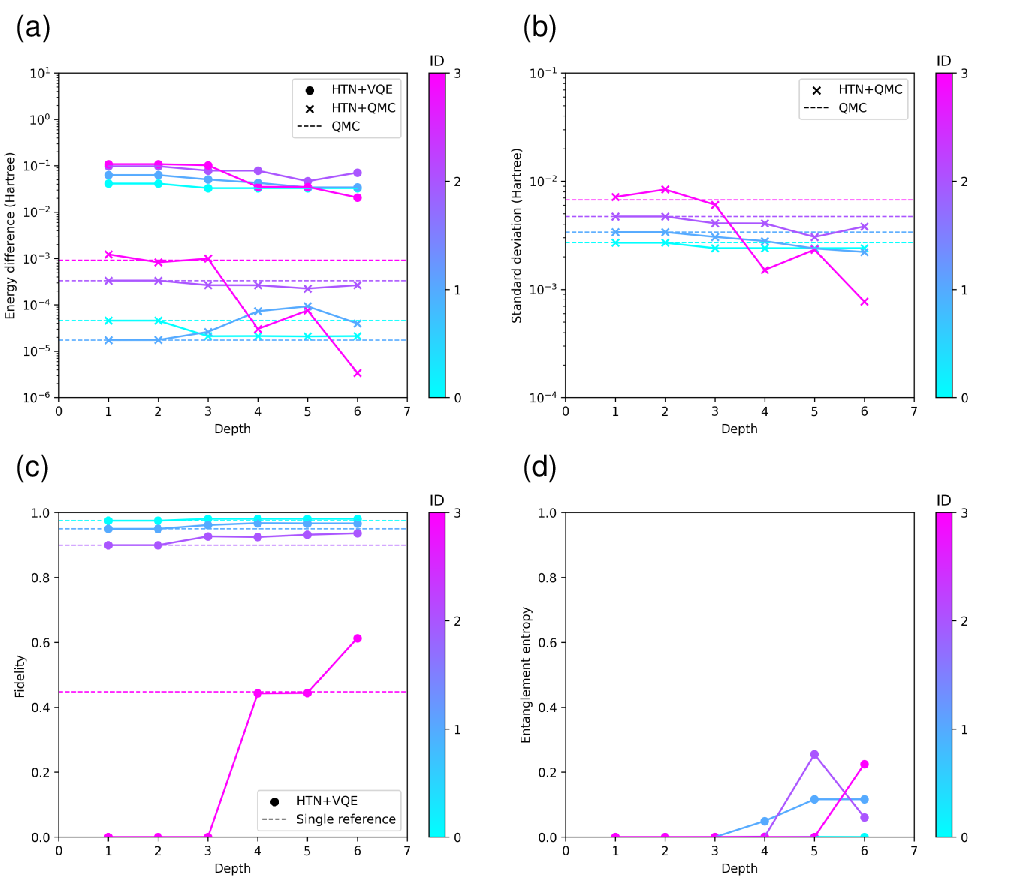}
\caption{Results of the hydrogen plane model of the HOMO-LUMO setting with $n=4, k=2$, and $d_H=1,2,\dots,6$. The values of ID are shown in color bars. The circle and cross marks denote the results of HTN+VQE and HTN+QMC, respectively. The dashed lines in (a) and (b) [(c)] denote the results of QMC [a single reference state]. (a) Energy difference. (b) Standard deviation. (c) Fidelity. (d) Bipartite entanglement entropy.}
 \label{fig: Benchmark_Hp_2column}
\end{figure*}

\begin{figure*}[]
 \includegraphics[width=0.8\textwidth]{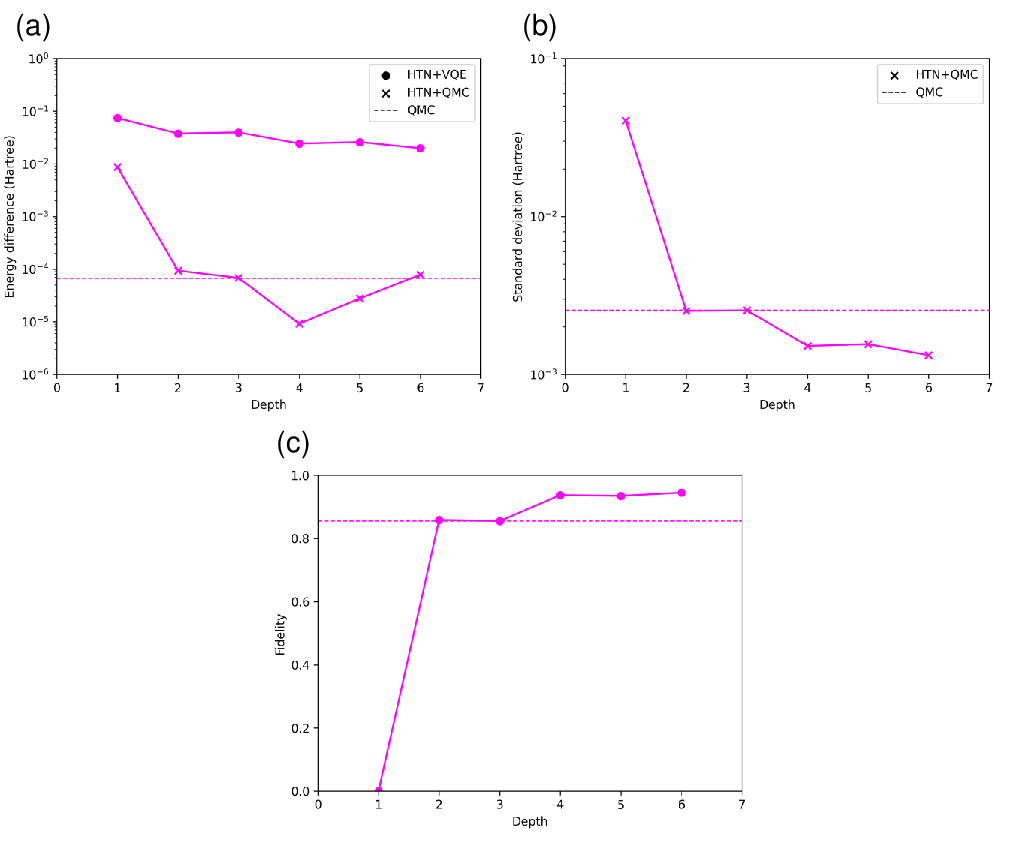}
\caption{Results of MABI of the HOMO-LUMO setting with $n=4, k=3$, and $d_H=1,2,\dots,6$. The circle and cross marks denote the results of HTN+VQE and HTN+QMC, respectively. The dashed lines in (a) and (b) [(c)] denote the results of QMC [a single reference state]. (a) Energy difference. (b) Standard deviation. (c) Fidelity.}
 \label{fig: Benchmark_MABI_2column}
\end{figure*}

\section{Benchmark results}
\label{sec: benchmark results}
We benchmark the models with circuit depth $d_H=1,2,\dots,6$.
As an example, we first explain the benchmark results of the Heisenberg chain model in the cluster setting with $k=2$, especially $J_\mathrm{inter}=0.2, 1.0$, and $2.0$ Hartree.
Figure~\ref{fig: Benchmark_Heisenberg_2column}(a), (b), and (c) show the energy difference, standard deviation, and fidelity versus circuit $d_H$, respectively. 
For HTN+VQE in Fig.~\ref{fig: Benchmark_Heisenberg_2column}(a), the energy difference is more than $10^{-1}$ Hartree when $J_\mathrm{inter}$ is large, e.g., for $J_\mathrm{inter} = 1.0$ Hartree, the difference is $4.3\times10^{-1}$ Hartree even with $d_H=6$.
On the other hand, all the values of the difference for HTN+QMC are less than $10^{-2}$ Hartree with $d_H \geq 2$.
In addition, the difference and standard deviation (Fig.~\ref{fig: Benchmark_Heisenberg_2column}(b)) for HTN+QMC with $d_H \geq 3$ are much smaller than those of QMC.

Figure~\ref{fig: Benchmark_Heisenberg_2column}(c) shows the fidelity of HTN+VQE, and the fidelity for the single reference state used as the QMC trial wave function is also shown.
The energy accuracy worsens as the fidelity decreases, for example of $d_H=4$, the fidelity with $J_\mathrm{inter}=0.2, 1.0$, and $2.0$ Hartree is 1.00, 0.92, and 0.49, respectively, and the energy difference with $J_\mathrm{inter} = 2.0$ for HTN+QMC in Fig.~\ref{fig: Benchmark_Heisenberg_2column}(a) is one order of magnitude larger than that of $J_\mathrm{inter}=0.2$ and 1.0.
Nevertheless, the standard deviation with $d_H=4$ and $J_\mathrm{inter}=2.0$ for HTN+QMC is one order of magnitude less than that for QMC, i.e., the accuracy of HTN+QMC is still higher than that of QMC even when the fidelity is low.
Therefore, even if the trial wave function prepared in HTN+VQE is not highly accurate, HTN+QMC can be expected to provide better accuracy than the classical calculation.
Note that we also calculate the bipartite entropy for the models in Fig.~\ref{fig: Benchmark_Heisenberg_2column}(d), which tends to increase with $J_\mathrm{inter}$.

Finally, we show the benchmark results for the Heisenberg chain model with $k=3$, the graphite model, the hydrogen plane model, and MABI as in Figs.~\ref{fig: Benchmark_Heisenbergk3_2column}, \ref{fig: Benchmark_Graphite_2column}, \ref{fig: Benchmark_Hp_2column}, and \ref{fig: Benchmark_MABI_2column}, respectively.
The results of bipartite entanglement entropy are shown for the models with $k=2$.
From these data, we found that $d_H=4$ would provide sufficient accuracy and adopted this value as the default.
Note that the entropy for graphite is almost zero, which reflects the small interactions between graphite layers and the fact that graphite is an easy model to decompose by layers.

\section{Detail of the calculation conditions on VQE and QMC}
\label{sec: conditions for VQE and QMC}
HTN+QMC codes were implemented by the Python, especially the NumPy~\cite{Harris2020-fh} and SciPy~\cite{Virtanen2020-rg} packages, and Qiskit~\cite{Treinish2023-co} and Qulacs~\cite{Suzuki2021-lj} were adopted as quantum circuit simulators.
The sequential least squares programming optimizer (SLSQP) in the SciPy package was used in the parameter optimizations.
In MABI, although there are six electrons in the ground state in the CASCI(6e,6o) problem, the number of electrons in the ground state of the corresponding qubit Hamiltonian is not six; this is because the hardware-efficient ansatz is not capable of imposing the same constraints as CASCI(6e,6o).
VQE/AC \cite{Gocho2023-uy} algorithm was therefore used to perform energy optimization with the constraints for the number of electrons and also employed in the pseudo-Hadamard test execution.
The ground state of MABI refers to the eigenstate with the smallest eigenvalue acquired by restricting the electron number to six.

In all the QMC methods (QMC, QC-QMC, and HTN+QMC), the initial number of walkers was one, and the single reference state was selected as the initial configuration.
In the hydrogen plane model (except for ID 3) and MABI, we checked out that the Hartree-Fock state was chosen as the single reference state.
The trial wave function was the single reference state, and the states obtained by VQE and HTN+VQE were used for QC-QMC and HTN+QMC, respectively.
The maximum number of iterations in all the types of QMC was 100,000 for the Heisenberg chain model and 10,000 for the others.

The conditions for FCIQMC are the following.
The initial value of the energy shift $S$ was set to the energy of the leading single reference state in the orthonormal basis state of the ground state (the ground state when constrained to six electrons in MABI).
The imaginary time increment $\Delta\tau$ was 0.001 $\mathrm{Hartree}^{-1}$ for the Heisenberg chain and graphite models, and 0.1 $\mathrm{Hartree}^{-1}$ for the hydrogen plane model and MABI.
The parameters for the variable shift were $A = 5$ and $\zeta = 0.1$, and $N_\mathrm{shift} = 10000$ for the Heisenberg models and $N_\mathrm{shift} = 1000$ for the others.

\section{Details of real device procedure}
\label{sec: details of real device execution}
We describe the previous studies for the overlap calculation techniques, and introduce the pseudo-Hadamard test, the algorithm for overlap calculation modified for executing HTN on real devices, together with the theoretical analysis on the noise resilience. 

\begin{figure}[]
 \centering
 \includegraphics[width=1\columnwidth]{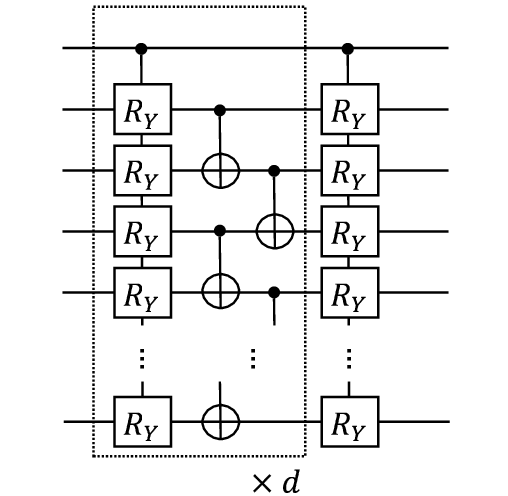}
\caption{Circuit of real amplitude ansatz for the pseudo-Hadamard test with no constraint. The topmost line represents an ancilla qubit and the other lines represent system qubits.}
\label{fig: RAansatz_nocons_1column.eps}

\end{figure}

\subsection{Studies on overlap calculations}
\label{sec: studies on overlap calculations}
Here we present our motivation for employing the pseudo-Hadamard test for overlap computation. There are studies on the overlap calculation techniques to avoid a controlled gate in the Hadamard test~\cite{Aharonov2009-aa, Huggins2020-fk, Baek2023-xb, Baek2023-xb, Lu2021-on, Xu2023-sn}, which is used to prepare a trial state.
However, overlap techniques in HTN+QMC require a condition that is executable on non-particle number preserving ansatz since the number of particles in the subsystem is not conserved in the corresponding lower tensor, even in a fermionic system. The previous techniques without the Hadamard test do not satisfy the condition.
In addition, the pseudo-Hadamard has preferable features such as not requiring projective measurements that fail with a certain probability~\cite{Xu2023-sn} and avoiding double the number of required qubits for a system size~\cite{Huggins2020-fk, Baek2023-xb}.
Drawbacks of the pseudo-Hadamard include an inaccurate overlap value when the constraints are incompletely satisfied and a reduction in the representation ability of ansatz due to the constraints. 
In the present study, the constraints are satisfied with a negligibly small error because a constrained optimization solver was used.
Numerical stability regarding these drawbacks is a future issue.
Note that a possible solution is to use an ansatz which does not require any constraints. 
For example, the ansatz shown in Fig.~\ref{fig: RAansatz_nocons_1column.eps}, in which each RY gate is replaced by a controlled-RY gate from Fig.~\ref{fig: RAansatz_1column}(a), can perform a pseudo-Hadamard test without any constraints since both the conditions in Eq.~\eqref{eq: overlap requirement} are satisfied in any parameters, where $d$ is the number of depth. However, we did not employ the ansatz in this study because of increasing the number of long-range two-qubit gates.

\begin{figure}[ht!]
 \centering
 \includegraphics[width=1\columnwidth]{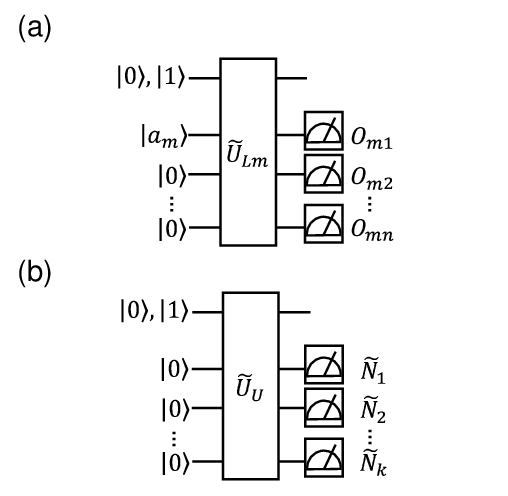}
\caption{Quantum circuits for calculating HTN+VQE to execute the pseudo-Hadamard test. The topmost line represents an ancilla qubit and the other lines represent system qubits. The circuit with the ancilla qubit set to $\ket{0}$ is used when calculating constraints. (a) Calculation for the lower tensor. (b) Calculation for the upper tensor.}
\label{fig: Circuit_RD_HTNVQE_1column.eps}

\end{figure}

\begin{figure}[ht!]
 \centering
 \includegraphics[width=1\columnwidth]{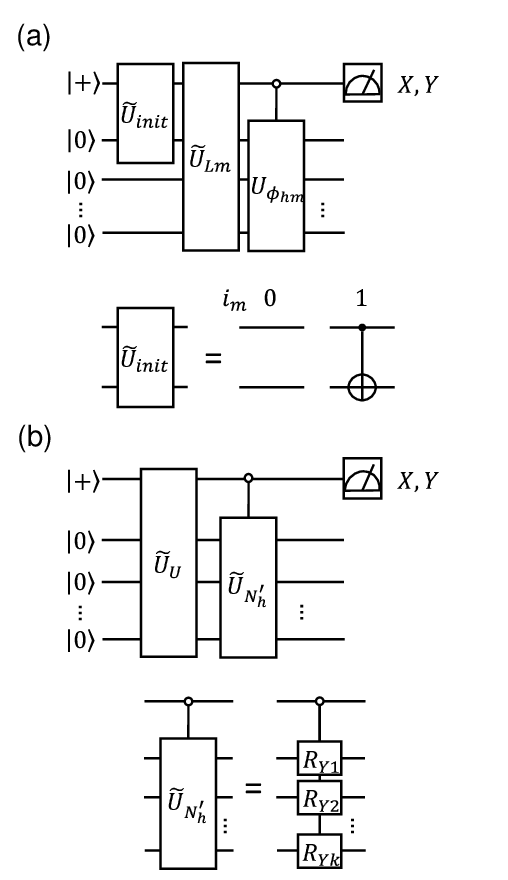}
\caption{Quantum circuits for calculating $T_h$ by using pseudo-Hadamard test. The topmost line represents an ancilla qubit and the other lines represent system qubits. (a) Calculation for the lower tensor. (b) Calculation for the upper tensor.}
\label{fig: Circuit_RD_HTNQMC_1column.eps}

\end{figure}

\begin{figure}[]
 \centering
 \includegraphics[width=1\columnwidth]{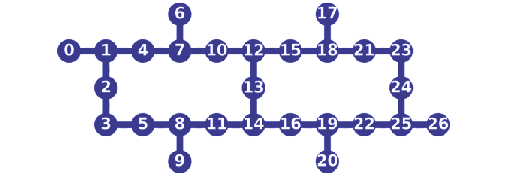}
\caption{Device topology of $ibmq\_kolkata$.}
\label{fig: Kolkata_1column.eps}

\end{figure}

\begin{figure}[]
 \centering
 \includegraphics[width=1\columnwidth]{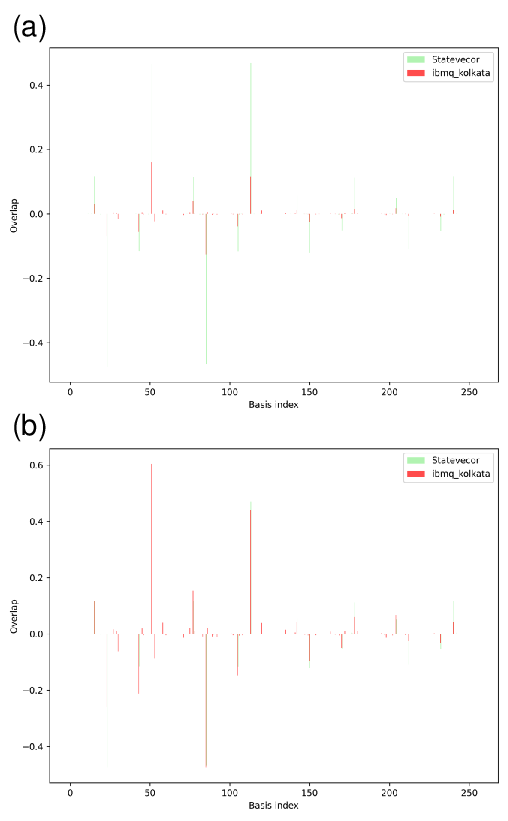}
\caption{Results of $T_h$ in the hydrogen plane model without and with normalization in the real device procedure. The light green and red bars represent the values in the statevector and real device procedure, respectively. (a) Result without normalization. (b) Result with normalization.}
\label{fig: Overlap_compare_normalization_1column.eps}
\end{figure}

\begin{table}[]
    \centering
    \caption{Values of $G_{mr}$ and mutual information (MI) in all the decomposition settings. The ``Decomposition'' column represents the decomposition setting for the models. $k$ denotes the number of subsystems. The unit of $G_{mr}$ is Hartree.}
    \label{tab: values of G}
    \begin{tabular}{c c c c c}
    \hline
        Model & Decomposition & $k$ & $G_{mr}$ & MI\\ \hline\hline
        \multirow{2}{*}{Heisenberg} & Cluster & 2 & 1.50 & 1.32\\ 
         & Even-Odd & 2 & 10.50 & 6.92\\ \hline
        \multirow{2}{*}{Graphite} & Horizontal & 2 & 0.02 & 0.03\\ 
         & Vertical & 2 & 0.65 & 7.85\\ \hline
        \multirow{3}{*}{Hydrogen plane} & HOMO-LUMO & 2 & 1.51 & 1.14\\ 
         & Alpha-Beta & 2 & 1.65 & 2.84\\ 
         & Occ-Unocc & 2 & 1.44 & 3.14\\ \hline
        \multirow{3}{*}{MABI} & HOMO-LUMO & 3 & 1.67 & 0.00\\ 
         & Alpha-Beta & 2 & 1.96 & 0.68\\ 
         & Occ-Unocc & 2 & 2.10 & 1.33\\ \hline
    \end{tabular}
\end{table}

\begin{figure*}[]
\begin{minipage}[b]{0.45\textwidth}
\centering    
\raisebox{1.5cm}{\includegraphics[width=\columnwidth]{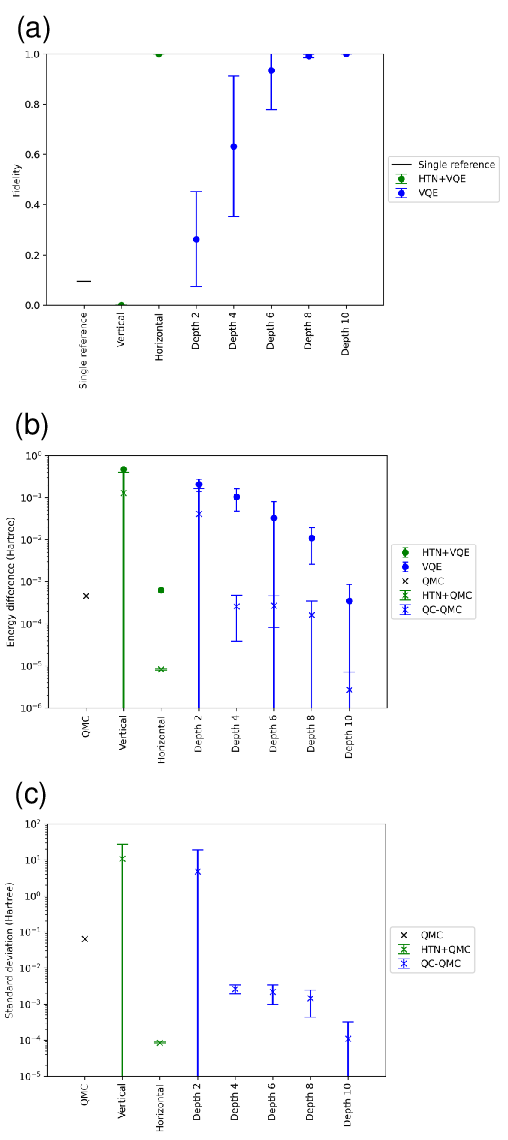}}
\caption{Results of the analysis for the graphite model. The average values and error bars were calculated over 10 different random seeds used for the initial parameters in VQE or HTN+VQE execution.
For reference, values obtained from the single reference state and those from QMC are shown by the black bar in (a) and cross in (b) and (c), respectively. 
The green (blue) circle and cross denote the results from the decomposition (no-decomposition) setting in HTN+VQE and HTN+QMC (VQE and QC-QMC), respectively. 
In HTN+VQE and HTN+QMC, $n=4$ and $k=2$ for all the settings.
(a) Fidelity. (b) Energy difference. (c) Standard deviation.}
\label{fig: Analysis_Gr_1column}

\end{minipage}
\hfill
\begin{minipage}[b]{0.45\textwidth}
\centering
\includegraphics[width=\columnwidth]{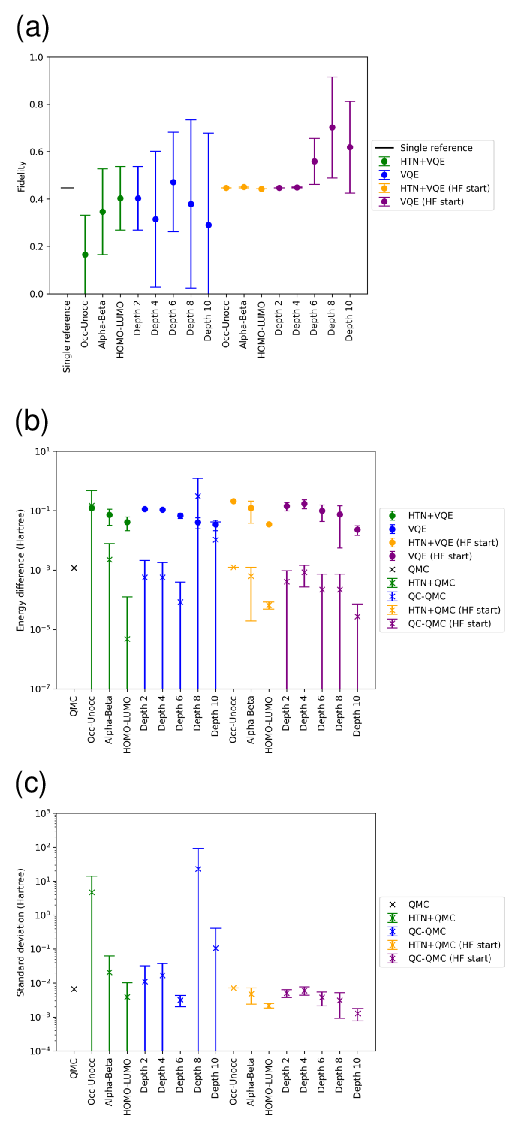}
\caption{Results of the analysis for the hydrogen plane model with ID 3. The average values and error bars were calculated over 10 different random seeds used for the initial parameters in VQE or HTN+VQE execution.
For reference, values obtained from the single reference state and those from QMC are shown by the black bar in (a) and cross in (b) and (c), respectively. 
The green circle, green cross, blue circle, and blue cross respectively denote the results from the decomposition, decomposition, no-decomposition, and no-decomposition settings in HTN+VQE, HTN+QMC, VQE, and QC-QMC, and the yellow circle, yellow cross, purple circle, and purple cross respectively denote those when the initial state is close to the Hartree-Fock state in HTN+VQE, HTN+VQE, VQE, and VQE.
In HTN+VQE and HTN+QMC, $n=4$ and $k=2$ for all the settings.
(a) Fidelity. (b) Energy difference. (c) Standard deviation.}
\label{fig: Analysis_Hp_1column}

\end{minipage}
\end{figure*}

\begin{figure}[]
 \centering
 \includegraphics[width=1\columnwidth]{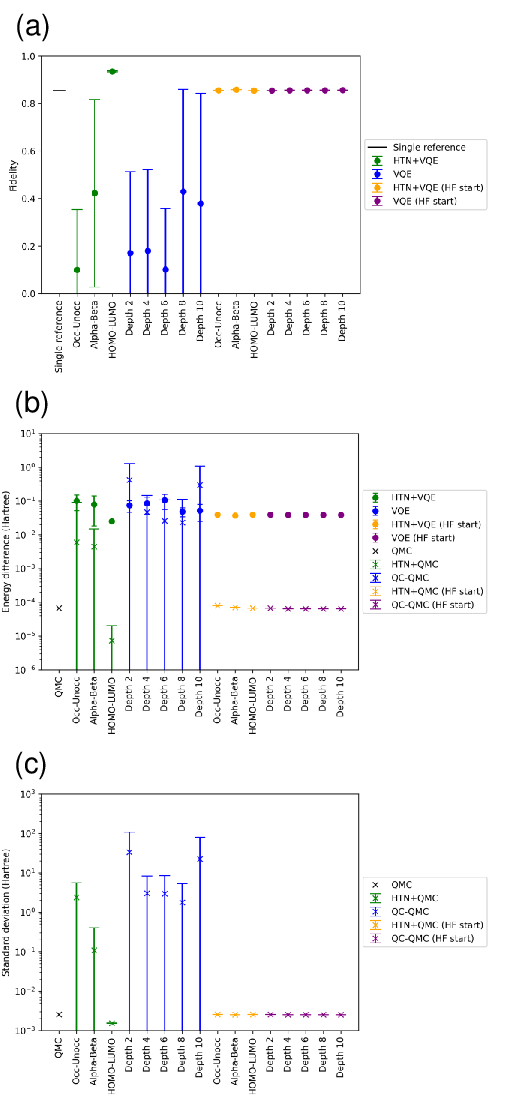}
\caption{Results of the analysis for MABI. The average values and error bars were calculated over 10 different random seeds used for the initial parameters in VQE or HTN+VQE execution.
For reference, values obtained from the single reference state and those from QMC are shown by the black bar in (a) and cross in (b) and (c), respectively. 
The green circle, green cross, blue circle, and blue cross respectively denote the results from the decomposition, decomposition, no-decomposition, and no-decomposition settings in HTN+VQE, HTN+QMC, VQE, and QC-QMC, and the yellow circle, yellow cross, purple circle, and purple cross respectively denote those when the initial state is close to the Hartree-Fock state in HTN+VQE, HTN+VQE, VQE, and VQE.
In HTN+VQE and HTN+QMC, $n=4$, $k=3$ for the HOMO-LUMO setting, and $n=6$, $k=2$ for the other two settings.
(a) Fidelity. (b) Energy difference. (c) Standard deviation.}
\label{fig: Analysis_MABI_1column}
\end{figure}

\subsection{Extension of the pseudo-Hadamard test to HTN}
\label{sec: extension of the overlap technique to HTN}
We start with the wave function in the form considering an ancilla qubit
\begin{equation}
    \begin{aligned}
        \ket{\Tilde{\psi}_\mathrm{HTN}} = \sum_{\Vec{i}} \Tilde{\psi}_{\Vec{i}} \bigotimes_m \ket{\Tilde{\varphi}^{i_m}},
    \end{aligned}
\end{equation}
where $\Vec{i} = i_1 i_2 \dots i_k$ represent a $k$-qubit binary string, $m \in \{1,2,\dots,k\}$, and $i_m \in \{0,1\}$.
The general formalism corresponding to Eq.~\eqref{Eq: general formalism for pseudo-Hadamard test} are
\begin{equation}
\begin{aligned}
\label{eq: HTN+VQE with constraint in general}
&\underset{\Tilde{U}_{Lm}, \Tilde{U}_{U}}{\min} \expval{H}{\Tilde{\psi}_\mathrm{HTN}}\\
&\mathrm{s.t.}\\
&\bra{1}\bra{i_m}\bra{0}^{\otimes n-1} \Tilde{U}_{Lm}^{\dag} (n_{m0} \otimes I^{\otimes n}) \Tilde{U}_{Lm} \ket{1}\ket{i_m}\ket{0}^{\otimes n-1} \\
&\quad = 1 \quad\forall m,\\
&\bra{0}\bra{0}^{\otimes n} \Tilde{U}_{Lm}^{\dag} (\sum_{r=0}^{n} I^{\otimes r} \otimes n_{mr} \otimes I^{\otimes n-r}) \Tilde{U}_{Lm} \ket{0}\ket{0}^{\otimes n}\\
&\quad = 0 \quad\forall m,\\
&\bra{1}\bra{0}^{\otimes k} \Tilde{U}_{U}^{\dag} (n_0 \otimes I^{\otimes k} ) \Tilde{U}_{U} \ket{1}\ket{0}^{\otimes k} = 1,\\
&\bra{0}\bra{0}^{\otimes k} \Tilde{U}_{U}^{\dag} (\sum_{m=0}^{k} I^{\otimes m} \otimes n_{m} \otimes I^{\otimes k-m}) \Tilde{U}_{U} \ket{0}\ket{0}^{\otimes k} = 0,
\end{aligned}
\end{equation}
and the simplified formulation corresponding to Eq.~\eqref{Eq: simple formalism for pseudo-Hadamard test} assuming the use of the ansatz as in Fig.~\ref{fig: RAansatz_1column}(b) is 
\begin{equation}
\begin{aligned}
\label{eq: HTN+VQE with constraint}
&\underset{\Tilde{U}_{Lm}, \Tilde{U}_{U}}{\min} \expval{H}{\Tilde{\psi}_\mathrm{HTN}}\\
&\mathrm{s.t.}\\
&\bra{0}\bra{0}^{\otimes n} \Tilde{U}_{Lm}^{\dag} (\sum_{r=1}^{n} I^{\otimes r} \otimes n_{mr} \otimes I^{\otimes n-r}) \Tilde{U}_{Lm} \ket{0}\ket{0}^{\otimes n}\\
&\quad = 0\quad\forall m,\\
&\bra{0}\bra{0}^{\otimes k} \Tilde{U}_{U}^{\dag} (\sum_{m=1}^{k} I^{\otimes m} \otimes n_{m} 
\otimes I^{\otimes k-m}) \Tilde{U}_{U} \ket{0}\ket{0}^{\otimes k}\\
&\quad = 0,
\end{aligned}
\end{equation}
where $n_{mr}$ and $n_{m}$ are the number operator on the $r$-th qubit of the $m$-th lower tensor and the $m$-th qubit of upper tensor, respectively, and $\Tilde{U}_{Lm}$ and $\Tilde{U}_{U}$ are unitary gates that are specified in detail in the next paragraph. The following discussion is based on the simplified formalism that was applied in the real device execution.

The expectation value of the observable can be rewritten in terms of the upper and lower tensors (both involving the ancilla) as
\begin{equation}
    \begin{aligned}
    \label{Eq: ExpectationPseudo1}
        &\expval{O}{\Tilde{\psi}_\mathrm{HTN}}  \\
        &= \sum_{\Vec{i'}~\Vec{i}} \Tilde{\psi}_{\Vec{i'}}^* \Tilde{\psi}_{\Vec{i}} \prod_{m=1}^k \mel{\Tilde{\varphi}^{i_m'}}{\bigotimes_{r=1}^n O_{mr}}{\Tilde{\varphi}^{i_m}}\\
        &= \sum_{\Vec{i'}~\Vec{i}} \Tilde{\psi}_{\Vec{i'}}^* \Tilde{\psi}_{\Vec{i}} \prod_{m=1}^k \bra{1}\bra*{\Tilde{\varphi}^{i_m'}} (I \otimes \bigotimes_{r=1}^n O_{mr}) \ket{1}\ket*{\Tilde{\varphi}^{i_m}}\\
        &= \sum_{\Vec{i'}~\Vec{i}} \Tilde{\psi}_{\Vec{i'}}^* \Tilde{\psi}_{\Vec{i}} \prod_{m=1}^k \Tilde{N}^{i_m' i_m}.
    \end{aligned}
\end{equation}
Figure~\ref{fig: Circuit_RD_HTNVQE_1column.eps}(a) show the quantum circuit for calculating $\Tilde{N}^{i_m' i_m}$ in the lower tensor assuming $\ket{1}\ket{\Tilde{\varphi}^{i_m}} = \Tilde{U}_{Lm}\ket{1}\ket{i_m}\ket{0}^{\otimes n-1}$.
We construct $\Tilde{N}_m = \mqty(\Tilde{N}^{00} & \Tilde{N}^{01} \\ \Tilde{N}^{10} & \Tilde{N}^{11})$ by combining the results from the lower tensors for the four initial states; $\ket{a_m}$ takes $\ket{0}, \ket{1}, \ket{+},$ and $\ket{y+}$, and from the corresponding measurement results denoted by $M^0, M^1, M^+,$ and $M^{y+}$, respectively, we obtain the matrix elements $\Tilde{N}^{00}=M^0, \Tilde{N}^{11}=M^1, \Tilde{N}^{01}= \frac{i-1}{2}M^0 + \frac{i-1}{2}M^1 + M^+ -iM^{y+},$ and $\Tilde{N}^{10} = \Tilde{N}^{01*}$~\cite{Kanno2021-zn}.
Equation~\eqref{Eq: ExpectationPseudo1} then becomes the evaluation in the upper tensor
\begin{equation}
    \begin{aligned}
        \expval{O}{\Tilde{\psi}_\mathrm{HTN}}&=\expval{\bigotimes_{m=1}^k \Tilde{N}_m}{\Tilde{\psi}}\\
        &=\bra*{1}\bra*{\Tilde{\psi}} (I \otimes \bigotimes_{m=1}^k \Tilde{N}_m) \ket*{1}\ket*{\Tilde{\psi}}
    \end{aligned}
    \label{Eq: ExpectationPseudo2}
\end{equation}
Figure~\ref{fig: Circuit_RD_HTNVQE_1column.eps}(b) shows the circuit for calculating the last expression assuming $\ket*{1}\ket*{\Tilde{\psi}} = \Tilde{U}_U\ket*{1}\ket*{0}^{\otimes k}$.
$\Tilde{N}_m$ is Hermitian and can be measured using the eigenvalue decomposition of $\Tilde{N}_m$ classically.

Now we move on to the overlap $T_h$ between the quantum state and orthonormal basis state, which is represented similarly to the observable as
\begin{equation}
\begin{aligned}
    T_h&=\braket{\Tilde{\psi}_\mathrm{HTN}}{\phi_h}\\
    &=\sum_{\Vec{i}} \Tilde{\psi}_{\Vec{i}}^* \prod_{m=1}^k\braket{\Tilde{\varphi}^{i_m}}{\phi_{hm}}\\
    &=\sum_{\Vec{i}} \Tilde{\psi}_{\Vec{i}}^* \prod_{m=1}^k \Tilde{N}^{i_{hm}}.
\end{aligned}
\end{equation}
The circuit for calculating $\Tilde{N}^{i_{hm}}$ in the lower tensor is shown in Fig.~\ref{fig: Circuit_RD_HTNQMC_1column.eps}(a).
Since there is no tensor index in $\ket{\phi_{hm}}$, only two different circuits are needed for $\Tilde{U}_{init}$ (the lower panel of the figure).
$U_{\phi_{hm}}$ is composed of CNOT gates from ancilla to target qubits which should be set to $\ket{1}$ in $\ket{\phi_{hm}}$.
Next, we explain the calculation of the upper tensor.
$\Tilde{N}^{i_{hm}}$ is a tensor with one leg, i.e., can be represented using a two-element vector as $\Tilde{N}^{i_{hm}} = \braket{i_{m}}{\Tilde{N}_{hm}}$.
$T_h$ is then written as
\begin{equation}
    \begin{aligned}
        T_h&=\sum_{\Vec{i}}\Tilde{\psi}_{\Vec{i}}^* \prod_{m=1}^k \braket{i_{m}}{\Tilde{N}_{hm}}\\
        &=\sqrt{\prod_{m=1}^k \braket{\Tilde{N}_{hm}}} \braket{\Tilde{\psi}}{\Tilde{N}^{'}_h}, 
    \end{aligned}
\end{equation}
where
\begin{equation}
    \begin{aligned}
        \ket{\Tilde{N}^{'}_h}&=\bigotimes_{m=1}^k \ket{\Tilde{N}^{'}_{hm}}\\
        &=\bigotimes_{m=1}^k \frac{\ket{\Tilde{N}_{hm}}}{\sqrt{\braket{\Tilde{N}_{hm}}}},\\
    \end{aligned}
\end{equation}
each wave function is represented as a normalized vector, i.e., 
\begin{equation}
    \begin{aligned}
        \frac{1}{\sqrt{\braket{\Tilde{N}_{hm}}}}
        \mqty(\braket*{0}{\Tilde{N}_{hm}} \\ \braket*{1}{\Tilde{N}_{hm}})
        =\mqty(\alpha_{hm} \\ \beta_{hm}),
    \end{aligned}
\end{equation}
$\sqrt{\prod_{m=1}^k \braket{\Tilde{N}_{hm}}}$ is a normalization constant, and $\alpha_{hm}^2+\beta_{hm}^2=1$.
The circuit for the upper tensor is shown in Fig.~\ref{fig: Circuit_RD_HTNQMC_1column.eps}(b), where $\ket{\Tilde{N}^{'}_h}$ is embedded by using controlled-RY gates (lower panel of the figure).
The rotation angle of $R_{Ym}$ is set to $\theta_{hm} = 2\mathrm{arccos}(\alpha_{hm})$, and is substituted for $-\theta_{hm}$ if $\beta_{hm} <0$.
Note that ancilla qubits are measured only $X$ basis in this study since the trial wave function is real.

Finally, the rest of the calculation conditions, which are not described in Section~\ref{sec: models and calculation condition}, are listed below.
We checked on the four-qubit Heisenberg chain model to confirm that the amplitudes close to those of the exact ground state were obtained by using the pseudo-Hadamard test when a highly accurate wave function is obtained in HTN+VQE.
We used at most five qubits in real device execution. We choose the initial qubit layout in the device as 12, 15, 18, 21, and 23, where the device topology is shown in Fig.~\ref{fig: Kolkata_1column.eps}. The number of shots was 4000.
We combined the constraints in Eq.~\eqref{eq: HTN+VQE with constraint} as  
$\{\sum_{m=1}^k \bra{0}\bra{0}^{\otimes n} \Tilde{U}_{Lm}^{\dag} (\sum_{r=1}^{n} I^{\otimes r} \otimes n_{mr} \otimes I^{\otimes n-r}) \Tilde{U}_{Lm} \ket{0}\ket{0}^{\otimes n}\} + \bra{0}\bra{0}^{\otimes k} \Tilde{U}_{U}^{\dag} (\sum_{m=1}^{k} I^{\otimes m} \otimes n_{m} 
\otimes I^{\otimes k-m}) \Tilde{U}_{U} \ket{0}\ket{0}^{\otimes k} = 0$ in the implementation because upper sides of each constraint in Eq.~\eqref{eq: HTN+VQE with constraint} is non-negative for any $\Tilde{U}_{Lm}$ and $\Tilde{U}_{U}$ because they consist of the expectation values for the particle number.
The overlaps are calculated only for the Slater determinants of four and six electrons in the hydrogen plane model and MABI, respectively, with the rest of the amplitudes set to zeros.
In the calculation of MABI, in addition to the constraints on the overlap calculation, an additional constraint enforcing the number of electrons to be six was imposed.

\subsection{Analysis of noise resilience}
\label{sec: analysis of noise resilience}
We first comment on the noise affection of HTN+VQE.
In the very recent study~\cite{Harada2023-il}, noise propagation in the HTN of a multi-layer quantum tree tensor was investigated in the Hamiltonian (and the observable) evaluation, where two-layer quantum tree tensor is chosen in our study.
They found that the magnitude of the desired expectation value decreases exponentially (excluding an identity shift) with the number of quantum tensors.
Therefore, although HTN can extend effective system size, despite quantum error suppression by problem decomposition and error mitigation for each tensor, exponential error increase with system size seems to be inevitable.
In order to perform HTN+VQE in practical cases, careful selection of the condition, such as ansatz and error mitigation techniques, will be needed, or it may be effective to avoid parameter optimization with quantum devices~\cite{Baek2023-xb, Okada2023-jm}.

On the other hand, we show that the QMC step is robust to the noise in the simple case used in Ref.~\cite{Huggins2022-ly}.
In the circuit of Fig.~\ref{fig: Circuit_RD_1column.eps}(b), the state before the measurement $\rho$ is represented as
\begin{equation}
    \begin{aligned}
        \rho &= \ket{\lambda}\bra{\lambda},\\
        \ket{\lambda} &= \frac{1}{\sqrt{2}}(\ket{0}\ket{\phi_h}+\ket{1}\ket*{\Tilde{\psi}}).
    \end{aligned}
\end{equation}
When a depolarizing channel applies to $\rho$, i.e., $\rho' = (1-p)\rho + p I^{\otimes \nu+1}$ where $p$ is an error rate, the expectation value after the measurement is given
\begin{equation}
    \begin{aligned}
        \mathrm{Tr}((X \otimes I^{\otimes \nu}) \rho') &= (1-p)\braket{\Tilde{\psi}}{\phi_h}\\
        &+p\mathrm{Tr}((X \otimes I^{\otimes \nu}) I^{\otimes \nu+1})\\
        &= (1-p)\braket{\Tilde{\psi}}{\phi_h}.
    \end{aligned}
\end{equation}
Since $T_h$ can be represented by substituting $\Tilde{U}_\psi = (\prod_m \Tilde{U}_{Lm}) \Tilde{U}_{U}$ (as described in Appendix~\ref{sec: tensor contraction procedure}), $U_{\phi_h} = \prod_m U_{\phi_{hm}}$, and $\nu = nk$ in Fig.~\ref{fig: Circuit_RD_1column.eps}(b), we can assume the above discussion holds true, where $\Tilde{U}_{Lm}$ operates the ancilla qubit and qubits corresponding to $m$-th subsystem, and $\Tilde{U}_{U}$ operates on the ancilla qubit and qubits corresponding the first qubit of each subsystem.
Then the ratio of $T_h$ for different orthonormal basis states is
\begin{equation}
    \begin{aligned}
        \frac{T_h}{T_{h'}} &= \frac{(1-p)\braket{\Tilde{\psi}_\mathrm{HTN}}{\phi_h}}{(1-p)\braket{\Tilde{\psi}_\mathrm{HTN}}{\phi_{h'}}}\\
        &=\frac{\braket{\Tilde{\psi}_\mathrm{HTN}}{\phi_h}}{\braket{\Tilde{\psi}_\mathrm{HTN}}{\phi_{h'}}}.
    \end{aligned}
\end{equation}
Since $E_\mathrm{proj}$ is an estimator that includes the normalization of the wave function in HTN, HTN+QMC is unaffected by depolarizing noise.
Figure~\ref{fig: Overlap_compare_normalization_1column.eps}(a) and (b) show the results of $T_h$ in the hydrogen plane model without and with the normalization, respectively.
The fidelity is increased from 0.04 to 0.60 through normalization, whereas the fidelity is 0.46 in the statevector.
As a result, the noise robustness of HTN+QMC, which was experimentally observed, was supported theoretically and numerically.

\section{Remaining results of the decomposition settings}
\label{sec: other data of comparison of the decomposition data}
We first consider a simple measure of decomposition, an average interaction strength between the subsystems $G_{mr}$, defined as
\begin{equation}
\begin{aligned}
G_{mr} = \frac{1}{k} \sum_a \abs{c_a} \delta_{amr},
\label{Eq: decomposition cost}
\end{aligned}
\end{equation}
where $\delta_{amr} = 1$ if Pauli $X, Y$, or $Z$ operator is included in $\bigotimes_{r} P_{amr}$ of the Hamiltonian, as defined in Eq.~\eqref{Eq: rewrite Hamiltonian1}, in more than two subsystems, and $\delta_{amr} = 0$ otherwise.
Table~\ref{tab: values of G} show the values of $G_{mr}$ for the Heisenberg model with $k=2$ and $J_\mathrm{inter}=1.0$, the graphite model, the hydrogen plane model with ID 3, and MABI. 
We confirmed that the results, especially in the physical models, are almost consistent with intuition. 
For example, $G_{mr}$ in the cluster and even-odd settings with the Heisenberg model are 1.5 and 10.5 Hartree, respectively, which corresponds to the results in Fig.~\ref{fig: Ana_Dist_Hei_2column.eps}(a), (b), and (c).
However, the chemical models, especially in the hydrogen plane model, disagree with intuition, indicating the need for more sophisticated measures in complicated systems. 

The mutual information, representing the degree of dependence between random variables, could be used as the alternative measure~\cite{Zhang2020-vv, Zhang2022-mu}. 
The mutual information for two and three subsystems $I(A;B)$ and $I(A;B;C)$ are defined as
\begin{equation}
    \begin{aligned}
        I(A;B) = H(A) + H(B) - H(A,B),\\
\end{aligned}
\end{equation}
and
\begin{equation}
    \begin{aligned}        
        I(A;B;C) = I(A;B) - I(A;B|C),\\
    \end{aligned}
\end{equation}
respectively, where $I(A; B|C)=I(A|C) - I(A|B, C), I(A|C) = H(C, A) - H(C), I(A|B, C) = H(B, C, A) - H(B, C)$, and $H(\chi)$ is von Neumann entropy calculated from the reduced density matrix for the subsystem $\chi$.
We show the values using the exact ground state in the ``MI'' column of Table~\ref{tab: values of G}, where we transformed the wave function, which was originally constructed in a NumPy array to the reduced density matrix using PennyLane~\cite{Bergholm2018-cn}.
The tendency of the mutual information seems to be more consistent with intuition than that of $G_{mr}$.
Although the exact ground states cannot be obtained in realistic situations, in a previous study~\cite{Zhang2020-vv}, the classical construction of approximate value using the density matrix renormalization group is proposed.

Figures~\ref{fig: Analysis_Gr_1column}, \ref{fig: Analysis_Hp_1column}, and \ref{fig: Analysis_MABI_1column} show the results of the analysis for the graphite model, the hydrogen plane model with ID 3, and MABI, respectively.
We also executed the VQE and HTN+VQE using the initial state close to the Hartree-Fock state;
in the particle preserving ansatz, the Hartree-Fock state can be achieved by starting from the state corresponding to the Hartree-Fock state and setting all parameters to zero. However, it is non-trivial in hardware-efficient ansatz because the ansatz does not preserve the particle number even when all parameters are zero in general.
Nonetheless, by using the property that the unitary matrix $U$ corresponding to the real amplitude ansatz with all parameters set to zero satisfies $U\ket{0}^{\otimes nk} = \ket{0}^{\otimes nk}$, the orthonormal state $\ket{\phi_{h}}$ can be constructed in the state preparation for VQE by applying Pauli $X$ gates on the appropriate qubits after $U$.
We show that the same procedure of the state construction is also possible in HTN+VQE as follows.
In case that the basis state corresponds to a computational basis state $\ket{\phi_{h}} = \bigotimes_{m,r} \ket{j_{mr}(h)} = (\bigotimes_{m,r} X^{j_{mr}(h)}) \ket{0}^{\bigotimes nk}$, the state can be constucted by substituting $U_{Lm}$ for $\bigotimes_{r} X^{j_{mr}(h)} U_{Lm}$ in HTN, where $j_{mr}(h) \in \{0,1\}$: assuming $\ket{\varphi^{i_m}} = \bigotimes_{r} X^{j_{mr}(h)} U_{Lm}\ket{i_m}\ket{0}^{\otimes n-1}$, $\psi_{\Vec{i}} = \mel{\Vec{i}}{U_U}{0}^{\otimes k}$, $U_{Lm}\ket{0}^{\otimes n} = \ket{0}^{\otimes n}$, and $U_U \ket{0}^{\otimes k} = \ket{0}^{\otimes k}$, $\ket{\psi_\mathrm{HTN}}$ in Eq.~(\ref{Eq: tensornetwork0_re}) becomes
\begin{equation}
    \begin{aligned}
    \label{Eq: HF in HTN}
\ket{\psi_\mathrm{HTN}} &=\sum_{\Vec{i}} \psi_{\Vec{i}} \bigotimes_m \ket{\varphi^{i_m}}\\
&= \bigotimes_m \ket{\varphi^{0}}\\
&= (\bigotimes_{m, r} X^{j_{mr}(h)}) \ket{0}^{\otimes nk}.
    \end{aligned}
\end{equation}

We performed the VQE and HTN+VQE from the initial states close to the Hartree-Fock state by using the initial parameters which are chosen randomly from $[0, 0.01)$, and the results for the hydrogen plane model with ID 3 and MABI are shown in Figs.~\ref{fig: Analysis_Hp_1column} and~\ref{fig: Analysis_MABI_1column}, respectively.

\putbib[bibHTNQMC]
\end{bibunit}


\begin{thebibliography}{72}%
\makeatletter
\providecommand \@ifxundefined [1]{%
 \@ifx{#1\undefined}
}%
\providecommand \@ifnum [1]{%
 \ifnum #1\expandafter \@firstoftwo
 \else \expandafter \@secondoftwo
 \fi
}%
\providecommand \@ifx [1]{%
 \ifx #1\expandafter \@firstoftwo
 \else \expandafter \@secondoftwo
 \fi
}%
\providecommand \natexlab [1]{#1}%
\providecommand \enquote  [1]{``#1''}%
\providecommand \bibnamefont  [1]{#1}%
\providecommand \bibfnamefont [1]{#1}%
\providecommand \citenamefont [1]{#1}%
\providecommand \href@noop [0]{\@secondoftwo}%
\providecommand \href [0]{\begingroup \@sanitize@url \@href}%
\providecommand \@href[1]{\@@startlink{#1}\@@href}%
\providecommand \@@href[1]{\endgroup#1\@@endlink}%
\providecommand \@sanitize@url [0]{\catcode `\\12\catcode `\$12\catcode `\&12\catcode `\#12\catcode `\^12\catcode `\_12\catcode `\%12\relax}%
\providecommand \@@startlink[1]{}%
\providecommand \@@endlink[0]{}%
\providecommand \url  [0]{\begingroup\@sanitize@url \@url }%
\providecommand \@url [1]{\endgroup\@href {#1}{\urlprefix }}%
\providecommand \urlprefix  [0]{URL }%
\providecommand \Eprint [0]{\href }%
\providecommand \doibase [0]{http://dx.doi.org/}%
\providecommand \selectlanguage [0]{\@gobble}%
\providecommand \bibinfo  [0]{\@secondoftwo}%
\providecommand \bibfield  [0]{\@secondoftwo}%
\providecommand \translation [1]{[#1]}%
\providecommand \BibitemOpen [0]{}%
\providecommand \bibitemStop [0]{}%
\providecommand \bibitemNoStop [0]{.\EOS\space}%
\providecommand \EOS [0]{\spacefactor3000\relax}%
\providecommand \BibitemShut  [1]{\csname bibitem#1\endcsname}%
\let\auto@bib@innerbib\@empty
\bibitem [{\citenamefont {Ceder}\ \emph {et~al.}(1998)\citenamefont {Ceder}, \citenamefont {Chiang}, \citenamefont {Sadoway}, \citenamefont {Aydinol}, \citenamefont {Jang},\ and\ \citenamefont {Huang}}]{Ceder1998-re}%
  \BibitemOpen
  \bibfield  {author} {\bibinfo {author} {\bibfnamefont {G.}~\bibnamefont {Ceder}}, \bibinfo {author} {\bibfnamefont {Y.-M.}\ \bibnamefont {Chiang}}, \bibinfo {author} {\bibfnamefont {D.~R.}\ \bibnamefont {Sadoway}}, \bibinfo {author} {\bibfnamefont {M.~K.}\ \bibnamefont {Aydinol}}, \bibinfo {author} {\bibfnamefont {Y.-I.}\ \bibnamefont {Jang}}, \ and\ \bibinfo {author} {\bibfnamefont {B.}~\bibnamefont {Huang}},\ }\href@noop {} {\bibfield  {journal} {\bibinfo  {journal} {Nature}\ }\textbf {\bibinfo {volume} {392}},\ \bibinfo {pages} {694} (\bibinfo {year} {1998})}\BibitemShut {NoStop}%
\bibitem [{\citenamefont {Gao}\ \emph {et~al.}(2021)\citenamefont {Gao}, \citenamefont {Nakamura}, \citenamefont {Gujarati}, \citenamefont {Jones}, \citenamefont {Rice}, \citenamefont {Wood}, \citenamefont {Pistoia}, \citenamefont {Garcia},\ and\ \citenamefont {Yamamoto}}]{Gao2021-hr}%
  \BibitemOpen
  \bibfield  {author} {\bibinfo {author} {\bibfnamefont {Q.}~\bibnamefont {Gao}}, \bibinfo {author} {\bibfnamefont {H.}~\bibnamefont {Nakamura}}, \bibinfo {author} {\bibfnamefont {T.~P.}\ \bibnamefont {Gujarati}}, \bibinfo {author} {\bibfnamefont {G.~O.}\ \bibnamefont {Jones}}, \bibinfo {author} {\bibfnamefont {J.~E.}\ \bibnamefont {Rice}}, \bibinfo {author} {\bibfnamefont {S.~P.}\ \bibnamefont {Wood}}, \bibinfo {author} {\bibfnamefont {M.}~\bibnamefont {Pistoia}}, \bibinfo {author} {\bibfnamefont {J.~M.}\ \bibnamefont {Garcia}}, \ and\ \bibinfo {author} {\bibfnamefont {N.}~\bibnamefont {Yamamoto}},\ }\href@noop {} {\bibfield  {journal} {\bibinfo  {journal} {J. Phys. Chem. A}\ }\textbf {\bibinfo {volume} {125}},\ \bibinfo {pages} {1827} (\bibinfo {year} {2021})}\BibitemShut {NoStop}%
\bibitem [{\citenamefont {N{\o}rskov}\ \emph {et~al.}(2009)\citenamefont {N{\o}rskov}, \citenamefont {Bligaard}, \citenamefont {Rossmeisl},\ and\ \citenamefont {Christensen}}]{Norskov2009-fr}%
  \BibitemOpen
  \bibfield  {author} {\bibinfo {author} {\bibfnamefont {J.~K.}\ \bibnamefont {N{\o}rskov}}, \bibinfo {author} {\bibfnamefont {T.}~\bibnamefont {Bligaard}}, \bibinfo {author} {\bibfnamefont {J.}~\bibnamefont {Rossmeisl}}, \ and\ \bibinfo {author} {\bibfnamefont {C.~H.}\ \bibnamefont {Christensen}},\ }\href@noop {} {\bibfield  {journal} {\bibinfo  {journal} {Nat. Chem.}\ }\textbf {\bibinfo {volume} {1}},\ \bibinfo {pages} {37} (\bibinfo {year} {2009})}\BibitemShut {NoStop}%
\bibitem [{\citenamefont {Turro}(1991)}]{Turro1991-qw}%
  \BibitemOpen
  \bibfield  {author} {\bibinfo {author} {\bibfnamefont {N.~J.}\ \bibnamefont {Turro}},\ }\href@noop {} {\emph {\bibinfo {title} {Modern Molecular Photochemistry}}}\ (\bibinfo  {publisher} {University Science Books},\ \bibinfo {year} {1991})\BibitemShut {NoStop}%
\bibitem [{\citenamefont {Michl}\ and\ \citenamefont {Bonacic-Koutecky}(1990)}]{Michl1990-pn}%
  \BibitemOpen
  \bibfield  {author} {\bibinfo {author} {\bibfnamefont {J.}~\bibnamefont {Michl}}\ and\ \bibinfo {author} {\bibfnamefont {V.}~\bibnamefont {Bonacic-Koutecky}},\ }\href@noop {} {\emph {\bibinfo {title} {Electronic Aspects of Organic Photochemistry}}}\ (\bibinfo  {publisher} {Wiley},\ \bibinfo {year} {1990})\BibitemShut {NoStop}%
\bibitem [{\citenamefont {Bauer}\ \emph {et~al.}(2020)\citenamefont {Bauer}, \citenamefont {Bravyi}, \citenamefont {Motta},\ and\ \citenamefont {Kin-Lic~Chan}}]{Bauer2020-ug}%
  \BibitemOpen
  \bibfield  {author} {\bibinfo {author} {\bibfnamefont {B.}~\bibnamefont {Bauer}}, \bibinfo {author} {\bibfnamefont {S.}~\bibnamefont {Bravyi}}, \bibinfo {author} {\bibfnamefont {M.}~\bibnamefont {Motta}}, \ and\ \bibinfo {author} {\bibfnamefont {G.}~\bibnamefont {Kin-Lic~Chan}},\ }\href@noop {} {\bibfield  {journal} {\bibinfo  {journal} {Chem. Rev.}\ }\textbf {\bibinfo {volume} {120}},\ \bibinfo {pages} {12685} (\bibinfo {year} {2020})}\BibitemShut {NoStop}%
\bibitem [{\citenamefont {Preskill}(2018)}]{Preskill2018-sc}%
  \BibitemOpen
  \bibfield  {author} {\bibinfo {author} {\bibfnamefont {J.}~\bibnamefont {Preskill}},\ }\href@noop {} {\bibfield  {journal} {\bibinfo  {journal} {Quantum}\ }\textbf {\bibinfo {volume} {2}},\ \bibinfo {pages} {79} (\bibinfo {year} {2018})}\BibitemShut {NoStop}%
\bibitem [{\citenamefont {Bharti}\ \emph {et~al.}(2022)\citenamefont {Bharti}, \citenamefont {Cervera-Lierta}, \citenamefont {Kyaw}, \citenamefont {Haug}, \citenamefont {Alperin-Lea}, \citenamefont {Anand}, \citenamefont {Degroote}, \citenamefont {Heimonen}, \citenamefont {Kottmann}, \citenamefont {Menke}, \citenamefont {Mok}, \citenamefont {Sim}, \citenamefont {Kwek},\ and\ \citenamefont {Aspuru-Guzik}}]{Bharti2022-ov}%
  \BibitemOpen
  \bibfield  {author} {\bibinfo {author} {\bibfnamefont {K.}~\bibnamefont {Bharti}}, \bibinfo {author} {\bibfnamefont {A.}~\bibnamefont {Cervera-Lierta}}, \bibinfo {author} {\bibfnamefont {T.~H.}\ \bibnamefont {Kyaw}}, \bibinfo {author} {\bibfnamefont {T.}~\bibnamefont {Haug}}, \bibinfo {author} {\bibfnamefont {S.}~\bibnamefont {Alperin-Lea}}, \bibinfo {author} {\bibfnamefont {A.}~\bibnamefont {Anand}}, \bibinfo {author} {\bibfnamefont {M.}~\bibnamefont {Degroote}}, \bibinfo {author} {\bibfnamefont {H.}~\bibnamefont {Heimonen}}, \bibinfo {author} {\bibfnamefont {J.~S.}\ \bibnamefont {Kottmann}}, \bibinfo {author} {\bibfnamefont {T.}~\bibnamefont {Menke}}, \bibinfo {author} {\bibfnamefont {W.-K.}\ \bibnamefont {Mok}}, \bibinfo {author} {\bibfnamefont {S.}~\bibnamefont {Sim}}, \bibinfo {author} {\bibfnamefont {L.-C.}\ \bibnamefont {Kwek}}, \ and\ \bibinfo {author} {\bibfnamefont {A.}~\bibnamefont {Aspuru-Guzik}},\ }\href@noop {} {\bibfield  {journal} {\bibinfo  {journal} {Rev. Mod. Phys.}\ }\textbf {\bibinfo
  {volume} {94}},\ \bibinfo {pages} {015004} (\bibinfo {year} {2022})}\BibitemShut {NoStop}%
\bibitem [{\citenamefont {Cerezo}\ \emph {et~al.}(2021{\natexlab{a}})\citenamefont {Cerezo}, \citenamefont {Arrasmith}, \citenamefont {Babbush}, \citenamefont {Benjamin}, \citenamefont {Endo}, \citenamefont {Fujii}, \citenamefont {McClean}, \citenamefont {Mitarai}, \citenamefont {Yuan}, \citenamefont {Cincio},\ and\ \citenamefont {Coles}}]{Cerezo2021-dy}%
  \BibitemOpen
  \bibfield  {author} {\bibinfo {author} {\bibfnamefont {M.}~\bibnamefont {Cerezo}}, \bibinfo {author} {\bibfnamefont {A.}~\bibnamefont {Arrasmith}}, \bibinfo {author} {\bibfnamefont {R.}~\bibnamefont {Babbush}}, \bibinfo {author} {\bibfnamefont {S.~C.}\ \bibnamefont {Benjamin}}, \bibinfo {author} {\bibfnamefont {S.}~\bibnamefont {Endo}}, \bibinfo {author} {\bibfnamefont {K.}~\bibnamefont {Fujii}}, \bibinfo {author} {\bibfnamefont {J.~R.}\ \bibnamefont {McClean}}, \bibinfo {author} {\bibfnamefont {K.}~\bibnamefont {Mitarai}}, \bibinfo {author} {\bibfnamefont {X.}~\bibnamefont {Yuan}}, \bibinfo {author} {\bibfnamefont {L.}~\bibnamefont {Cincio}}, \ and\ \bibinfo {author} {\bibfnamefont {P.~J.}\ \bibnamefont {Coles}},\ }\href@noop {} {\bibfield  {journal} {\bibinfo  {journal} {Nature Reviews Physics}\ }\textbf {\bibinfo {volume} {3}},\ \bibinfo {pages} {625} (\bibinfo {year} {2021}{\natexlab{a}})}\BibitemShut {NoStop}%
\bibitem [{\citenamefont {Peruzzo}\ \emph {et~al.}(2014)\citenamefont {Peruzzo}, \citenamefont {McClean}, \citenamefont {Shadbolt}, \citenamefont {Yung}, \citenamefont {Zhou}, \citenamefont {Love}, \citenamefont {Aspuru-Guzik},\ and\ \citenamefont {O'Brien}}]{Peruzzo2014-kp}%
  \BibitemOpen
  \bibfield  {author} {\bibinfo {author} {\bibfnamefont {A.}~\bibnamefont {Peruzzo}}, \bibinfo {author} {\bibfnamefont {J.}~\bibnamefont {McClean}}, \bibinfo {author} {\bibfnamefont {P.}~\bibnamefont {Shadbolt}}, \bibinfo {author} {\bibfnamefont {M.-H.}\ \bibnamefont {Yung}}, \bibinfo {author} {\bibfnamefont {X.-Q.}\ \bibnamefont {Zhou}}, \bibinfo {author} {\bibfnamefont {P.~J.}\ \bibnamefont {Love}}, \bibinfo {author} {\bibfnamefont {A.}~\bibnamefont {Aspuru-Guzik}}, \ and\ \bibinfo {author} {\bibfnamefont {J.~L.}\ \bibnamefont {O'Brien}},\ }\href@noop {} {\bibfield  {journal} {\bibinfo  {journal} {Nat. Commun.}\ }\textbf {\bibinfo {volume} {5}},\ \bibinfo {pages} {4213} (\bibinfo {year} {2014})}\BibitemShut {NoStop}%
\bibitem [{\citenamefont {Yu.~Kitaev}(1995)}]{Yu_Kitaev1995-cf}%
  \BibitemOpen
  \bibfield  {author} {\bibinfo {author} {\bibfnamefont {A.}~\bibnamefont {Yu.~Kitaev}},\ }\href@noop {} {\  (\bibinfo {year} {1995})},\ \Eprint {http://arxiv.org/abs/quant-ph/9511026} {arXiv:quant-ph/9511026 [quant-ph]} \BibitemShut {NoStop}%
\bibitem [{\citenamefont {Stilck~Fran{\c c}a}\ and\ \citenamefont {Garc{\'\i}a-Patr{\'o}n}(2021)}]{Stilck_Franca2021-qx}%
  \BibitemOpen
  \bibfield  {author} {\bibinfo {author} {\bibfnamefont {D.}~\bibnamefont {Stilck~Fran{\c c}a}}\ and\ \bibinfo {author} {\bibfnamefont {R.}~\bibnamefont {Garc{\'\i}a-Patr{\'o}n}},\ }\href@noop {} {\bibfield  {journal} {\bibinfo  {journal} {Nat. Phys.}\ }\textbf {\bibinfo {volume} {17}},\ \bibinfo {pages} {1221} (\bibinfo {year} {2021})}\BibitemShut {NoStop}%
\bibitem [{\citenamefont {McClean}\ \emph {et~al.}(2018)\citenamefont {McClean}, \citenamefont {Boixo}, \citenamefont {Smelyanskiy}, \citenamefont {Babbush},\ and\ \citenamefont {Neven}}]{McClean2018-pn}%
  \BibitemOpen
  \bibfield  {author} {\bibinfo {author} {\bibfnamefont {J.~R.}\ \bibnamefont {McClean}}, \bibinfo {author} {\bibfnamefont {S.}~\bibnamefont {Boixo}}, \bibinfo {author} {\bibfnamefont {V.~N.}\ \bibnamefont {Smelyanskiy}}, \bibinfo {author} {\bibfnamefont {R.}~\bibnamefont {Babbush}}, \ and\ \bibinfo {author} {\bibfnamefont {H.}~\bibnamefont {Neven}},\ }\href@noop {} {\bibfield  {journal} {\bibinfo  {journal} {Nat. Commun.}\ }\textbf {\bibinfo {volume} {9}},\ \bibinfo {pages} {4812} (\bibinfo {year} {2018})}\BibitemShut {NoStop}%
\bibitem [{\citenamefont {Grant}\ \emph {et~al.}(2019)\citenamefont {Grant}, \citenamefont {Wossnig}, \citenamefont {Ostaszewski},\ and\ \citenamefont {Benedetti}}]{Grant2019-ul}%
  \BibitemOpen
  \bibfield  {author} {\bibinfo {author} {\bibfnamefont {E.}~\bibnamefont {Grant}}, \bibinfo {author} {\bibfnamefont {L.}~\bibnamefont {Wossnig}}, \bibinfo {author} {\bibfnamefont {M.}~\bibnamefont {Ostaszewski}}, \ and\ \bibinfo {author} {\bibfnamefont {M.}~\bibnamefont {Benedetti}},\ }\href@noop {} {\bibfield  {journal} {\bibinfo  {journal} {Quantum}\ }\textbf {\bibinfo {volume} {3}},\ \bibinfo {pages} {214} (\bibinfo {year} {2019})}\BibitemShut {NoStop}%
\bibitem [{\citenamefont {Skolik}\ \emph {et~al.}(2021)\citenamefont {Skolik}, \citenamefont {McClean}, \citenamefont {Mohseni}, \citenamefont {van~der Smagt},\ and\ \citenamefont {Leib}}]{Skolik2021-ju}%
  \BibitemOpen
  \bibfield  {author} {\bibinfo {author} {\bibfnamefont {A.}~\bibnamefont {Skolik}}, \bibinfo {author} {\bibfnamefont {J.~R.}\ \bibnamefont {McClean}}, \bibinfo {author} {\bibfnamefont {M.}~\bibnamefont {Mohseni}}, \bibinfo {author} {\bibfnamefont {P.}~\bibnamefont {van~der Smagt}}, \ and\ \bibinfo {author} {\bibfnamefont {M.}~\bibnamefont {Leib}},\ }\href@noop {} {\bibfield  {journal} {\bibinfo  {journal} {Quantum Machine Intelligence}\ }\textbf {\bibinfo {volume} {3}},\ \bibinfo {pages} {5} (\bibinfo {year} {2021})}\BibitemShut {NoStop}%
\bibitem [{\citenamefont {Cerezo}\ \emph {et~al.}(2021{\natexlab{b}})\citenamefont {Cerezo}, \citenamefont {Sone}, \citenamefont {Volkoff}, \citenamefont {Cincio},\ and\ \citenamefont {Coles}}]{Cerezo2021-kj}%
  \BibitemOpen
  \bibfield  {author} {\bibinfo {author} {\bibfnamefont {M.}~\bibnamefont {Cerezo}}, \bibinfo {author} {\bibfnamefont {A.}~\bibnamefont {Sone}}, \bibinfo {author} {\bibfnamefont {T.}~\bibnamefont {Volkoff}}, \bibinfo {author} {\bibfnamefont {L.}~\bibnamefont {Cincio}}, \ and\ \bibinfo {author} {\bibfnamefont {P.~J.}\ \bibnamefont {Coles}},\ }\href@noop {} {\bibfield  {journal} {\bibinfo  {journal} {Nat. Commun.}\ }\textbf {\bibinfo {volume} {12}},\ \bibinfo {pages} {1791} (\bibinfo {year} {2021}{\natexlab{b}})}\BibitemShut {NoStop}%
\bibitem [{\citenamefont {Kanno}\ \emph {et~al.}(2023)\citenamefont {Kanno}, \citenamefont {Kohda}, \citenamefont {Imai}, \citenamefont {Koh}, \citenamefont {Mitarai}, \citenamefont {Mizukami},\ and\ \citenamefont {Nakagawa}}]{Kanno2023-ip}%
  \BibitemOpen
  \bibfield  {author} {\bibinfo {author} {\bibfnamefont {K.}~\bibnamefont {Kanno}}, \bibinfo {author} {\bibfnamefont {M.}~\bibnamefont {Kohda}}, \bibinfo {author} {\bibfnamefont {R.}~\bibnamefont {Imai}}, \bibinfo {author} {\bibfnamefont {S.}~\bibnamefont {Koh}}, \bibinfo {author} {\bibfnamefont {K.}~\bibnamefont {Mitarai}}, \bibinfo {author} {\bibfnamefont {W.}~\bibnamefont {Mizukami}}, \ and\ \bibinfo {author} {\bibfnamefont {Y.~O.}\ \bibnamefont {Nakagawa}},\ }\href@noop {} {\  (\bibinfo {year} {2023})},\ \Eprint {http://arxiv.org/abs/2302.11320} {arXiv:2302.11320 [quant-ph]} \BibitemShut {NoStop}%
\bibitem [{\citenamefont {Huggins}\ \emph {et~al.}(2022)\citenamefont {Huggins}, \citenamefont {O'Gorman}, \citenamefont {Rubin}, \citenamefont {Reichman}, \citenamefont {Babbush},\ and\ \citenamefont {Lee}}]{Huggins2022-ly}%
  \BibitemOpen
  \bibfield  {author} {\bibinfo {author} {\bibfnamefont {W.~J.}\ \bibnamefont {Huggins}}, \bibinfo {author} {\bibfnamefont {B.~A.}\ \bibnamefont {O'Gorman}}, \bibinfo {author} {\bibfnamefont {N.~C.}\ \bibnamefont {Rubin}}, \bibinfo {author} {\bibfnamefont {D.~R.}\ \bibnamefont {Reichman}}, \bibinfo {author} {\bibfnamefont {R.}~\bibnamefont {Babbush}}, \ and\ \bibinfo {author} {\bibfnamefont {J.}~\bibnamefont {Lee}},\ }\href@noop {} {\bibfield  {journal} {\bibinfo  {journal} {Nature}\ }\textbf {\bibinfo {volume} {603}},\ \bibinfo {pages} {416} (\bibinfo {year} {2022})}\BibitemShut {NoStop}%
\bibitem [{\citenamefont {Yang}\ \emph {et~al.}(2021)\citenamefont {Yang}, \citenamefont {Lu},\ and\ \citenamefont {Li}}]{Yang2021-ie}%
  \BibitemOpen
  \bibfield  {author} {\bibinfo {author} {\bibfnamefont {Y.}~\bibnamefont {Yang}}, \bibinfo {author} {\bibfnamefont {B.-N.}\ \bibnamefont {Lu}}, \ and\ \bibinfo {author} {\bibfnamefont {Y.}~\bibnamefont {Li}},\ }\href@noop {} {\bibfield  {journal} {\bibinfo  {journal} {PRX Quantum}\ }\textbf {\bibinfo {volume} {2}},\ \bibinfo {pages} {040361} (\bibinfo {year} {2021})}\BibitemShut {NoStop}%
\bibitem [{\citenamefont {Tan}\ \emph {et~al.}(2022)\citenamefont {Tan}, \citenamefont {Bhowmick},\ and\ \citenamefont {Sengupta}}]{Tan2022-fh}%
  \BibitemOpen
  \bibfield  {author} {\bibinfo {author} {\bibfnamefont {K.~C.}\ \bibnamefont {Tan}}, \bibinfo {author} {\bibfnamefont {D.}~\bibnamefont {Bhowmick}}, \ and\ \bibinfo {author} {\bibfnamefont {P.}~\bibnamefont {Sengupta}},\ }\href@noop {} {\bibfield  {journal} {\bibinfo  {journal} {npj Quantum Information}\ }\textbf {\bibinfo {volume} {8}},\ \bibinfo {pages} {1} (\bibinfo {year} {2022})}\BibitemShut {NoStop}%
\bibitem [{\citenamefont {Xu}\ and\ \citenamefont {Li}(2023)}]{Xu2023-sn}%
  \BibitemOpen
  \bibfield  {author} {\bibinfo {author} {\bibfnamefont {X.}~\bibnamefont {Xu}}\ and\ \bibinfo {author} {\bibfnamefont {Y.}~\bibnamefont {Li}},\ }\href@noop {} {\bibfield  {journal} {\bibinfo  {journal} {Quantum}\ }\textbf {\bibinfo {volume} {7}},\ \bibinfo {pages} {1072} (\bibinfo {year} {2023})}\BibitemShut {NoStop}%
\bibitem [{\citenamefont {Zhang}\ \emph {et~al.}(2022)\citenamefont {Zhang}, \citenamefont {Huang}, \citenamefont {Sun}, \citenamefont {Lv},\ and\ \citenamefont {Yuan}}]{Zhang2022-lf}%
  \BibitemOpen
  \bibfield  {author} {\bibinfo {author} {\bibfnamefont {Y.}~\bibnamefont {Zhang}}, \bibinfo {author} {\bibfnamefont {Y.}~\bibnamefont {Huang}}, \bibinfo {author} {\bibfnamefont {J.}~\bibnamefont {Sun}}, \bibinfo {author} {\bibfnamefont {D.}~\bibnamefont {Lv}}, \ and\ \bibinfo {author} {\bibfnamefont {X.}~\bibnamefont {Yuan}},\ }\href@noop {} {\  (\bibinfo {year} {2022})},\ \Eprint {http://arxiv.org/abs/2206.10431} {arXiv:2206.10431 [quant-ph]} \BibitemShut {NoStop}%
\bibitem [{\citenamefont {Layden}\ \emph {et~al.}(2023)\citenamefont {Layden}, \citenamefont {Mazzola}, \citenamefont {Mishmash}, \citenamefont {Motta}, \citenamefont {Wocjan}, \citenamefont {Kim},\ and\ \citenamefont {Sheldon}}]{Layden2023-ke}%
  \BibitemOpen
  \bibfield  {author} {\bibinfo {author} {\bibfnamefont {D.}~\bibnamefont {Layden}}, \bibinfo {author} {\bibfnamefont {G.}~\bibnamefont {Mazzola}}, \bibinfo {author} {\bibfnamefont {R.~V.}\ \bibnamefont {Mishmash}}, \bibinfo {author} {\bibfnamefont {M.}~\bibnamefont {Motta}}, \bibinfo {author} {\bibfnamefont {P.}~\bibnamefont {Wocjan}}, \bibinfo {author} {\bibfnamefont {J.-S.}\ \bibnamefont {Kim}}, \ and\ \bibinfo {author} {\bibfnamefont {S.}~\bibnamefont {Sheldon}},\ }\href@noop {} {\bibfield  {journal} {\bibinfo  {journal} {Nature}\ }\textbf {\bibinfo {volume} {619}},\ \bibinfo {pages} {282} (\bibinfo {year} {2023})}\BibitemShut {NoStop}%
\bibitem [{\citenamefont {Lee}\ \emph {et~al.}(2022)\citenamefont {Lee}, \citenamefont {Reichman}, \citenamefont {Babbush}, \citenamefont {Rubin}, \citenamefont {Malone}, \citenamefont {O'Gorman},\ and\ \citenamefont {Huggins}}]{Lee2022-lq}%
  \BibitemOpen
  \bibfield  {author} {\bibinfo {author} {\bibfnamefont {J.}~\bibnamefont {Lee}}, \bibinfo {author} {\bibfnamefont {D.~R.}\ \bibnamefont {Reichman}}, \bibinfo {author} {\bibfnamefont {R.}~\bibnamefont {Babbush}}, \bibinfo {author} {\bibfnamefont {N.~C.}\ \bibnamefont {Rubin}}, \bibinfo {author} {\bibfnamefont {F.~D.}\ \bibnamefont {Malone}}, \bibinfo {author} {\bibfnamefont {B.}~\bibnamefont {O'Gorman}}, \ and\ \bibinfo {author} {\bibfnamefont {W.~J.}\ \bibnamefont {Huggins}},\ }\href@noop {} {\  (\bibinfo {year} {2022})},\ \Eprint {http://arxiv.org/abs/2207.13776} {arXiv:2207.13776 [quant-ph]} \BibitemShut {NoStop}%
\bibitem [{\citenamefont {Austin}\ \emph {et~al.}(2012)\citenamefont {Austin}, \citenamefont {Zubarev},\ and\ \citenamefont {Lester}}]{Austin2012-qi}%
  \BibitemOpen
  \bibfield  {author} {\bibinfo {author} {\bibfnamefont {B.~M.}\ \bibnamefont {Austin}}, \bibinfo {author} {\bibfnamefont {D.~Y.}\ \bibnamefont {Zubarev}}, \ and\ \bibinfo {author} {\bibfnamefont {W.~A.}\ \bibnamefont {Lester}, \bibfnamefont {Jr}},\ }\href@noop {} {\bibfield  {journal} {\bibinfo  {journal} {Chem. Rev.}\ }\textbf {\bibinfo {volume} {112}},\ \bibinfo {pages} {263} (\bibinfo {year} {2012})}\BibitemShut {NoStop}%
\bibitem [{\citenamefont {Al-Hamdani}\ \emph {et~al.}(2021)\citenamefont {Al-Hamdani}, \citenamefont {Nagy}, \citenamefont {Zen}, \citenamefont {Barton}, \citenamefont {K{\'a}llay}, \citenamefont {Brandenburg},\ and\ \citenamefont {Tkatchenko}}]{Al-Hamdani2021-ns}%
  \BibitemOpen
  \bibfield  {author} {\bibinfo {author} {\bibfnamefont {Y.~S.}\ \bibnamefont {Al-Hamdani}}, \bibinfo {author} {\bibfnamefont {P.~R.}\ \bibnamefont {Nagy}}, \bibinfo {author} {\bibfnamefont {A.}~\bibnamefont {Zen}}, \bibinfo {author} {\bibfnamefont {D.}~\bibnamefont {Barton}}, \bibinfo {author} {\bibfnamefont {M.}~\bibnamefont {K{\'a}llay}}, \bibinfo {author} {\bibfnamefont {J.~G.}\ \bibnamefont {Brandenburg}}, \ and\ \bibinfo {author} {\bibfnamefont {A.}~\bibnamefont {Tkatchenko}},\ }\href@noop {} {\bibfield  {journal} {\bibinfo  {journal} {Nat. Commun.}\ }\textbf {\bibinfo {volume} {12}},\ \bibinfo {pages} {3927} (\bibinfo {year} {2021})}\BibitemShut {NoStop}%
\bibitem [{\citenamefont {McMillan}(1965)}]{McMillan1965-dd}%
  \BibitemOpen
  \bibfield  {author} {\bibinfo {author} {\bibfnamefont {W.~L.}\ \bibnamefont {McMillan}},\ }\href@noop {} {\bibfield  {journal} {\bibinfo  {journal} {Phys. Rev.}\ }\textbf {\bibinfo {volume} {138}},\ \bibinfo {pages} {A442} (\bibinfo {year} {1965})}\BibitemShut {NoStop}%
\bibitem [{\citenamefont {Ceperley}\ \emph {et~al.}(1977)\citenamefont {Ceperley}, \citenamefont {Chester},\ and\ \citenamefont {Kalos}}]{Ceperley1977-yz}%
  \BibitemOpen
  \bibfield  {author} {\bibinfo {author} {\bibfnamefont {D.}~\bibnamefont {Ceperley}}, \bibinfo {author} {\bibfnamefont {G.~V.}\ \bibnamefont {Chester}}, \ and\ \bibinfo {author} {\bibfnamefont {M.~H.}\ \bibnamefont {Kalos}},\ }\href@noop {} {\bibfield  {journal} {\bibinfo  {journal} {Phys. Rev.}\ }\textbf {\bibinfo {volume} {16}},\ \bibinfo {pages} {3081} (\bibinfo {year} {1977})}\BibitemShut {NoStop}%
\bibitem [{\citenamefont {Blankenbecler}\ \emph {et~al.}(1981)\citenamefont {Blankenbecler}, \citenamefont {Scalapino},\ and\ \citenamefont {Sugar}}]{Blankenbecler1981-is}%
  \BibitemOpen
  \bibfield  {author} {\bibinfo {author} {\bibfnamefont {R.}~\bibnamefont {Blankenbecler}}, \bibinfo {author} {\bibfnamefont {D.~J.}\ \bibnamefont {Scalapino}}, \ and\ \bibinfo {author} {\bibfnamefont {R.~L.}\ \bibnamefont {Sugar}},\ }\href@noop {} {\bibfield  {journal} {\bibinfo  {journal} {Phys. Rev. D}\ }\textbf {\bibinfo {volume} {24}},\ \bibinfo {pages} {2278} (\bibinfo {year} {1981})}\BibitemShut {NoStop}%
\bibitem [{\citenamefont {Sugiyama}\ and\ \citenamefont {Koonin}(1986)}]{Sugiyama1986-xm}%
  \BibitemOpen
  \bibfield  {author} {\bibinfo {author} {\bibfnamefont {G.}~\bibnamefont {Sugiyama}}\ and\ \bibinfo {author} {\bibfnamefont {S.~E.}\ \bibnamefont {Koonin}},\ }\href@noop {} {\bibfield  {journal} {\bibinfo  {journal} {Ann. Phys.}\ }\textbf {\bibinfo {volume} {168}},\ \bibinfo {pages} {1} (\bibinfo {year} {1986})}\BibitemShut {NoStop}%
\bibitem [{\citenamefont {Booth}\ \emph {et~al.}(2009)\citenamefont {Booth}, \citenamefont {Thom},\ and\ \citenamefont {Alavi}}]{Booth2009-lv}%
  \BibitemOpen
  \bibfield  {author} {\bibinfo {author} {\bibfnamefont {G.~H.}\ \bibnamefont {Booth}}, \bibinfo {author} {\bibfnamefont {A.~J.~W.}\ \bibnamefont {Thom}}, \ and\ \bibinfo {author} {\bibfnamefont {A.}~\bibnamefont {Alavi}},\ }\href@noop {} {\bibfield  {journal} {\bibinfo  {journal} {J. Chem. Phys.}\ }\textbf {\bibinfo {volume} {131}},\ \bibinfo {pages} {054106} (\bibinfo {year} {2009})}\BibitemShut {NoStop}%
\bibitem [{\citenamefont {Takeshita}\ \emph {et~al.}(2020)\citenamefont {Takeshita}, \citenamefont {Rubin}, \citenamefont {Jiang}, \citenamefont {Lee}, \citenamefont {Babbush},\ and\ \citenamefont {McClean}}]{Takeshita2020-ec}%
  \BibitemOpen
  \bibfield  {author} {\bibinfo {author} {\bibfnamefont {T.}~\bibnamefont {Takeshita}}, \bibinfo {author} {\bibfnamefont {N.~C.}\ \bibnamefont {Rubin}}, \bibinfo {author} {\bibfnamefont {Z.}~\bibnamefont {Jiang}}, \bibinfo {author} {\bibfnamefont {E.}~\bibnamefont {Lee}}, \bibinfo {author} {\bibfnamefont {R.}~\bibnamefont {Babbush}}, \ and\ \bibinfo {author} {\bibfnamefont {J.~R.}\ \bibnamefont {McClean}},\ }\href@noop {} {\bibfield  {journal} {\bibinfo  {journal} {Phys. Rev. X}\ }\textbf {\bibinfo {volume} {10}},\ \bibinfo {pages} {011004} (\bibinfo {year} {2020})}\BibitemShut {NoStop}%
\bibitem [{\citenamefont {Roos}(2007)}]{Roos2007-cm}%
  \BibitemOpen
  \bibfield  {author} {\bibinfo {author} {\bibfnamefont {B.~O.}\ \bibnamefont {Roos}},\ }in\ \href@noop {} {\emph {\bibinfo {booktitle} {Advances in Chemical Physics}}},\ \bibinfo {series and number} {Advances in chemical physics}\ (\bibinfo  {publisher} {John Wiley \& Sons, Inc.},\ \bibinfo {address} {Hoboken, NJ, USA},\ \bibinfo {year} {2007})\ pp.\ \bibinfo {pages} {399--445}\BibitemShut {NoStop}%
\bibitem [{\citenamefont {Yamazaki}\ \emph {et~al.}(2018)\citenamefont {Yamazaki}, \citenamefont {Matsuura}, \citenamefont {Narimani}, \citenamefont {Saidmuradov},\ and\ \citenamefont {Zaribafiyan}}]{Yamazaki2018-pu}%
  \BibitemOpen
  \bibfield  {author} {\bibinfo {author} {\bibfnamefont {T.}~\bibnamefont {Yamazaki}}, \bibinfo {author} {\bibfnamefont {S.}~\bibnamefont {Matsuura}}, \bibinfo {author} {\bibfnamefont {A.}~\bibnamefont {Narimani}}, \bibinfo {author} {\bibfnamefont {A.}~\bibnamefont {Saidmuradov}}, \ and\ \bibinfo {author} {\bibfnamefont {A.}~\bibnamefont {Zaribafiyan}},\ }\href@noop {} {\  (\bibinfo {year} {2018})},\ \Eprint {http://arxiv.org/abs/1806.01305} {arXiv:1806.01305 [quant-ph]} \BibitemShut {NoStop}%
\bibitem [{\citenamefont {Fujii}\ \emph {et~al.}(2022)\citenamefont {Fujii}, \citenamefont {Mizuta}, \citenamefont {Ueda}, \citenamefont {Mitarai}, \citenamefont {Mizukami},\ and\ \citenamefont {Nakagawa}}]{Fujii2022-bq}%
  \BibitemOpen
  \bibfield  {author} {\bibinfo {author} {\bibfnamefont {K.}~\bibnamefont {Fujii}}, \bibinfo {author} {\bibfnamefont {K.}~\bibnamefont {Mizuta}}, \bibinfo {author} {\bibfnamefont {H.}~\bibnamefont {Ueda}}, \bibinfo {author} {\bibfnamefont {K.}~\bibnamefont {Mitarai}}, \bibinfo {author} {\bibfnamefont {W.}~\bibnamefont {Mizukami}}, \ and\ \bibinfo {author} {\bibfnamefont {Y.~O.}\ \bibnamefont {Nakagawa}},\ }\href@noop {} {\bibfield  {journal} {\bibinfo  {journal} {PRX Quantum}\ }\textbf {\bibinfo {volume} {3}},\ \bibinfo {pages} {010346} (\bibinfo {year} {2022})}\BibitemShut {NoStop}%
\bibitem [{\citenamefont {Kawashima}\ \emph {et~al.}(2021)\citenamefont {Kawashima}, \citenamefont {Lloyd}, \citenamefont {Coons}, \citenamefont {Nam}, \citenamefont {Matsuura}, \citenamefont {Garza}, \citenamefont {Johri}, \citenamefont {Huntington}, \citenamefont {Senicourt}, \citenamefont {Maksymov}, \citenamefont {Nguyen}, \citenamefont {Kim}, \citenamefont {Alidoust}, \citenamefont {Zaribafiyan},\ and\ \citenamefont {Yamazaki}}]{Kawashima2021-os}%
  \BibitemOpen
  \bibfield  {author} {\bibinfo {author} {\bibfnamefont {Y.}~\bibnamefont {Kawashima}}, \bibinfo {author} {\bibfnamefont {E.}~\bibnamefont {Lloyd}}, \bibinfo {author} {\bibfnamefont {M.~P.}\ \bibnamefont {Coons}}, \bibinfo {author} {\bibfnamefont {Y.}~\bibnamefont {Nam}}, \bibinfo {author} {\bibfnamefont {S.}~\bibnamefont {Matsuura}}, \bibinfo {author} {\bibfnamefont {A.~J.}\ \bibnamefont {Garza}}, \bibinfo {author} {\bibfnamefont {S.}~\bibnamefont {Johri}}, \bibinfo {author} {\bibfnamefont {L.}~\bibnamefont {Huntington}}, \bibinfo {author} {\bibfnamefont {V.}~\bibnamefont {Senicourt}}, \bibinfo {author} {\bibfnamefont {A.~O.}\ \bibnamefont {Maksymov}}, \bibinfo {author} {\bibfnamefont {J.~H.~V.}\ \bibnamefont {Nguyen}}, \bibinfo {author} {\bibfnamefont {J.}~\bibnamefont {Kim}}, \bibinfo {author} {\bibfnamefont {N.}~\bibnamefont {Alidoust}}, \bibinfo {author} {\bibfnamefont {A.}~\bibnamefont {Zaribafiyan}}, \ and\ \bibinfo {author} {\bibfnamefont {T.}~\bibnamefont {Yamazaki}},\ }\href@noop {} {\bibfield
  {journal} {\bibinfo  {journal} {Communications Physics}\ }\textbf {\bibinfo {volume} {4}},\ \bibinfo {pages} {1} (\bibinfo {year} {2021})}\BibitemShut {NoStop}%
\bibitem [{\citenamefont {Greene-Diniz}\ \emph {et~al.}(2022)\citenamefont {Greene-Diniz}, \citenamefont {Manrique}, \citenamefont {Sennane}, \citenamefont {Magnin}, \citenamefont {Shishenina}, \citenamefont {Cordier}, \citenamefont {Llewellyn}, \citenamefont {Krompiec}, \citenamefont {Ran{\v c}i{\'c}},\ and\ \citenamefont {Ramo}}]{Greene-Diniz2022-xo}%
  \BibitemOpen
  \bibfield  {author} {\bibinfo {author} {\bibfnamefont {G.}~\bibnamefont {Greene-Diniz}}, \bibinfo {author} {\bibfnamefont {D.~Z.}\ \bibnamefont {Manrique}}, \bibinfo {author} {\bibfnamefont {W.}~\bibnamefont {Sennane}}, \bibinfo {author} {\bibfnamefont {Y.}~\bibnamefont {Magnin}}, \bibinfo {author} {\bibfnamefont {E.}~\bibnamefont {Shishenina}}, \bibinfo {author} {\bibfnamefont {P.}~\bibnamefont {Cordier}}, \bibinfo {author} {\bibfnamefont {P.}~\bibnamefont {Llewellyn}}, \bibinfo {author} {\bibfnamefont {M.}~\bibnamefont {Krompiec}}, \bibinfo {author} {\bibfnamefont {M.~J.}\ \bibnamefont {Ran{\v c}i{\'c}}}, \ and\ \bibinfo {author} {\bibfnamefont {D.~M.}\ \bibnamefont {Ramo}},\ }\href@noop {} {\bibfield  {journal} {\bibinfo  {journal} {EPJ Quantum Technology}\ }\textbf {\bibinfo {volume} {9}},\ \bibinfo {pages} {37} (\bibinfo {year} {2022})}\BibitemShut {NoStop}%
\bibitem [{\citenamefont {Cao}\ \emph {et~al.}(2022)\citenamefont {Cao}, \citenamefont {Sun}, \citenamefont {Yuan}, \citenamefont {Hu}, \citenamefont {Pham},\ and\ \citenamefont {Lv}}]{Cao2022-hn}%
  \BibitemOpen
  \bibfield  {author} {\bibinfo {author} {\bibfnamefont {C.}~\bibnamefont {Cao}}, \bibinfo {author} {\bibfnamefont {J.}~\bibnamefont {Sun}}, \bibinfo {author} {\bibfnamefont {X.}~\bibnamefont {Yuan}}, \bibinfo {author} {\bibfnamefont {H.-S.}\ \bibnamefont {Hu}}, \bibinfo {author} {\bibfnamefont {H.~Q.}\ \bibnamefont {Pham}}, \ and\ \bibinfo {author} {\bibfnamefont {D.}~\bibnamefont {Lv}},\ }\href@noop {} {\  (\bibinfo {year} {2022})},\ \Eprint {http://arxiv.org/abs/2209.03202} {arXiv:2209.03202 [quant-ph]} \BibitemShut {NoStop}%
\bibitem [{\citenamefont {Peng}\ \emph {et~al.}(2020)\citenamefont {Peng}, \citenamefont {Harrow}, \citenamefont {Ozols},\ and\ \citenamefont {Wu}}]{Peng2020-xc}%
  \BibitemOpen
  \bibfield  {author} {\bibinfo {author} {\bibfnamefont {T.}~\bibnamefont {Peng}}, \bibinfo {author} {\bibfnamefont {A.~W.}\ \bibnamefont {Harrow}}, \bibinfo {author} {\bibfnamefont {M.}~\bibnamefont {Ozols}}, \ and\ \bibinfo {author} {\bibfnamefont {X.}~\bibnamefont {Wu}},\ }\href@noop {} {\bibfield  {journal} {\bibinfo  {journal} {Phys. Rev. Lett.}\ }\textbf {\bibinfo {volume} {125}},\ \bibinfo {pages} {150504} (\bibinfo {year} {2020})}\BibitemShut {NoStop}%
\bibitem [{\citenamefont {Harada}\ \emph {et~al.}(2023)\citenamefont {Harada}, \citenamefont {Wada},\ and\ \citenamefont {Yamamoto}}]{Harada2023-ym}%
  \BibitemOpen
  \bibfield  {author} {\bibinfo {author} {\bibfnamefont {H.}~\bibnamefont {Harada}}, \bibinfo {author} {\bibfnamefont {K.}~\bibnamefont {Wada}}, \ and\ \bibinfo {author} {\bibfnamefont {N.}~\bibnamefont {Yamamoto}},\ }\href@noop {} {\  (\bibinfo {year} {2023})},\ \Eprint {http://arxiv.org/abs/2303.07340} {arXiv:2303.07340 [quant-ph]} \BibitemShut {NoStop}%
\bibitem [{\citenamefont {Sun}\ \emph {et~al.}(2022)\citenamefont {Sun}, \citenamefont {Endo}, \citenamefont {Lin}, \citenamefont {Hayden}, \citenamefont {Vedral},\ and\ \citenamefont {Yuan}}]{Sun2022-bh}%
  \BibitemOpen
  \bibfield  {author} {\bibinfo {author} {\bibfnamefont {J.}~\bibnamefont {Sun}}, \bibinfo {author} {\bibfnamefont {S.}~\bibnamefont {Endo}}, \bibinfo {author} {\bibfnamefont {H.}~\bibnamefont {Lin}}, \bibinfo {author} {\bibfnamefont {P.}~\bibnamefont {Hayden}}, \bibinfo {author} {\bibfnamefont {V.}~\bibnamefont {Vedral}}, \ and\ \bibinfo {author} {\bibfnamefont {X.}~\bibnamefont {Yuan}},\ }\href@noop {} {\bibfield  {journal} {\bibinfo  {journal} {Phys. Rev. Lett.}\ }\textbf {\bibinfo {volume} {129}},\ \bibinfo {pages} {120505} (\bibinfo {year} {2022})}\BibitemShut {NoStop}%
\bibitem [{\citenamefont {Huggins}\ \emph {et~al.}(2019)\citenamefont {Huggins}, \citenamefont {Patil}, \citenamefont {Mitchell}, \citenamefont {Birgitta~Whaley},\ and\ \citenamefont {Miles~Stoudenmire}}]{Huggins2019-qa}%
  \BibitemOpen
  \bibfield  {author} {\bibinfo {author} {\bibfnamefont {W.}~\bibnamefont {Huggins}}, \bibinfo {author} {\bibfnamefont {P.}~\bibnamefont {Patil}}, \bibinfo {author} {\bibfnamefont {B.}~\bibnamefont {Mitchell}}, \bibinfo {author} {\bibfnamefont {K.}~\bibnamefont {Birgitta~Whaley}}, \ and\ \bibinfo {author} {\bibfnamefont {E.}~\bibnamefont {Miles~Stoudenmire}},\ }\href@noop {} {\bibfield  {journal} {\bibinfo  {journal} {Quantum Sci. Technol.}\ }\textbf {\bibinfo {volume} {4}},\ \bibinfo {pages} {024001} (\bibinfo {year} {2019})}\BibitemShut {NoStop}%
\bibitem [{\citenamefont {Yuan}\ \emph {et~al.}(2021)\citenamefont {Yuan}, \citenamefont {Sun}, \citenamefont {Liu}, \citenamefont {Zhao},\ and\ \citenamefont {Zhou}}]{Yuan2021-ih}%
  \BibitemOpen
  \bibfield  {author} {\bibinfo {author} {\bibfnamefont {X.}~\bibnamefont {Yuan}}, \bibinfo {author} {\bibfnamefont {J.}~\bibnamefont {Sun}}, \bibinfo {author} {\bibfnamefont {J.}~\bibnamefont {Liu}}, \bibinfo {author} {\bibfnamefont {Q.}~\bibnamefont {Zhao}}, \ and\ \bibinfo {author} {\bibfnamefont {Y.}~\bibnamefont {Zhou}},\ }\href@noop {} {\bibfield  {journal} {\bibinfo  {journal} {Phys. Rev. Lett.}\ }\textbf {\bibinfo {volume} {127}},\ \bibinfo {pages} {040501} (\bibinfo {year} {2021})}\BibitemShut {NoStop}%
\bibitem [{\citenamefont {Schollw{\"o}ck}(2011)}]{Schollwock2011-im}%
  \BibitemOpen
  \bibfield  {author} {\bibinfo {author} {\bibfnamefont {U.}~\bibnamefont {Schollw{\"o}ck}},\ }\href@noop {} {\bibfield  {journal} {\bibinfo  {journal} {Ann. Phys.}\ }\textbf {\bibinfo {volume} {326}},\ \bibinfo {pages} {96} (\bibinfo {year} {2011})}\BibitemShut {NoStop}%
\bibitem [{\citenamefont {Verstraete}\ and\ \citenamefont {Cirac}(2004)}]{Verstraete2004-lj}%
  \BibitemOpen
  \bibfield  {author} {\bibinfo {author} {\bibfnamefont {F.}~\bibnamefont {Verstraete}}\ and\ \bibinfo {author} {\bibfnamefont {J.~I.}\ \bibnamefont {Cirac}},\ }\href@noop {} {\  (\bibinfo {year} {2004})},\ \Eprint {http://arxiv.org/abs/cond-mat/0407066} {arXiv:cond-mat/0407066 [cond-mat.str-el]} \BibitemShut {NoStop}%
\bibitem [{\citenamefont {Shi}\ \emph {et~al.}(2006)\citenamefont {Shi}, \citenamefont {Duan},\ and\ \citenamefont {Vidal}}]{Shi2006-he}%
  \BibitemOpen
  \bibfield  {author} {\bibinfo {author} {\bibfnamefont {Y.-Y.}\ \bibnamefont {Shi}}, \bibinfo {author} {\bibfnamefont {L.-M.}\ \bibnamefont {Duan}}, \ and\ \bibinfo {author} {\bibfnamefont {G.}~\bibnamefont {Vidal}},\ }\href@noop {} {\bibfield  {journal} {\bibinfo  {journal} {Phys. Rev. A}\ }\textbf {\bibinfo {volume} {74}},\ \bibinfo {pages} {022320} (\bibinfo {year} {2006})}\BibitemShut {NoStop}%
\bibitem [{\citenamefont {Eddins}\ \emph {et~al.}(2022)\citenamefont {Eddins}, \citenamefont {Motta}, \citenamefont {Gujarati}, \citenamefont {Bravyi}, \citenamefont {Mezzacapo}, \citenamefont {Hadfield},\ and\ \citenamefont {Sheldon}}]{Eddins2022-ny}%
  \BibitemOpen
  \bibfield  {author} {\bibinfo {author} {\bibfnamefont {A.}~\bibnamefont {Eddins}}, \bibinfo {author} {\bibfnamefont {M.}~\bibnamefont {Motta}}, \bibinfo {author} {\bibfnamefont {T.~P.}\ \bibnamefont {Gujarati}}, \bibinfo {author} {\bibfnamefont {S.}~\bibnamefont {Bravyi}}, \bibinfo {author} {\bibfnamefont {A.}~\bibnamefont {Mezzacapo}}, \bibinfo {author} {\bibfnamefont {C.}~\bibnamefont {Hadfield}}, \ and\ \bibinfo {author} {\bibfnamefont {S.}~\bibnamefont {Sheldon}},\ }\href@noop {} {\bibfield  {journal} {\bibinfo  {journal} {PRX Quantum}\ }\textbf {\bibinfo {volume} {3}},\ \bibinfo {pages} {010309} (\bibinfo {year} {2022})}\BibitemShut {NoStop}%
\bibitem [{\citenamefont {Motta}\ \emph {et~al.}(2023)\citenamefont {Motta}, \citenamefont {Jones}, \citenamefont {Rice}, \citenamefont {Gujarati}, \citenamefont {Sakuma}, \citenamefont {Liepuoniute}, \citenamefont {Garcia},\ and\ \citenamefont {Ohnishi}}]{Motta2023-mg}%
  \BibitemOpen
  \bibfield  {author} {\bibinfo {author} {\bibfnamefont {M.}~\bibnamefont {Motta}}, \bibinfo {author} {\bibfnamefont {G.~O.}\ \bibnamefont {Jones}}, \bibinfo {author} {\bibfnamefont {J.~E.}\ \bibnamefont {Rice}}, \bibinfo {author} {\bibfnamefont {T.~P.}\ \bibnamefont {Gujarati}}, \bibinfo {author} {\bibfnamefont {R.}~\bibnamefont {Sakuma}}, \bibinfo {author} {\bibfnamefont {I.}~\bibnamefont {Liepuoniute}}, \bibinfo {author} {\bibfnamefont {J.~M.}\ \bibnamefont {Garcia}}, \ and\ \bibinfo {author} {\bibfnamefont {Y.-Y.}\ \bibnamefont {Ohnishi}},\ }\href@noop {} {\bibfield  {journal} {\bibinfo  {journal} {Chem. Sci.}\ }\textbf {\bibinfo {volume} {14}},\ \bibinfo {pages} {2915} (\bibinfo {year} {2023})}\BibitemShut {NoStop}%
\bibitem [{\citenamefont {Momma}\ and\ \citenamefont {Izumi}(2011)}]{Momma2011-ck}%
  \BibitemOpen
  \bibfield  {author} {\bibinfo {author} {\bibfnamefont {K.}~\bibnamefont {Momma}}\ and\ \bibinfo {author} {\bibfnamefont {F.}~\bibnamefont {Izumi}},\ }\href@noop {} {\bibfield  {journal} {\bibinfo  {journal} {J. Appl. Crystallogr.}\ }\textbf {\bibinfo {volume} {44}},\ \bibinfo {pages} {1272} (\bibinfo {year} {2011})}\BibitemShut {NoStop}%
\bibitem [{\citenamefont {Motta}\ \emph {et~al.}(2019)\citenamefont {Motta}, \citenamefont {Sun}, \citenamefont {Tan}, \citenamefont {O'Rourke}, \citenamefont {Ye}, \citenamefont {Minnich}, \citenamefont {Brand{\~a}o},\ and\ \citenamefont {Chan}}]{Motta2019-sf}%
  \BibitemOpen
  \bibfield  {author} {\bibinfo {author} {\bibfnamefont {M.}~\bibnamefont {Motta}}, \bibinfo {author} {\bibfnamefont {C.}~\bibnamefont {Sun}}, \bibinfo {author} {\bibfnamefont {A.~T.~K.}\ \bibnamefont {Tan}}, \bibinfo {author} {\bibfnamefont {M.~J.}\ \bibnamefont {O'Rourke}}, \bibinfo {author} {\bibfnamefont {E.}~\bibnamefont {Ye}}, \bibinfo {author} {\bibfnamefont {A.~J.}\ \bibnamefont {Minnich}}, \bibinfo {author} {\bibfnamefont {F.~G. S.~L.}\ \bibnamefont {Brand{\~a}o}}, \ and\ \bibinfo {author} {\bibfnamefont {G.~K.-L.}\ \bibnamefont {Chan}},\ }\href@noop {} {\bibfield  {journal} {\bibinfo  {journal} {Nat. Phys.}\ }\textbf {\bibinfo {volume} {16}},\ \bibinfo {pages} {205} (\bibinfo {year} {2019})}\BibitemShut {NoStop}%
\bibitem [{\citenamefont {Thinius}\ \emph {et~al.}(2014)\citenamefont {Thinius}, \citenamefont {Islam}, \citenamefont {Heitjans},\ and\ \citenamefont {Bredow}}]{Thinius2014-xz}%
  \BibitemOpen
  \bibfield  {author} {\bibinfo {author} {\bibfnamefont {S.}~\bibnamefont {Thinius}}, \bibinfo {author} {\bibfnamefont {M.~M.}\ \bibnamefont {Islam}}, \bibinfo {author} {\bibfnamefont {P.}~\bibnamefont {Heitjans}}, \ and\ \bibinfo {author} {\bibfnamefont {T.}~\bibnamefont {Bredow}},\ }\href@noop {} {\bibfield  {journal} {\bibinfo  {journal} {J. Phys. Chem. C Nanomater. Interfaces}\ }\textbf {\bibinfo {volume} {118}},\ \bibinfo {pages} {2273} (\bibinfo {year} {2014})}\BibitemShut {NoStop}%
\bibitem [{\citenamefont {Kobayashi}\ \emph {et~al.}(2017)\citenamefont {Kobayashi}, \citenamefont {Okajima}, \citenamefont {Sotome}, \citenamefont {Yanai}, \citenamefont {Mutoh}, \citenamefont {Yoneda}, \citenamefont {Shigeta}, \citenamefont {Sakamoto}, \citenamefont {Miyasaka},\ and\ \citenamefont {Abe}}]{Kobayashi2017-ty}%
  \BibitemOpen
  \bibfield  {author} {\bibinfo {author} {\bibfnamefont {Y.}~\bibnamefont {Kobayashi}}, \bibinfo {author} {\bibfnamefont {H.}~\bibnamefont {Okajima}}, \bibinfo {author} {\bibfnamefont {H.}~\bibnamefont {Sotome}}, \bibinfo {author} {\bibfnamefont {T.}~\bibnamefont {Yanai}}, \bibinfo {author} {\bibfnamefont {K.}~\bibnamefont {Mutoh}}, \bibinfo {author} {\bibfnamefont {Y.}~\bibnamefont {Yoneda}}, \bibinfo {author} {\bibfnamefont {Y.}~\bibnamefont {Shigeta}}, \bibinfo {author} {\bibfnamefont {A.}~\bibnamefont {Sakamoto}}, \bibinfo {author} {\bibfnamefont {H.}~\bibnamefont {Miyasaka}}, \ and\ \bibinfo {author} {\bibfnamefont {J.}~\bibnamefont {Abe}},\ }\href@noop {} {\bibfield  {journal} {\bibinfo  {journal} {J. Am. Chem. Soc.}\ }\textbf {\bibinfo {volume} {139}},\ \bibinfo {pages} {6382} (\bibinfo {year} {2017})}\BibitemShut {NoStop}%
\bibitem [{\citenamefont {Amsler}\ \emph {et~al.}(2023)\citenamefont {Amsler}, \citenamefont {Deglmann}, \citenamefont {Degroote}, \citenamefont {Kaicher}, \citenamefont {Kiser}, \citenamefont {K{\"u}hn}, \citenamefont {Kumar}, \citenamefont {Maier}, \citenamefont {Samsonidze}, \citenamefont {Schroeder}, \citenamefont {Streif}, \citenamefont {Vodola},\ and\ \citenamefont {Wever}}]{Amsler2023-vh}%
  \BibitemOpen
  \bibfield  {author} {\bibinfo {author} {\bibfnamefont {M.}~\bibnamefont {Amsler}}, \bibinfo {author} {\bibfnamefont {P.}~\bibnamefont {Deglmann}}, \bibinfo {author} {\bibfnamefont {M.}~\bibnamefont {Degroote}}, \bibinfo {author} {\bibfnamefont {M.~P.}\ \bibnamefont {Kaicher}}, \bibinfo {author} {\bibfnamefont {M.}~\bibnamefont {Kiser}}, \bibinfo {author} {\bibfnamefont {M.}~\bibnamefont {K{\"u}hn}}, \bibinfo {author} {\bibfnamefont {C.}~\bibnamefont {Kumar}}, \bibinfo {author} {\bibfnamefont {A.}~\bibnamefont {Maier}}, \bibinfo {author} {\bibfnamefont {G.}~\bibnamefont {Samsonidze}}, \bibinfo {author} {\bibfnamefont {A.}~\bibnamefont {Schroeder}}, \bibinfo {author} {\bibfnamefont {M.}~\bibnamefont {Streif}}, \bibinfo {author} {\bibfnamefont {D.}~\bibnamefont {Vodola}}, \ and\ \bibinfo {author} {\bibfnamefont {C.}~\bibnamefont {Wever}},\ }\href@noop {} {\  (\bibinfo {year} {2023})},\ \Eprint {http://arxiv.org/abs/2301.11838} {arXiv:2301.11838 [quant-ph]} \BibitemShut {NoStop}%
\bibitem [{\citenamefont {Huggins}\ \emph {et~al.}(2020)\citenamefont {Huggins}, \citenamefont {Lee}, \citenamefont {Baek}, \citenamefont {O'Gorman},\ and\ \citenamefont {Birgitta~Whaley}}]{Huggins2020-fk}%
  \BibitemOpen
  \bibfield  {author} {\bibinfo {author} {\bibfnamefont {W.~J.}\ \bibnamefont {Huggins}}, \bibinfo {author} {\bibfnamefont {J.}~\bibnamefont {Lee}}, \bibinfo {author} {\bibfnamefont {U.}~\bibnamefont {Baek}}, \bibinfo {author} {\bibfnamefont {B.}~\bibnamefont {O'Gorman}}, \ and\ \bibinfo {author} {\bibfnamefont {K.}~\bibnamefont {Birgitta~Whaley}},\ }\href@noop {} {\bibfield  {journal} {\bibinfo  {journal} {New J. Phys.}\ }\textbf {\bibinfo {volume} {22}},\ \bibinfo {pages} {073009} (\bibinfo {year} {2020})}\BibitemShut {NoStop}%
\bibitem [{\citenamefont {Ibe}\ \emph {et~al.}(2022)\citenamefont {Ibe}, \citenamefont {Nakagawa}, \citenamefont {Earnest}, \citenamefont {Yamamoto}, \citenamefont {Mitarai}, \citenamefont {Gao},\ and\ \citenamefont {Kobayashi}}]{Ibe2022-jp}%
  \BibitemOpen
  \bibfield  {author} {\bibinfo {author} {\bibfnamefont {Y.}~\bibnamefont {Ibe}}, \bibinfo {author} {\bibfnamefont {Y.~O.}\ \bibnamefont {Nakagawa}}, \bibinfo {author} {\bibfnamefont {N.}~\bibnamefont {Earnest}}, \bibinfo {author} {\bibfnamefont {T.}~\bibnamefont {Yamamoto}}, \bibinfo {author} {\bibfnamefont {K.}~\bibnamefont {Mitarai}}, \bibinfo {author} {\bibfnamefont {Q.}~\bibnamefont {Gao}}, \ and\ \bibinfo {author} {\bibfnamefont {T.}~\bibnamefont {Kobayashi}},\ }\href@noop {} {\bibfield  {journal} {\bibinfo  {journal} {Phys. Rev. Res.}\ }\textbf {\bibinfo {volume} {4}},\ \bibinfo {pages} {013173} (\bibinfo {year} {2022})}\BibitemShut {NoStop}%
\bibitem [{\citenamefont {Sawaya}\ and\ \citenamefont {Huh}(2023)}]{Sawaya2023-wx}%
  \BibitemOpen
  \bibfield  {author} {\bibinfo {author} {\bibfnamefont {N.~P.~D.}\ \bibnamefont {Sawaya}}\ and\ \bibinfo {author} {\bibfnamefont {J.}~\bibnamefont {Huh}},\ }\href@noop {} {\bibfield  {journal} {\bibinfo  {journal} {Adv. Quantum Technol.}\ } (\bibinfo {year} {2023})}\BibitemShut {NoStop}%
\bibitem [{\citenamefont {Haghshenas}\ \emph {et~al.}(2022)\citenamefont {Haghshenas}, \citenamefont {Gray}, \citenamefont {Potter},\ and\ \citenamefont {Chan}}]{Haghshenas2022-zn}%
  \BibitemOpen
  \bibfield  {author} {\bibinfo {author} {\bibfnamefont {R.}~\bibnamefont {Haghshenas}}, \bibinfo {author} {\bibfnamefont {J.}~\bibnamefont {Gray}}, \bibinfo {author} {\bibfnamefont {A.~C.}\ \bibnamefont {Potter}}, \ and\ \bibinfo {author} {\bibfnamefont {G.~K.-L.}\ \bibnamefont {Chan}},\ }\href@noop {} {\bibfield  {journal} {\bibinfo  {journal} {Phys. Rev. X}\ }\textbf {\bibinfo {volume} {12}},\ \bibinfo {pages} {011047} (\bibinfo {year} {2022})}\BibitemShut {NoStop}%
\bibitem [{\citenamefont {Matthews}(2021)}]{Matthews2021-pl}%
  \BibitemOpen
  \bibfield  {author} {\bibinfo {author} {\bibfnamefont {D.}~\bibnamefont {Matthews}},\ }\href@noop {} {\bibfield  {journal} {\bibinfo  {journal} {Nature}\ }\textbf {\bibinfo {volume} {591}},\ \bibinfo {pages} {166} (\bibinfo {year} {2021})}\BibitemShut {NoStop}%
\bibitem [{\citenamefont {Jim{\'e}nez-Hoyos}\ and\ \citenamefont {Scuseria}(2015)}]{Jimenez-Hoyos2015-sh}%
  \BibitemOpen
  \bibfield  {author} {\bibinfo {author} {\bibfnamefont {C.~A.}\ \bibnamefont {Jim{\'e}nez-Hoyos}}\ and\ \bibinfo {author} {\bibfnamefont {G.~E.}\ \bibnamefont {Scuseria}},\ }\href@noop {} {\bibfield  {journal} {\bibinfo  {journal} {Phys. Rev. B Condens. Matter}\ }\textbf {\bibinfo {volume} {92}},\ \bibinfo {pages} {085101} (\bibinfo {year} {2015})}\BibitemShut {NoStop}%
\bibitem [{\citenamefont {Abraham}\ and\ \citenamefont {Mayhall}(2020)}]{Abraham2020-ih}%
  \BibitemOpen
  \bibfield  {author} {\bibinfo {author} {\bibfnamefont {V.}~\bibnamefont {Abraham}}\ and\ \bibinfo {author} {\bibfnamefont {N.~J.}\ \bibnamefont {Mayhall}},\ }\href@noop {} {\bibfield  {journal} {\bibinfo  {journal} {J. Chem. Theory Comput.}\ }\textbf {\bibinfo {volume} {16}},\ \bibinfo {pages} {6098} (\bibinfo {year} {2020})}\BibitemShut {NoStop}%
\bibitem [{\citenamefont {Li}(2021)}]{Li2021-uk}%
  \BibitemOpen
  \bibfield  {author} {\bibinfo {author} {\bibfnamefont {Z.}~\bibnamefont {Li}},\ }\href@noop {} {\bibfield  {journal} {\bibinfo  {journal} {Electron. Struct.}\ }\textbf {\bibinfo {volume} {3}},\ \bibinfo {pages} {014001} (\bibinfo {year} {2021})}\BibitemShut {NoStop}%
\bibitem [{\citenamefont {Parker}\ \emph {et~al.}(2013)\citenamefont {Parker}, \citenamefont {Seideman}, \citenamefont {Ratner},\ and\ \citenamefont {Shiozaki}}]{Parker2013-di}%
  \BibitemOpen
  \bibfield  {author} {\bibinfo {author} {\bibfnamefont {S.~M.}\ \bibnamefont {Parker}}, \bibinfo {author} {\bibfnamefont {T.}~\bibnamefont {Seideman}}, \bibinfo {author} {\bibfnamefont {M.~A.}\ \bibnamefont {Ratner}}, \ and\ \bibinfo {author} {\bibfnamefont {T.}~\bibnamefont {Shiozaki}},\ }\href@noop {} {\bibfield  {journal} {\bibinfo  {journal} {J. Chem. Phys.}\ }\textbf {\bibinfo {volume} {139}},\ \bibinfo {pages} {021108} (\bibinfo {year} {2013})}\BibitemShut {NoStop}%
\bibitem [{\citenamefont {Taube}\ and\ \citenamefont {Bartlett}(2006)}]{Taube2006-jj}%
  \BibitemOpen
  \bibfield  {author} {\bibinfo {author} {\bibfnamefont {A.~G.}\ \bibnamefont {Taube}}\ and\ \bibinfo {author} {\bibfnamefont {R.~J.}\ \bibnamefont {Bartlett}},\ }\href@noop {} {\bibfield  {journal} {\bibinfo  {journal} {Int. J. Quantum Chem.}\ }\textbf {\bibinfo {volume} {106}},\ \bibinfo {pages} {3393} (\bibinfo {year} {2006})}\BibitemShut {NoStop}%
\bibitem [{\citenamefont {Wecker}\ \emph {et~al.}(2015)\citenamefont {Wecker}, \citenamefont {Hastings},\ and\ \citenamefont {Troyer}}]{Wecker2015-fu}%
  \BibitemOpen
  \bibfield  {author} {\bibinfo {author} {\bibfnamefont {D.}~\bibnamefont {Wecker}}, \bibinfo {author} {\bibfnamefont {M.~B.}\ \bibnamefont {Hastings}}, \ and\ \bibinfo {author} {\bibfnamefont {M.}~\bibnamefont {Troyer}},\ }\href@noop {} {\bibfield  {journal} {\bibinfo  {journal} {Phys. Rev. A}\ }\textbf {\bibinfo {volume} {92}},\ \bibinfo {pages} {042303} (\bibinfo {year} {2015})}\BibitemShut {NoStop}%
\bibitem [{\citenamefont {Kandala}\ \emph {et~al.}(2017)\citenamefont {Kandala}, \citenamefont {Mezzacapo}, \citenamefont {Temme}, \citenamefont {Takita}, \citenamefont {Brink}, \citenamefont {Chow},\ and\ \citenamefont {Gambetta}}]{Kandala2017-lh}%
  \BibitemOpen
  \bibfield  {author} {\bibinfo {author} {\bibfnamefont {A.}~\bibnamefont {Kandala}}, \bibinfo {author} {\bibfnamefont {A.}~\bibnamefont {Mezzacapo}}, \bibinfo {author} {\bibfnamefont {K.}~\bibnamefont {Temme}}, \bibinfo {author} {\bibfnamefont {M.}~\bibnamefont {Takita}}, \bibinfo {author} {\bibfnamefont {M.}~\bibnamefont {Brink}}, \bibinfo {author} {\bibfnamefont {J.~M.}\ \bibnamefont {Chow}}, \ and\ \bibinfo {author} {\bibfnamefont {J.~M.}\ \bibnamefont {Gambetta}},\ }\href@noop {} {\bibfield  {journal} {\bibinfo  {journal} {Nature}\ }\textbf {\bibinfo {volume} {549}},\ \bibinfo {pages} {242} (\bibinfo {year} {2017})}\BibitemShut {NoStop}%
\bibitem [{\citenamefont {Tilly}\ \emph {et~al.}(2022)\citenamefont {Tilly}, \citenamefont {Chen}, \citenamefont {Cao}, \citenamefont {Picozzi}, \citenamefont {Setia}, \citenamefont {Li}, \citenamefont {Grant}, \citenamefont {Wossnig}, \citenamefont {Rungger}, \citenamefont {Booth},\ and\ \citenamefont {Tennyson}}]{Tilly2022-pb}%
  \BibitemOpen
  \bibfield  {author} {\bibinfo {author} {\bibfnamefont {J.}~\bibnamefont {Tilly}}, \bibinfo {author} {\bibfnamefont {H.}~\bibnamefont {Chen}}, \bibinfo {author} {\bibfnamefont {S.}~\bibnamefont {Cao}}, \bibinfo {author} {\bibfnamefont {D.}~\bibnamefont {Picozzi}}, \bibinfo {author} {\bibfnamefont {K.}~\bibnamefont {Setia}}, \bibinfo {author} {\bibfnamefont {Y.}~\bibnamefont {Li}}, \bibinfo {author} {\bibfnamefont {E.}~\bibnamefont {Grant}}, \bibinfo {author} {\bibfnamefont {L.}~\bibnamefont {Wossnig}}, \bibinfo {author} {\bibfnamefont {I.}~\bibnamefont {Rungger}}, \bibinfo {author} {\bibfnamefont {G.~H.}\ \bibnamefont {Booth}}, \ and\ \bibinfo {author} {\bibfnamefont {J.}~\bibnamefont {Tennyson}},\ }\href@noop {} {\bibfield  {journal} {\bibinfo  {journal} {Phys. Rep.}\ }\textbf {\bibinfo {volume} {986}},\ \bibinfo {pages} {1} (\bibinfo {year} {2022})}\BibitemShut {NoStop}%
\bibitem [{\citenamefont {Jordan}\ and\ \citenamefont {Wigner}(1928)}]{Jordan1928-hf}%
  \BibitemOpen
  \bibfield  {author} {\bibinfo {author} {\bibfnamefont {P.}~\bibnamefont {Jordan}}\ and\ \bibinfo {author} {\bibfnamefont {E.}~\bibnamefont {Wigner}},\ }\href@noop {} {\bibfield  {journal} {\bibinfo  {journal} {Zeitschrift f{\"u}r Physik}\ }\textbf {\bibinfo {volume} {47}},\ \bibinfo {pages} {631} (\bibinfo {year} {1928})}\BibitemShut {NoStop}%
\bibitem [{\citenamefont {Apaja}(2018)}]{Apaja2018-fi}%
  \BibitemOpen
  \bibfield  {author} {\bibinfo {author} {\bibfnamefont {V.}~\bibnamefont {Apaja}},\ }\href@noop {} {\enquote {\bibinfo {title} {Quantum monte carlo},}\ }\bibinfo {howpublished} {\url{http://users.jyu.fi/~veapaja/QMC/MC-lecture.pdf}} (\bibinfo {year} {2018}),\ \bibinfo {note} {accessed: 2023-3-9}\BibitemShut {NoStop}%
\bibitem [{\citenamefont {Huang}\ \emph {et~al.}(2020)\citenamefont {Huang}, \citenamefont {Kueng},\ and\ \citenamefont {Preskill}}]{Huang2020-ld}%
  \BibitemOpen
  \bibfield  {author} {\bibinfo {author} {\bibfnamefont {H.-Y.}\ \bibnamefont {Huang}}, \bibinfo {author} {\bibfnamefont {R.}~\bibnamefont {Kueng}}, \ and\ \bibinfo {author} {\bibfnamefont {J.}~\bibnamefont {Preskill}},\ }\href@noop {} {\bibfield  {journal} {\bibinfo  {journal} {Nat. Phys.}\ }\textbf {\bibinfo {volume} {16}},\ \bibinfo {pages} {1050} (\bibinfo {year} {2020})}\BibitemShut {NoStop}%
\bibitem [{\citenamefont {Zhao}\ \emph {et~al.}(2021)\citenamefont {Zhao}, \citenamefont {Rubin},\ and\ \citenamefont {Miyake}}]{Zhao2021-kw}%
  \BibitemOpen
  \bibfield  {author} {\bibinfo {author} {\bibfnamefont {A.}~\bibnamefont {Zhao}}, \bibinfo {author} {\bibfnamefont {N.~C.}\ \bibnamefont {Rubin}}, \ and\ \bibinfo {author} {\bibfnamefont {A.}~\bibnamefont {Miyake}},\ }\href@noop {} {\bibfield  {journal} {\bibinfo  {journal} {Phys. Rev. Lett.}\ }\textbf {\bibinfo {volume} {127}},\ \bibinfo {pages} {110504} (\bibinfo {year} {2021})}\BibitemShut {NoStop}%
\bibitem [{\citenamefont {Kanno}\ \emph {et~al.}(2021)\citenamefont {Kanno}, \citenamefont {Endo}, \citenamefont {Suzuki},\ and\ \citenamefont {Tokunaga}}]{Kanno2021-zn}%
  \BibitemOpen
  \bibfield  {author} {\bibinfo {author} {\bibfnamefont {S.}~\bibnamefont {Kanno}}, \bibinfo {author} {\bibfnamefont {S.}~\bibnamefont {Endo}}, \bibinfo {author} {\bibfnamefont {Y.}~\bibnamefont {Suzuki}}, \ and\ \bibinfo {author} {\bibfnamefont {Y.}~\bibnamefont {Tokunaga}},\ }\href@noop {} {\bibfield  {journal} {\bibinfo  {journal} {Phys. Rev. A}\ }\textbf {\bibinfo {volume} {104}},\ \bibinfo {pages} {042424} (\bibinfo {year} {2021})}\BibitemShut {NoStop}%
\bibitem [{\citenamefont {Kiser}\ \emph {et~al.}(2023)\citenamefont {Kiser}, \citenamefont {Schroeder}, \citenamefont {Anselmetti}, \citenamefont {Kumar}, \citenamefont {Moll}, \citenamefont {Streif},\ and\ \citenamefont {Vodola}}]{Kiser2023-hi}%
  \BibitemOpen
  \bibfield  {author} {\bibinfo {author} {\bibfnamefont {M.}~\bibnamefont {Kiser}}, \bibinfo {author} {\bibfnamefont {A.}~\bibnamefont {Schroeder}}, \bibinfo {author} {\bibfnamefont {G.-L.~R.}\ \bibnamefont {Anselmetti}}, \bibinfo {author} {\bibfnamefont {C.}~\bibnamefont {Kumar}}, \bibinfo {author} {\bibfnamefont {N.}~\bibnamefont {Moll}}, \bibinfo {author} {\bibfnamefont {M.}~\bibnamefont {Streif}}, \ and\ \bibinfo {author} {\bibfnamefont {D.}~\bibnamefont {Vodola}},\ }\href@noop {} {\  (\bibinfo {year} {2023})},\ \Eprint {http://arxiv.org/abs/2312.09872} {arXiv:2312.09872 [quant-ph]} \BibitemShut {NoStop}%
\end{thebibliography}%


\begin{thebibliography}{44}%
\makeatletter
\providecommand \@ifxundefined [1]{%
 \@ifx{#1\undefined}
}%
\providecommand \@ifnum [1]{%
 \ifnum #1\expandafter \@firstoftwo
 \else \expandafter \@secondoftwo
 \fi
}%
\providecommand \@ifx [1]{%
 \ifx #1\expandafter \@firstoftwo
 \else \expandafter \@secondoftwo
 \fi
}%
\providecommand \natexlab [1]{#1}%
\providecommand \enquote  [1]{``#1''}%
\providecommand \bibnamefont  [1]{#1}%
\providecommand \bibfnamefont [1]{#1}%
\providecommand \citenamefont [1]{#1}%
\providecommand \href@noop [0]{\@secondoftwo}%
\providecommand \href [0]{\begingroup \@sanitize@url \@href}%
\providecommand \@href[1]{\@@startlink{#1}\@@href}%
\providecommand \@@href[1]{\endgroup#1\@@endlink}%
\providecommand \@sanitize@url [0]{\catcode `\\12\catcode `\$12\catcode `\&12\catcode `\#12\catcode `\^12\catcode `\_12\catcode `\%12\relax}%
\providecommand \@@startlink[1]{}%
\providecommand \@@endlink[0]{}%
\providecommand \url  [0]{\begingroup\@sanitize@url \@url }%
\providecommand \@url [1]{\endgroup\@href {#1}{\urlprefix }}%
\providecommand \urlprefix  [0]{URL }%
\providecommand \Eprint [0]{\href }%
\providecommand \doibase [0]{http://dx.doi.org/}%
\providecommand \selectlanguage [0]{\@gobble}%
\providecommand \bibinfo  [0]{\@secondoftwo}%
\providecommand \bibfield  [0]{\@secondoftwo}%
\providecommand \translation [1]{[#1]}%
\providecommand \BibitemOpen [0]{}%
\providecommand \bibitemStop [0]{}%
\providecommand \bibitemNoStop [0]{.\EOS\space}%
\providecommand \EOS [0]{\spacefactor3000\relax}%
\providecommand \BibitemShut  [1]{\csname bibitem#1\endcsname}%
\let\auto@bib@innerbib\@empty
\bibitem [{\citenamefont {Booth}\ \emph {et~al.}(2009)\citenamefont {Booth}, \citenamefont {Thom},\ and\ \citenamefont {Alavi}}]{Booth2009-lv}%
  \BibitemOpen
  \bibfield  {author} {\bibinfo {author} {\bibfnamefont {G.~H.}\ \bibnamefont {Booth}}, \bibinfo {author} {\bibfnamefont {A.~J.~W.}\ \bibnamefont {Thom}}, \ and\ \bibinfo {author} {\bibfnamefont {A.}~\bibnamefont {Alavi}},\ }\href@noop {} {\bibfield  {journal} {\bibinfo  {journal} {J. Chem. Phys.}\ }\textbf {\bibinfo {volume} {131}},\ \bibinfo {pages} {054106} (\bibinfo {year} {2009})}\BibitemShut {NoStop}%
\bibitem [{\citenamefont {Cleland}\ \emph {et~al.}(2010)\citenamefont {Cleland}, \citenamefont {Booth},\ and\ \citenamefont {Alavi}}]{Cleland2010-au}%
  \BibitemOpen
  \bibfield  {author} {\bibinfo {author} {\bibfnamefont {D.}~\bibnamefont {Cleland}}, \bibinfo {author} {\bibfnamefont {G.~H.}\ \bibnamefont {Booth}}, \ and\ \bibinfo {author} {\bibfnamefont {A.}~\bibnamefont {Alavi}},\ }\href@noop {} {\bibfield  {journal} {\bibinfo  {journal} {J. Chem. Phys.}\ }\textbf {\bibinfo {volume} {132}},\ \bibinfo {pages} {041103} (\bibinfo {year} {2010})}\BibitemShut {NoStop}%
\bibitem [{\citenamefont {Petruzielo}\ \emph {et~al.}(2012)\citenamefont {Petruzielo}, \citenamefont {Holmes}, \citenamefont {Changlani}, \citenamefont {Nightingale},\ and\ \citenamefont {Umrigar}}]{Petruzielo2012-rk}%
  \BibitemOpen
  \bibfield  {author} {\bibinfo {author} {\bibfnamefont {F.~R.}\ \bibnamefont {Petruzielo}}, \bibinfo {author} {\bibfnamefont {A.~A.}\ \bibnamefont {Holmes}}, \bibinfo {author} {\bibfnamefont {H.~J.}\ \bibnamefont {Changlani}}, \bibinfo {author} {\bibfnamefont {M.~P.}\ \bibnamefont {Nightingale}}, \ and\ \bibinfo {author} {\bibfnamefont {C.~J.}\ \bibnamefont {Umrigar}},\ }\href@noop {} {\bibfield  {journal} {\bibinfo  {journal} {Phys. Rev. Lett.}\ }\textbf {\bibinfo {volume} {109}},\ \bibinfo {pages} {230201} (\bibinfo {year} {2012})}\BibitemShut {NoStop}%
\bibitem [{\citenamefont {Guther}\ \emph {et~al.}(2020)\citenamefont {Guther}, \citenamefont {Anderson}, \citenamefont {Blunt}, \citenamefont {Bogdanov}, \citenamefont {Cleland}, \citenamefont {Dattani}, \citenamefont {Dobrautz}, \citenamefont {Ghanem}, \citenamefont {Jeszenszki}, \citenamefont {Liebermann}, \citenamefont {Manni}, \citenamefont {Lozovoi}, \citenamefont {Luo}, \citenamefont {Ma}, \citenamefont {Merz}, \citenamefont {Overy}, \citenamefont {Rampp}, \citenamefont {Samanta}, \citenamefont {Schwarz}, \citenamefont {Shepherd}, \citenamefont {Smart}, \citenamefont {Vitale}, \citenamefont {Weser}, \citenamefont {Booth},\ and\ \citenamefont {Alavi}}]{Guther2020-um}%
  \BibitemOpen
  \bibfield  {author} {\bibinfo {author} {\bibfnamefont {K.}~\bibnamefont {Guther}}, \bibinfo {author} {\bibfnamefont {R.~J.}\ \bibnamefont {Anderson}}, \bibinfo {author} {\bibfnamefont {N.~S.}\ \bibnamefont {Blunt}}, \bibinfo {author} {\bibfnamefont {N.~A.}\ \bibnamefont {Bogdanov}}, \bibinfo {author} {\bibfnamefont {D.}~\bibnamefont {Cleland}}, \bibinfo {author} {\bibfnamefont {N.}~\bibnamefont {Dattani}}, \bibinfo {author} {\bibfnamefont {W.}~\bibnamefont {Dobrautz}}, \bibinfo {author} {\bibfnamefont {K.}~\bibnamefont {Ghanem}}, \bibinfo {author} {\bibfnamefont {P.}~\bibnamefont {Jeszenszki}}, \bibinfo {author} {\bibfnamefont {N.}~\bibnamefont {Liebermann}}, \bibinfo {author} {\bibfnamefont {G.~L.}\ \bibnamefont {Manni}}, \bibinfo {author} {\bibfnamefont {A.~Y.}\ \bibnamefont {Lozovoi}}, \bibinfo {author} {\bibfnamefont {H.}~\bibnamefont {Luo}}, \bibinfo {author} {\bibfnamefont {D.}~\bibnamefont {Ma}}, \bibinfo {author} {\bibfnamefont {F.}~\bibnamefont {Merz}}, \bibinfo {author} {\bibfnamefont
  {C.}~\bibnamefont {Overy}}, \bibinfo {author} {\bibfnamefont {M.}~\bibnamefont {Rampp}}, \bibinfo {author} {\bibfnamefont {P.~K.}\ \bibnamefont {Samanta}}, \bibinfo {author} {\bibfnamefont {L.~R.}\ \bibnamefont {Schwarz}}, \bibinfo {author} {\bibfnamefont {J.~J.}\ \bibnamefont {Shepherd}}, \bibinfo {author} {\bibfnamefont {S.~D.}\ \bibnamefont {Smart}}, \bibinfo {author} {\bibfnamefont {E.}~\bibnamefont {Vitale}}, \bibinfo {author} {\bibfnamefont {O.}~\bibnamefont {Weser}}, \bibinfo {author} {\bibfnamefont {G.~H.}\ \bibnamefont {Booth}}, \ and\ \bibinfo {author} {\bibfnamefont {A.}~\bibnamefont {Alavi}},\ }\href@noop {} {\bibfield  {journal} {\bibinfo  {journal} {J. Chem. Phys.}\ }\textbf {\bibinfo {volume} {153}},\ \bibinfo {pages} {034107} (\bibinfo {year} {2020})}\BibitemShut {NoStop}%
\bibitem [{\citenamefont {Kanno}\ \emph {et~al.}(2021)\citenamefont {Kanno}, \citenamefont {Endo}, \citenamefont {Suzuki},\ and\ \citenamefont {Tokunaga}}]{Kanno2021-zn}%
  \BibitemOpen
  \bibfield  {author} {\bibinfo {author} {\bibfnamefont {S.}~\bibnamefont {Kanno}}, \bibinfo {author} {\bibfnamefont {S.}~\bibnamefont {Endo}}, \bibinfo {author} {\bibfnamefont {Y.}~\bibnamefont {Suzuki}}, \ and\ \bibinfo {author} {\bibfnamefont {Y.}~\bibnamefont {Tokunaga}},\ }\href@noop {} {\bibfield  {journal} {\bibinfo  {journal} {Phys. Rev. A}\ }\textbf {\bibinfo {volume} {104}},\ \bibinfo {pages} {042424} (\bibinfo {year} {2021})}\BibitemShut {NoStop}%
\bibitem [{\citenamefont {Yuan}\ \emph {et~al.}(2021)\citenamefont {Yuan}, \citenamefont {Sun}, \citenamefont {Liu}, \citenamefont {Zhao},\ and\ \citenamefont {Zhou}}]{Yuan2021-ih}%
  \BibitemOpen
  \bibfield  {author} {\bibinfo {author} {\bibfnamefont {X.}~\bibnamefont {Yuan}}, \bibinfo {author} {\bibfnamefont {J.}~\bibnamefont {Sun}}, \bibinfo {author} {\bibfnamefont {J.}~\bibnamefont {Liu}}, \bibinfo {author} {\bibfnamefont {Q.}~\bibnamefont {Zhao}}, \ and\ \bibinfo {author} {\bibfnamefont {Y.}~\bibnamefont {Zhou}},\ }\href@noop {} {\bibfield  {journal} {\bibinfo  {journal} {Phys. Rev. Lett.}\ }\textbf {\bibinfo {volume} {127}},\ \bibinfo {pages} {040501} (\bibinfo {year} {2021})}\BibitemShut {NoStop}%
\bibitem [{\citenamefont {Mart{\'\i}n}\ \emph {et~al.}(2022)\citenamefont {Mart{\'\i}n}, \citenamefont {Plekhanov},\ and\ \citenamefont {Lubasch}}]{Martin2022-pj}%
  \BibitemOpen
  \bibfield  {author} {\bibinfo {author} {\bibfnamefont {E.~C.}\ \bibnamefont {Mart{\'\i}n}}, \bibinfo {author} {\bibfnamefont {K.}~\bibnamefont {Plekhanov}}, \ and\ \bibinfo {author} {\bibfnamefont {M.}~\bibnamefont {Lubasch}},\ }\href@noop {} {\  (\bibinfo {year} {2022})},\ \Eprint {http://arxiv.org/abs/2209.00292} {arXiv:2209.00292 [quant-ph]} \BibitemShut {NoStop}%
\bibitem [{\citenamefont {Zhang}\ \emph {et~al.}(2022{\natexlab{a}})\citenamefont {Zhang}, \citenamefont {Huang}, \citenamefont {Sun}, \citenamefont {Lv},\ and\ \citenamefont {Yuan}}]{Zhang2022-lf}%
  \BibitemOpen
  \bibfield  {author} {\bibinfo {author} {\bibfnamefont {Y.}~\bibnamefont {Zhang}}, \bibinfo {author} {\bibfnamefont {Y.}~\bibnamefont {Huang}}, \bibinfo {author} {\bibfnamefont {J.}~\bibnamefont {Sun}}, \bibinfo {author} {\bibfnamefont {D.}~\bibnamefont {Lv}}, \ and\ \bibinfo {author} {\bibfnamefont {X.}~\bibnamefont {Yuan}},\ }\href@noop {} {\  (\bibinfo {year} {2022}{\natexlab{a}})},\ \Eprint {http://arxiv.org/abs/2206.10431} {arXiv:2206.10431 [quant-ph]} \BibitemShut {NoStop}%
\bibitem [{\citenamefont {Huggins}\ \emph {et~al.}(2022)\citenamefont {Huggins}, \citenamefont {O'Gorman}, \citenamefont {Rubin}, \citenamefont {Reichman}, \citenamefont {Babbush},\ and\ \citenamefont {Lee}}]{Huggins2022-ly}%
  \BibitemOpen
  \bibfield  {author} {\bibinfo {author} {\bibfnamefont {W.~J.}\ \bibnamefont {Huggins}}, \bibinfo {author} {\bibfnamefont {B.~A.}\ \bibnamefont {O'Gorman}}, \bibinfo {author} {\bibfnamefont {N.~C.}\ \bibnamefont {Rubin}}, \bibinfo {author} {\bibfnamefont {D.~R.}\ \bibnamefont {Reichman}}, \bibinfo {author} {\bibfnamefont {R.}~\bibnamefont {Babbush}}, \ and\ \bibinfo {author} {\bibfnamefont {J.}~\bibnamefont {Lee}},\ }\href@noop {} {\bibfield  {journal} {\bibinfo  {journal} {Nature}\ }\textbf {\bibinfo {volume} {603}},\ \bibinfo {pages} {416} (\bibinfo {year} {2022})}\BibitemShut {NoStop}%
\bibitem [{\citenamefont {Lee}\ \emph {et~al.}(2022)\citenamefont {Lee}, \citenamefont {Reichman}, \citenamefont {Babbush}, \citenamefont {Rubin}, \citenamefont {Malone}, \citenamefont {O'Gorman},\ and\ \citenamefont {Huggins}}]{Lee2022-lq}%
  \BibitemOpen
  \bibfield  {author} {\bibinfo {author} {\bibfnamefont {J.}~\bibnamefont {Lee}}, \bibinfo {author} {\bibfnamefont {D.~R.}\ \bibnamefont {Reichman}}, \bibinfo {author} {\bibfnamefont {R.}~\bibnamefont {Babbush}}, \bibinfo {author} {\bibfnamefont {N.~C.}\ \bibnamefont {Rubin}}, \bibinfo {author} {\bibfnamefont {F.~D.}\ \bibnamefont {Malone}}, \bibinfo {author} {\bibfnamefont {B.}~\bibnamefont {O'Gorman}}, \ and\ \bibinfo {author} {\bibfnamefont {W.~J.}\ \bibnamefont {Huggins}},\ }\href@noop {} {\  (\bibinfo {year} {2022})},\ \Eprint {http://arxiv.org/abs/2207.13776} {arXiv:2207.13776 [quant-ph]} \BibitemShut {NoStop}%
\bibitem [{\citenamefont {Xu}\ and\ \citenamefont {Li}(2023)}]{Xu2023-sn}%
  \BibitemOpen
  \bibfield  {author} {\bibinfo {author} {\bibfnamefont {X.}~\bibnamefont {Xu}}\ and\ \bibinfo {author} {\bibfnamefont {Y.}~\bibnamefont {Li}},\ }\href@noop {} {\bibfield  {journal} {\bibinfo  {journal} {Quantum}\ }\textbf {\bibinfo {volume} {7}},\ \bibinfo {pages} {1072} (\bibinfo {year} {2023})}\BibitemShut {NoStop}%
\bibitem [{\citenamefont {Giannozzi}\ \emph {et~al.}(2009)\citenamefont {Giannozzi}, \citenamefont {Baroni}, \citenamefont {Bonini}, \citenamefont {Calandra}, \citenamefont {Car}, \citenamefont {Cavazzoni}, \citenamefont {Ceresoli}, \citenamefont {Chiarotti}, \citenamefont {Cococcioni}, \citenamefont {Dabo}, \citenamefont {Dal~Corso}, \citenamefont {de~Gironcoli}, \citenamefont {Fabris}, \citenamefont {Fratesi}, \citenamefont {Gebauer}, \citenamefont {Gerstmann}, \citenamefont {Gougoussis}, \citenamefont {Kokalj}, \citenamefont {Lazzeri}, \citenamefont {Martin-Samos}, \citenamefont {Marzari}, \citenamefont {Mauri}, \citenamefont {Mazzarello}, \citenamefont {Paolini}, \citenamefont {Pasquarello}, \citenamefont {Paulatto}, \citenamefont {Sbraccia}, \citenamefont {Scandolo}, \citenamefont {Sclauzero}, \citenamefont {Seitsonen}, \citenamefont {Smogunov}, \citenamefont {Umari},\ and\ \citenamefont {Wentzcovitch}}]{Giannozzi2009-zq}%
  \BibitemOpen
  \bibfield  {author} {\bibinfo {author} {\bibfnamefont {P.}~\bibnamefont {Giannozzi}}, \bibinfo {author} {\bibfnamefont {S.}~\bibnamefont {Baroni}}, \bibinfo {author} {\bibfnamefont {N.}~\bibnamefont {Bonini}}, \bibinfo {author} {\bibfnamefont {M.}~\bibnamefont {Calandra}}, \bibinfo {author} {\bibfnamefont {R.}~\bibnamefont {Car}}, \bibinfo {author} {\bibfnamefont {C.}~\bibnamefont {Cavazzoni}}, \bibinfo {author} {\bibfnamefont {D.}~\bibnamefont {Ceresoli}}, \bibinfo {author} {\bibfnamefont {G.~L.}\ \bibnamefont {Chiarotti}}, \bibinfo {author} {\bibfnamefont {M.}~\bibnamefont {Cococcioni}}, \bibinfo {author} {\bibfnamefont {I.}~\bibnamefont {Dabo}}, \bibinfo {author} {\bibfnamefont {A.}~\bibnamefont {Dal~Corso}}, \bibinfo {author} {\bibfnamefont {S.}~\bibnamefont {de~Gironcoli}}, \bibinfo {author} {\bibfnamefont {S.}~\bibnamefont {Fabris}}, \bibinfo {author} {\bibfnamefont {G.}~\bibnamefont {Fratesi}}, \bibinfo {author} {\bibfnamefont {R.}~\bibnamefont {Gebauer}}, \bibinfo {author} {\bibfnamefont
  {U.}~\bibnamefont {Gerstmann}}, \bibinfo {author} {\bibfnamefont {C.}~\bibnamefont {Gougoussis}}, \bibinfo {author} {\bibfnamefont {A.}~\bibnamefont {Kokalj}}, \bibinfo {author} {\bibfnamefont {M.}~\bibnamefont {Lazzeri}}, \bibinfo {author} {\bibfnamefont {L.}~\bibnamefont {Martin-Samos}}, \bibinfo {author} {\bibfnamefont {N.}~\bibnamefont {Marzari}}, \bibinfo {author} {\bibfnamefont {F.}~\bibnamefont {Mauri}}, \bibinfo {author} {\bibfnamefont {R.}~\bibnamefont {Mazzarello}}, \bibinfo {author} {\bibfnamefont {S.}~\bibnamefont {Paolini}}, \bibinfo {author} {\bibfnamefont {A.}~\bibnamefont {Pasquarello}}, \bibinfo {author} {\bibfnamefont {L.}~\bibnamefont {Paulatto}}, \bibinfo {author} {\bibfnamefont {C.}~\bibnamefont {Sbraccia}}, \bibinfo {author} {\bibfnamefont {S.}~\bibnamefont {Scandolo}}, \bibinfo {author} {\bibfnamefont {G.}~\bibnamefont {Sclauzero}}, \bibinfo {author} {\bibfnamefont {A.~P.}\ \bibnamefont {Seitsonen}}, \bibinfo {author} {\bibfnamefont {A.}~\bibnamefont {Smogunov}}, \bibinfo {author}
  {\bibfnamefont {P.}~\bibnamefont {Umari}}, \ and\ \bibinfo {author} {\bibfnamefont {R.~M.}\ \bibnamefont {Wentzcovitch}},\ }\href@noop {} {\bibfield  {journal} {\bibinfo  {journal} {J. Phys. Condens. Matter}\ }\textbf {\bibinfo {volume} {21}},\ \bibinfo {pages} {395502} (\bibinfo {year} {2009})}\BibitemShut {NoStop}%
\bibitem [{\citenamefont {Giannozzi}\ \emph {et~al.}(2017)\citenamefont {Giannozzi}, \citenamefont {Andreussi}, \citenamefont {Brumme}, \citenamefont {Bunau}, \citenamefont {Buongiorno~Nardelli}, \citenamefont {Calandra}, \citenamefont {Car}, \citenamefont {Cavazzoni}, \citenamefont {Ceresoli}, \citenamefont {Cococcioni}, \citenamefont {Colonna}, \citenamefont {Carnimeo}, \citenamefont {Dal~Corso}, \citenamefont {de~Gironcoli}, \citenamefont {Delugas}, \citenamefont {DiStasio}, \citenamefont {Ferretti}, \citenamefont {Floris}, \citenamefont {Fratesi}, \citenamefont {Fugallo}, \citenamefont {Gebauer}, \citenamefont {Gerstmann}, \citenamefont {Giustino}, \citenamefont {Gorni}, \citenamefont {Jia}, \citenamefont {Kawamura}, \citenamefont {Ko}, \citenamefont {Kokalj}, \citenamefont {K{\"u}{\c c}{\"u}kbenli}, \citenamefont {Lazzeri}, \citenamefont {Marsili}, \citenamefont {Marzari}, \citenamefont {Mauri}, \citenamefont {Nguyen}, \citenamefont {Nguyen}, \citenamefont {Otero-de-la Roza}, \citenamefont {Paulatto},
  \citenamefont {Ponc{\'e}}, \citenamefont {Rocca}, \citenamefont {Sabatini}, \citenamefont {Santra}, \citenamefont {Schlipf}, \citenamefont {Seitsonen}, \citenamefont {Smogunov}, \citenamefont {Timrov}, \citenamefont {Thonhauser}, \citenamefont {Umari}, \citenamefont {Vast}, \citenamefont {Wu},\ and\ \citenamefont {Baroni}}]{Giannozzi2017-jz}%
  \BibitemOpen
  \bibfield  {author} {\bibinfo {author} {\bibfnamefont {P.}~\bibnamefont {Giannozzi}}, \bibinfo {author} {\bibfnamefont {O.}~\bibnamefont {Andreussi}}, \bibinfo {author} {\bibfnamefont {T.}~\bibnamefont {Brumme}}, \bibinfo {author} {\bibfnamefont {O.}~\bibnamefont {Bunau}}, \bibinfo {author} {\bibfnamefont {M.}~\bibnamefont {Buongiorno~Nardelli}}, \bibinfo {author} {\bibfnamefont {M.}~\bibnamefont {Calandra}}, \bibinfo {author} {\bibfnamefont {R.}~\bibnamefont {Car}}, \bibinfo {author} {\bibfnamefont {C.}~\bibnamefont {Cavazzoni}}, \bibinfo {author} {\bibfnamefont {D.}~\bibnamefont {Ceresoli}}, \bibinfo {author} {\bibfnamefont {M.}~\bibnamefont {Cococcioni}}, \bibinfo {author} {\bibfnamefont {N.}~\bibnamefont {Colonna}}, \bibinfo {author} {\bibfnamefont {I.}~\bibnamefont {Carnimeo}}, \bibinfo {author} {\bibfnamefont {A.}~\bibnamefont {Dal~Corso}}, \bibinfo {author} {\bibfnamefont {S.}~\bibnamefont {de~Gironcoli}}, \bibinfo {author} {\bibfnamefont {P.}~\bibnamefont {Delugas}}, \bibinfo {author} {\bibfnamefont
  {R.~A.}\ \bibnamefont {DiStasio}, \bibfnamefont {Jr}}, \bibinfo {author} {\bibfnamefont {A.}~\bibnamefont {Ferretti}}, \bibinfo {author} {\bibfnamefont {A.}~\bibnamefont {Floris}}, \bibinfo {author} {\bibfnamefont {G.}~\bibnamefont {Fratesi}}, \bibinfo {author} {\bibfnamefont {G.}~\bibnamefont {Fugallo}}, \bibinfo {author} {\bibfnamefont {R.}~\bibnamefont {Gebauer}}, \bibinfo {author} {\bibfnamefont {U.}~\bibnamefont {Gerstmann}}, \bibinfo {author} {\bibfnamefont {F.}~\bibnamefont {Giustino}}, \bibinfo {author} {\bibfnamefont {T.}~\bibnamefont {Gorni}}, \bibinfo {author} {\bibfnamefont {J.}~\bibnamefont {Jia}}, \bibinfo {author} {\bibfnamefont {M.}~\bibnamefont {Kawamura}}, \bibinfo {author} {\bibfnamefont {H.-Y.}\ \bibnamefont {Ko}}, \bibinfo {author} {\bibfnamefont {A.}~\bibnamefont {Kokalj}}, \bibinfo {author} {\bibfnamefont {E.}~\bibnamefont {K{\"u}{\c c}{\"u}kbenli}}, \bibinfo {author} {\bibfnamefont {M.}~\bibnamefont {Lazzeri}}, \bibinfo {author} {\bibfnamefont {M.}~\bibnamefont {Marsili}}, \bibinfo
  {author} {\bibfnamefont {N.}~\bibnamefont {Marzari}}, \bibinfo {author} {\bibfnamefont {F.}~\bibnamefont {Mauri}}, \bibinfo {author} {\bibfnamefont {N.~L.}\ \bibnamefont {Nguyen}}, \bibinfo {author} {\bibfnamefont {H.-V.}\ \bibnamefont {Nguyen}}, \bibinfo {author} {\bibfnamefont {A.}~\bibnamefont {Otero-de-la Roza}}, \bibinfo {author} {\bibfnamefont {L.}~\bibnamefont {Paulatto}}, \bibinfo {author} {\bibfnamefont {S.}~\bibnamefont {Ponc{\'e}}}, \bibinfo {author} {\bibfnamefont {D.}~\bibnamefont {Rocca}}, \bibinfo {author} {\bibfnamefont {R.}~\bibnamefont {Sabatini}}, \bibinfo {author} {\bibfnamefont {B.}~\bibnamefont {Santra}}, \bibinfo {author} {\bibfnamefont {M.}~\bibnamefont {Schlipf}}, \bibinfo {author} {\bibfnamefont {A.~P.}\ \bibnamefont {Seitsonen}}, \bibinfo {author} {\bibfnamefont {A.}~\bibnamefont {Smogunov}}, \bibinfo {author} {\bibfnamefont {I.}~\bibnamefont {Timrov}}, \bibinfo {author} {\bibfnamefont {T.}~\bibnamefont {Thonhauser}}, \bibinfo {author} {\bibfnamefont {P.}~\bibnamefont {Umari}},
  \bibinfo {author} {\bibfnamefont {N.}~\bibnamefont {Vast}}, \bibinfo {author} {\bibfnamefont {X.}~\bibnamefont {Wu}}, \ and\ \bibinfo {author} {\bibfnamefont {S.}~\bibnamefont {Baroni}},\ }\href@noop {} {\bibfield  {journal} {\bibinfo  {journal} {J. Phys. Condens. Matter}\ }\textbf {\bibinfo {volume} {29}},\ \bibinfo {pages} {465901} (\bibinfo {year} {2017})}\BibitemShut {NoStop}%
\bibitem [{\citenamefont {Giannozzi}\ \emph {et~al.}(2020)\citenamefont {Giannozzi}, \citenamefont {Baseggio}, \citenamefont {Bonf{\`a}}, \citenamefont {Brunato}, \citenamefont {Car}, \citenamefont {Carnimeo}, \citenamefont {Cavazzoni}, \citenamefont {de~Gironcoli}, \citenamefont {Delugas}, \citenamefont {Ferrari~Ruffino}, \citenamefont {Ferretti}, \citenamefont {Marzari}, \citenamefont {Timrov}, \citenamefont {Urru},\ and\ \citenamefont {Baroni}}]{Giannozzi2020-yi}%
  \BibitemOpen
  \bibfield  {author} {\bibinfo {author} {\bibfnamefont {P.}~\bibnamefont {Giannozzi}}, \bibinfo {author} {\bibfnamefont {O.}~\bibnamefont {Baseggio}}, \bibinfo {author} {\bibfnamefont {P.}~\bibnamefont {Bonf{\`a}}}, \bibinfo {author} {\bibfnamefont {D.}~\bibnamefont {Brunato}}, \bibinfo {author} {\bibfnamefont {R.}~\bibnamefont {Car}}, \bibinfo {author} {\bibfnamefont {I.}~\bibnamefont {Carnimeo}}, \bibinfo {author} {\bibfnamefont {C.}~\bibnamefont {Cavazzoni}}, \bibinfo {author} {\bibfnamefont {S.}~\bibnamefont {de~Gironcoli}}, \bibinfo {author} {\bibfnamefont {P.}~\bibnamefont {Delugas}}, \bibinfo {author} {\bibfnamefont {F.}~\bibnamefont {Ferrari~Ruffino}}, \bibinfo {author} {\bibfnamefont {A.}~\bibnamefont {Ferretti}}, \bibinfo {author} {\bibfnamefont {N.}~\bibnamefont {Marzari}}, \bibinfo {author} {\bibfnamefont {I.}~\bibnamefont {Timrov}}, \bibinfo {author} {\bibfnamefont {A.}~\bibnamefont {Urru}}, \ and\ \bibinfo {author} {\bibfnamefont {S.}~\bibnamefont {Baroni}},\ }\href@noop {} {\bibfield
  {journal} {\bibinfo  {journal} {J. Chem. Phys.}\ }\textbf {\bibinfo {volume} {152}},\ \bibinfo {pages} {154105} (\bibinfo {year} {2020})}\BibitemShut {NoStop}%
\bibitem [{\citenamefont {Perdew}\ \emph {et~al.}(1996)\citenamefont {Perdew}, \citenamefont {Burke},\ and\ \citenamefont {Ernzerhof}}]{Perdew1996-qc}%
  \BibitemOpen
  \bibfield  {author} {\bibinfo {author} {\bibfnamefont {J.~P.}\ \bibnamefont {Perdew}}, \bibinfo {author} {\bibfnamefont {K.}~\bibnamefont {Burke}}, \ and\ \bibinfo {author} {\bibfnamefont {M.}~\bibnamefont {Ernzerhof}},\ }\href@noop {} {\bibfield  {journal} {\bibinfo  {journal} {Phys. Rev. Lett.}\ }\textbf {\bibinfo {volume} {77}},\ \bibinfo {pages} {3865} (\bibinfo {year} {1996})}\BibitemShut {NoStop}%
\bibitem [{\citenamefont {Hamann}\ \emph {et~al.}(1979)\citenamefont {Hamann}, \citenamefont {Schl{\"u}ter},\ and\ \citenamefont {Chiang}}]{Hamann1979-ti}%
  \BibitemOpen
  \bibfield  {author} {\bibinfo {author} {\bibfnamefont {D.~R.}\ \bibnamefont {Hamann}}, \bibinfo {author} {\bibfnamefont {M.}~\bibnamefont {Schl{\"u}ter}}, \ and\ \bibinfo {author} {\bibfnamefont {C.}~\bibnamefont {Chiang}},\ }\href@noop {} {\bibfield  {journal} {\bibinfo  {journal} {Phys. Rev. Lett.}\ }\textbf {\bibinfo {volume} {43}},\ \bibinfo {pages} {1494} (\bibinfo {year} {1979})}\BibitemShut {NoStop}%
\bibitem [{\citenamefont {Hamann}(2013)}]{Hamann2013-tx}%
  \BibitemOpen
  \bibfield  {author} {\bibinfo {author} {\bibfnamefont {D.~R.}\ \bibnamefont {Hamann}},\ }\href@noop {} {\bibfield  {journal} {\bibinfo  {journal} {Phys. Rev. B Condens. Matter}\ }\textbf {\bibinfo {volume} {88}},\ \bibinfo {pages} {085117} (\bibinfo {year} {2013})}\BibitemShut {NoStop}%
\bibitem [{\citenamefont {Marzari}\ and\ \citenamefont {Vanderbilt}(1997)}]{Marzari1997-sz}%
  \BibitemOpen
  \bibfield  {author} {\bibinfo {author} {\bibfnamefont {N.}~\bibnamefont {Marzari}}\ and\ \bibinfo {author} {\bibfnamefont {D.}~\bibnamefont {Vanderbilt}},\ }\href@noop {} {\bibfield  {journal} {\bibinfo  {journal} {Phys. Rev. B Condens. Matter}\ }\textbf {\bibinfo {volume} {56}},\ \bibinfo {pages} {12847} (\bibinfo {year} {1997})}\BibitemShut {NoStop}%
\bibitem [{\citenamefont {Souza}\ \emph {et~al.}(2001)\citenamefont {Souza}, \citenamefont {Marzari},\ and\ \citenamefont {Vanderbilt}}]{Souza2001-un}%
  \BibitemOpen
  \bibfield  {author} {\bibinfo {author} {\bibfnamefont {I.}~\bibnamefont {Souza}}, \bibinfo {author} {\bibfnamefont {N.}~\bibnamefont {Marzari}}, \ and\ \bibinfo {author} {\bibfnamefont {D.}~\bibnamefont {Vanderbilt}},\ }\href@noop {} {\bibfield  {journal} {\bibinfo  {journal} {Phys. Rev. B Condens. Matter}\ }\textbf {\bibinfo {volume} {65}},\ \bibinfo {pages} {035109} (\bibinfo {year} {2001})}\BibitemShut {NoStop}%
\bibitem [{\citenamefont {Aryasetiawan}\ \emph {et~al.}(2004)\citenamefont {Aryasetiawan}, \citenamefont {Imada}, \citenamefont {Georges}, \citenamefont {Kotliar}, \citenamefont {Biermann},\ and\ \citenamefont {Lichtenstein}}]{Aryasetiawan2004-sq}%
  \BibitemOpen
  \bibfield  {author} {\bibinfo {author} {\bibfnamefont {F.}~\bibnamefont {Aryasetiawan}}, \bibinfo {author} {\bibfnamefont {M.}~\bibnamefont {Imada}}, \bibinfo {author} {\bibfnamefont {A.}~\bibnamefont {Georges}}, \bibinfo {author} {\bibfnamefont {G.}~\bibnamefont {Kotliar}}, \bibinfo {author} {\bibfnamefont {S.}~\bibnamefont {Biermann}}, \ and\ \bibinfo {author} {\bibfnamefont {A.~I.}\ \bibnamefont {Lichtenstein}},\ }\href@noop {} {\bibfield  {journal} {\bibinfo  {journal} {Phys. Rev. B Condens. Matter}\ }\textbf {\bibinfo {volume} {70}},\ \bibinfo {pages} {195104} (\bibinfo {year} {2004})}\BibitemShut {NoStop}%
\bibitem [{\citenamefont {Fujiwara}\ \emph {et~al.}(2003)\citenamefont {Fujiwara}, \citenamefont {Yamamoto},\ and\ \citenamefont {Ishii}}]{Fujiwara2003-xg}%
  \BibitemOpen
  \bibfield  {author} {\bibinfo {author} {\bibfnamefont {T.}~\bibnamefont {Fujiwara}}, \bibinfo {author} {\bibfnamefont {S.}~\bibnamefont {Yamamoto}}, \ and\ \bibinfo {author} {\bibfnamefont {Y.}~\bibnamefont {Ishii}},\ }\href@noop {} {\bibfield  {journal} {\bibinfo  {journal} {J. Phys. Soc. Jpn.}\ }\textbf {\bibinfo {volume} {72}},\ \bibinfo {pages} {777} (\bibinfo {year} {2003})}\BibitemShut {NoStop}%
\bibitem [{\citenamefont {Nakamura}\ \emph {et~al.}(2008)\citenamefont {Nakamura}, \citenamefont {Arita},\ and\ \citenamefont {Imada}}]{Nakamura2008-hg}%
  \BibitemOpen
  \bibfield  {author} {\bibinfo {author} {\bibfnamefont {K.}~\bibnamefont {Nakamura}}, \bibinfo {author} {\bibfnamefont {R.}~\bibnamefont {Arita}}, \ and\ \bibinfo {author} {\bibfnamefont {M.}~\bibnamefont {Imada}},\ }\href@noop {} {\bibfield  {journal} {\bibinfo  {journal} {J. Phys. Soc. Jpn.}\ }\textbf {\bibinfo {volume} {77}},\ \bibinfo {pages} {093711} (\bibinfo {year} {2008})}\BibitemShut {NoStop}%
\bibitem [{\citenamefont {Nohara}\ \emph {et~al.}(2009)\citenamefont {Nohara}, \citenamefont {Yamamoto},\ and\ \citenamefont {Fujiwara}}]{Nohara2009-rm}%
  \BibitemOpen
  \bibfield  {author} {\bibinfo {author} {\bibfnamefont {Y.}~\bibnamefont {Nohara}}, \bibinfo {author} {\bibfnamefont {S.}~\bibnamefont {Yamamoto}}, \ and\ \bibinfo {author} {\bibfnamefont {T.}~\bibnamefont {Fujiwara}},\ }\href@noop {} {\bibfield  {journal} {\bibinfo  {journal} {Phys. Rev. B Condens. Matter Mater. Phys.}\ }\textbf {\bibinfo {volume} {79}},\ \bibinfo {pages} {195110} (\bibinfo {year} {2009})}\BibitemShut {NoStop}%
\bibitem [{\citenamefont {Nakamura}\ \emph {et~al.}(2009)\citenamefont {Nakamura}, \citenamefont {Yoshimoto}, \citenamefont {Kosugi}, \citenamefont {Arita},\ and\ \citenamefont {Imada}}]{Nakamura2009-te}%
  \BibitemOpen
  \bibfield  {author} {\bibinfo {author} {\bibfnamefont {K.}~\bibnamefont {Nakamura}}, \bibinfo {author} {\bibfnamefont {Y.}~\bibnamefont {Yoshimoto}}, \bibinfo {author} {\bibfnamefont {T.}~\bibnamefont {Kosugi}}, \bibinfo {author} {\bibfnamefont {R.}~\bibnamefont {Arita}}, \ and\ \bibinfo {author} {\bibfnamefont {M.}~\bibnamefont {Imada}},\ }\href@noop {} {\bibfield  {journal} {\bibinfo  {journal} {J. Phys. Soc. Jpn.}\ }\textbf {\bibinfo {volume} {78}},\ \bibinfo {pages} {083710} (\bibinfo {year} {2009})}\BibitemShut {NoStop}%
\bibitem [{\citenamefont {Nakamura}\ \emph {et~al.}(2016)\citenamefont {Nakamura}, \citenamefont {Nohara}, \citenamefont {Yosimoto},\ and\ \citenamefont {Nomura}}]{Nakamura2016-vj}%
  \BibitemOpen
  \bibfield  {author} {\bibinfo {author} {\bibfnamefont {K.}~\bibnamefont {Nakamura}}, \bibinfo {author} {\bibfnamefont {Y.}~\bibnamefont {Nohara}}, \bibinfo {author} {\bibfnamefont {Y.}~\bibnamefont {Yosimoto}}, \ and\ \bibinfo {author} {\bibfnamefont {Y.}~\bibnamefont {Nomura}},\ }\href@noop {} {\bibfield  {journal} {\bibinfo  {journal} {Phys. Rev. B Condens. Matter}\ }\textbf {\bibinfo {volume} {93}},\ \bibinfo {pages} {085124} (\bibinfo {year} {2016})}\BibitemShut {NoStop}%
\bibitem [{\citenamefont {Nakamura}\ \emph {et~al.}(2021)\citenamefont {Nakamura}, \citenamefont {Yoshimoto}, \citenamefont {Nomura}, \citenamefont {Tadano}, \citenamefont {Kawamura}, \citenamefont {Kosugi}, \citenamefont {Yoshimi}, \citenamefont {Misawa},\ and\ \citenamefont {Motoyama}}]{Nakamura2021-sh}%
  \BibitemOpen
  \bibfield  {author} {\bibinfo {author} {\bibfnamefont {K.}~\bibnamefont {Nakamura}}, \bibinfo {author} {\bibfnamefont {Y.}~\bibnamefont {Yoshimoto}}, \bibinfo {author} {\bibfnamefont {Y.}~\bibnamefont {Nomura}}, \bibinfo {author} {\bibfnamefont {T.}~\bibnamefont {Tadano}}, \bibinfo {author} {\bibfnamefont {M.}~\bibnamefont {Kawamura}}, \bibinfo {author} {\bibfnamefont {T.}~\bibnamefont {Kosugi}}, \bibinfo {author} {\bibfnamefont {K.}~\bibnamefont {Yoshimi}}, \bibinfo {author} {\bibfnamefont {T.}~\bibnamefont {Misawa}}, \ and\ \bibinfo {author} {\bibfnamefont {Y.}~\bibnamefont {Motoyama}},\ }\href@noop {} {\bibfield  {journal} {\bibinfo  {journal} {Comput. Phys. Commun.}\ }\textbf {\bibinfo {volume} {261}},\ \bibinfo {pages} {107781} (\bibinfo {year} {2021})}\BibitemShut {NoStop}%
\bibitem [{\citenamefont {Werner}\ \emph {et~al.}(2012)\citenamefont {Werner}, \citenamefont {Knowles}, \citenamefont {Knizia}, \citenamefont {Manby},\ and\ \citenamefont {Sch{\"u}tz}}]{Werner2012-ky}%
  \BibitemOpen
  \bibfield  {author} {\bibinfo {author} {\bibfnamefont {H.-J.}\ \bibnamefont {Werner}}, \bibinfo {author} {\bibfnamefont {P.~J.}\ \bibnamefont {Knowles}}, \bibinfo {author} {\bibfnamefont {G.}~\bibnamefont {Knizia}}, \bibinfo {author} {\bibfnamefont {F.~R.}\ \bibnamefont {Manby}}, \ and\ \bibinfo {author} {\bibfnamefont {M.}~\bibnamefont {Sch{\"u}tz}},\ }\href@noop {} {\bibfield  {journal} {\bibinfo  {journal} {Wiley Interdiscip. Rev. Comput. Mol. Sci.}\ }\textbf {\bibinfo {volume} {2}},\ \bibinfo {pages} {242} (\bibinfo {year} {2012})}\BibitemShut {NoStop}%
\bibitem [{\citenamefont {Werner}\ \emph {et~al.}(2020)\citenamefont {Werner}, \citenamefont {Knowles}, \citenamefont {Manby}, \citenamefont {Black}, \citenamefont {Doll}, \citenamefont {He{\ss}elmann}, \citenamefont {Kats}, \citenamefont {K{\"o}hn}, \citenamefont {Korona}, \citenamefont {Kreplin}, \citenamefont {Ma}, \citenamefont {Miller}, \citenamefont {Mitrushchenkov}, \citenamefont {Peterson}, \citenamefont {Polyak}, \citenamefont {Rauhut},\ and\ \citenamefont {Sibaev}}]{Werner2020-ph}%
  \BibitemOpen
  \bibfield  {author} {\bibinfo {author} {\bibfnamefont {H.-J.}\ \bibnamefont {Werner}}, \bibinfo {author} {\bibfnamefont {P.~J.}\ \bibnamefont {Knowles}}, \bibinfo {author} {\bibfnamefont {F.~R.}\ \bibnamefont {Manby}}, \bibinfo {author} {\bibfnamefont {J.~A.}\ \bibnamefont {Black}}, \bibinfo {author} {\bibfnamefont {K.}~\bibnamefont {Doll}}, \bibinfo {author} {\bibfnamefont {A.}~\bibnamefont {He{\ss}elmann}}, \bibinfo {author} {\bibfnamefont {D.}~\bibnamefont {Kats}}, \bibinfo {author} {\bibfnamefont {A.}~\bibnamefont {K{\"o}hn}}, \bibinfo {author} {\bibfnamefont {T.}~\bibnamefont {Korona}}, \bibinfo {author} {\bibfnamefont {D.~A.}\ \bibnamefont {Kreplin}}, \bibinfo {author} {\bibfnamefont {Q.}~\bibnamefont {Ma}}, \bibinfo {author} {\bibfnamefont {T.~F.}\ \bibnamefont {Miller}, \bibfnamefont {3rd}}, \bibinfo {author} {\bibfnamefont {A.}~\bibnamefont {Mitrushchenkov}}, \bibinfo {author} {\bibfnamefont {K.~A.}\ \bibnamefont {Peterson}}, \bibinfo {author} {\bibfnamefont {I.}~\bibnamefont {Polyak}}, \bibinfo
  {author} {\bibfnamefont {G.}~\bibnamefont {Rauhut}}, \ and\ \bibinfo {author} {\bibfnamefont {M.}~\bibnamefont {Sibaev}},\ }\href@noop {} {\bibfield  {journal} {\bibinfo  {journal} {J. Chem. Phys.}\ }\textbf {\bibinfo {volume} {152}},\ \bibinfo {pages} {144107} (\bibinfo {year} {2020})}\BibitemShut {NoStop}%
\bibitem [{\citenamefont {Sun}\ \emph {et~al.}(2020)\citenamefont {Sun}, \citenamefont {Zhang}, \citenamefont {Banerjee}, \citenamefont {Bao}, \citenamefont {Barbry}, \citenamefont {Blunt}, \citenamefont {Bogdanov}, \citenamefont {Booth}, \citenamefont {Chen}, \citenamefont {Cui}, \citenamefont {Eriksen}, \citenamefont {Gao}, \citenamefont {Guo}, \citenamefont {Hermann}, \citenamefont {Hermes}, \citenamefont {Koh}, \citenamefont {Koval}, \citenamefont {Lehtola}, \citenamefont {Li}, \citenamefont {Liu}, \citenamefont {Mardirossian}, \citenamefont {McClain}, \citenamefont {Motta}, \citenamefont {Mussard}, \citenamefont {Pham}, \citenamefont {Pulkin}, \citenamefont {Purwanto}, \citenamefont {Robinson}, \citenamefont {Ronca}, \citenamefont {Sayfutyarova}, \citenamefont {Scheurer}, \citenamefont {Schurkus}, \citenamefont {Smith}, \citenamefont {Sun}, \citenamefont {Sun}, \citenamefont {Upadhyay}, \citenamefont {Wagner}, \citenamefont {Wang}, \citenamefont {White}, \citenamefont {Whitfield}, \citenamefont
  {Williamson}, \citenamefont {Wouters}, \citenamefont {Yang}, \citenamefont {Yu}, \citenamefont {Zhu}, \citenamefont {Berkelbach}, \citenamefont {Sharma}, \citenamefont {Sokolov},\ and\ \citenamefont {Chan}}]{Sun2020-sa}%
  \BibitemOpen
  \bibfield  {author} {\bibinfo {author} {\bibfnamefont {Q.}~\bibnamefont {Sun}}, \bibinfo {author} {\bibfnamefont {X.}~\bibnamefont {Zhang}}, \bibinfo {author} {\bibfnamefont {S.}~\bibnamefont {Banerjee}}, \bibinfo {author} {\bibfnamefont {P.}~\bibnamefont {Bao}}, \bibinfo {author} {\bibfnamefont {M.}~\bibnamefont {Barbry}}, \bibinfo {author} {\bibfnamefont {N.~S.}\ \bibnamefont {Blunt}}, \bibinfo {author} {\bibfnamefont {N.~A.}\ \bibnamefont {Bogdanov}}, \bibinfo {author} {\bibfnamefont {G.~H.}\ \bibnamefont {Booth}}, \bibinfo {author} {\bibfnamefont {J.}~\bibnamefont {Chen}}, \bibinfo {author} {\bibfnamefont {Z.-H.}\ \bibnamefont {Cui}}, \bibinfo {author} {\bibfnamefont {J.~J.}\ \bibnamefont {Eriksen}}, \bibinfo {author} {\bibfnamefont {Y.}~\bibnamefont {Gao}}, \bibinfo {author} {\bibfnamefont {S.}~\bibnamefont {Guo}}, \bibinfo {author} {\bibfnamefont {J.}~\bibnamefont {Hermann}}, \bibinfo {author} {\bibfnamefont {M.~R.}\ \bibnamefont {Hermes}}, \bibinfo {author} {\bibfnamefont {K.}~\bibnamefont {Koh}},
  \bibinfo {author} {\bibfnamefont {P.}~\bibnamefont {Koval}}, \bibinfo {author} {\bibfnamefont {S.}~\bibnamefont {Lehtola}}, \bibinfo {author} {\bibfnamefont {Z.}~\bibnamefont {Li}}, \bibinfo {author} {\bibfnamefont {J.}~\bibnamefont {Liu}}, \bibinfo {author} {\bibfnamefont {N.}~\bibnamefont {Mardirossian}}, \bibinfo {author} {\bibfnamefont {J.~D.}\ \bibnamefont {McClain}}, \bibinfo {author} {\bibfnamefont {M.}~\bibnamefont {Motta}}, \bibinfo {author} {\bibfnamefont {B.}~\bibnamefont {Mussard}}, \bibinfo {author} {\bibfnamefont {H.~Q.}\ \bibnamefont {Pham}}, \bibinfo {author} {\bibfnamefont {A.}~\bibnamefont {Pulkin}}, \bibinfo {author} {\bibfnamefont {W.}~\bibnamefont {Purwanto}}, \bibinfo {author} {\bibfnamefont {P.~J.}\ \bibnamefont {Robinson}}, \bibinfo {author} {\bibfnamefont {E.}~\bibnamefont {Ronca}}, \bibinfo {author} {\bibfnamefont {E.~R.}\ \bibnamefont {Sayfutyarova}}, \bibinfo {author} {\bibfnamefont {M.}~\bibnamefont {Scheurer}}, \bibinfo {author} {\bibfnamefont {H.~F.}\ \bibnamefont {Schurkus}},
  \bibinfo {author} {\bibfnamefont {J.~E.~T.}\ \bibnamefont {Smith}}, \bibinfo {author} {\bibfnamefont {C.}~\bibnamefont {Sun}}, \bibinfo {author} {\bibfnamefont {S.-N.}\ \bibnamefont {Sun}}, \bibinfo {author} {\bibfnamefont {S.}~\bibnamefont {Upadhyay}}, \bibinfo {author} {\bibfnamefont {L.~K.}\ \bibnamefont {Wagner}}, \bibinfo {author} {\bibfnamefont {X.}~\bibnamefont {Wang}}, \bibinfo {author} {\bibfnamefont {A.}~\bibnamefont {White}}, \bibinfo {author} {\bibfnamefont {J.~D.}\ \bibnamefont {Whitfield}}, \bibinfo {author} {\bibfnamefont {M.~J.}\ \bibnamefont {Williamson}}, \bibinfo {author} {\bibfnamefont {S.}~\bibnamefont {Wouters}}, \bibinfo {author} {\bibfnamefont {J.}~\bibnamefont {Yang}}, \bibinfo {author} {\bibfnamefont {J.~M.}\ \bibnamefont {Yu}}, \bibinfo {author} {\bibfnamefont {T.}~\bibnamefont {Zhu}}, \bibinfo {author} {\bibfnamefont {T.~C.}\ \bibnamefont {Berkelbach}}, \bibinfo {author} {\bibfnamefont {S.}~\bibnamefont {Sharma}}, \bibinfo {author} {\bibfnamefont {A.~Y.}\ \bibnamefont {Sokolov}},
  \ and\ \bibinfo {author} {\bibfnamefont {G.~K.-L.}\ \bibnamefont {Chan}},\ }\href@noop {} {\bibfield  {journal} {\bibinfo  {journal} {J. Chem. Phys.}\ }\textbf {\bibinfo {volume} {153}},\ \bibinfo {pages} {024109} (\bibinfo {year} {2020})}\BibitemShut {NoStop}%
\bibitem [{\citenamefont {{Jmol development team}}(2016)}]{Jmol_development_team2016-ly}%
  \BibitemOpen
  \bibfield  {author} {\bibinfo {author} {\bibnamefont {{Jmol development team}}},\ }\href@noop {} {\enquote {\bibinfo {title} {Jmol},}\ } (\bibinfo {year} {2016})\BibitemShut {NoStop}%
\bibitem [{\citenamefont {Harris}\ \emph {et~al.}(2020)\citenamefont {Harris}, \citenamefont {Millman}, \citenamefont {van~der Walt}, \citenamefont {Gommers}, \citenamefont {Virtanen}, \citenamefont {Cournapeau}, \citenamefont {Wieser}, \citenamefont {Taylor}, \citenamefont {Berg}, \citenamefont {Smith}, \citenamefont {Kern}, \citenamefont {Picus}, \citenamefont {Hoyer}, \citenamefont {van Kerkwijk}, \citenamefont {Brett}, \citenamefont {Haldane}, \citenamefont {Del~R{\'\i}o}, \citenamefont {Wiebe}, \citenamefont {Peterson}, \citenamefont {G{\'e}rard-Marchant}, \citenamefont {Sheppard}, \citenamefont {Reddy}, \citenamefont {Weckesser}, \citenamefont {Abbasi}, \citenamefont {Gohlke},\ and\ \citenamefont {Oliphant}}]{Harris2020-fh}%
  \BibitemOpen
  \bibfield  {author} {\bibinfo {author} {\bibfnamefont {C.~R.}\ \bibnamefont {Harris}}, \bibinfo {author} {\bibfnamefont {K.~J.}\ \bibnamefont {Millman}}, \bibinfo {author} {\bibfnamefont {S.~J.}\ \bibnamefont {van~der Walt}}, \bibinfo {author} {\bibfnamefont {R.}~\bibnamefont {Gommers}}, \bibinfo {author} {\bibfnamefont {P.}~\bibnamefont {Virtanen}}, \bibinfo {author} {\bibfnamefont {D.}~\bibnamefont {Cournapeau}}, \bibinfo {author} {\bibfnamefont {E.}~\bibnamefont {Wieser}}, \bibinfo {author} {\bibfnamefont {J.}~\bibnamefont {Taylor}}, \bibinfo {author} {\bibfnamefont {S.}~\bibnamefont {Berg}}, \bibinfo {author} {\bibfnamefont {N.~J.}\ \bibnamefont {Smith}}, \bibinfo {author} {\bibfnamefont {R.}~\bibnamefont {Kern}}, \bibinfo {author} {\bibfnamefont {M.}~\bibnamefont {Picus}}, \bibinfo {author} {\bibfnamefont {S.}~\bibnamefont {Hoyer}}, \bibinfo {author} {\bibfnamefont {M.~H.}\ \bibnamefont {van Kerkwijk}}, \bibinfo {author} {\bibfnamefont {M.}~\bibnamefont {Brett}}, \bibinfo {author} {\bibfnamefont
  {A.}~\bibnamefont {Haldane}}, \bibinfo {author} {\bibfnamefont {J.~F.}\ \bibnamefont {Del~R{\'\i}o}}, \bibinfo {author} {\bibfnamefont {M.}~\bibnamefont {Wiebe}}, \bibinfo {author} {\bibfnamefont {P.}~\bibnamefont {Peterson}}, \bibinfo {author} {\bibfnamefont {P.}~\bibnamefont {G{\'e}rard-Marchant}}, \bibinfo {author} {\bibfnamefont {K.}~\bibnamefont {Sheppard}}, \bibinfo {author} {\bibfnamefont {T.}~\bibnamefont {Reddy}}, \bibinfo {author} {\bibfnamefont {W.}~\bibnamefont {Weckesser}}, \bibinfo {author} {\bibfnamefont {H.}~\bibnamefont {Abbasi}}, \bibinfo {author} {\bibfnamefont {C.}~\bibnamefont {Gohlke}}, \ and\ \bibinfo {author} {\bibfnamefont {T.~E.}\ \bibnamefont {Oliphant}},\ }\href@noop {} {\bibfield  {journal} {\bibinfo  {journal} {Nature}\ }\textbf {\bibinfo {volume} {585}},\ \bibinfo {pages} {357} (\bibinfo {year} {2020})}\BibitemShut {NoStop}%
\bibitem [{\citenamefont {Virtanen}\ \emph {et~al.}(2020)\citenamefont {Virtanen}, \citenamefont {Gommers}, \citenamefont {Oliphant}, \citenamefont {Haberland}, \citenamefont {Reddy}, \citenamefont {Cournapeau}, \citenamefont {Burovski}, \citenamefont {Peterson}, \citenamefont {Weckesser}, \citenamefont {Bright}, \citenamefont {van~der Walt}, \citenamefont {Brett}, \citenamefont {Wilson}, \citenamefont {Millman}, \citenamefont {Mayorov}, \citenamefont {Nelson}, \citenamefont {Jones}, \citenamefont {Kern}, \citenamefont {Larson}, \citenamefont {Carey}, \citenamefont {Polat}, \citenamefont {Feng}, \citenamefont {Moore}, \citenamefont {VanderPlas}, \citenamefont {Laxalde}, \citenamefont {Perktold}, \citenamefont {Cimrman}, \citenamefont {Henriksen}, \citenamefont {Quintero}, \citenamefont {Harris}, \citenamefont {Archibald}, \citenamefont {Ribeiro}, \citenamefont {Pedregosa}, \citenamefont {van Mulbregt},\ and\ \citenamefont {{SciPy 1.0 Contributors}}}]{Virtanen2020-rg}%
  \BibitemOpen
  \bibfield  {author} {\bibinfo {author} {\bibfnamefont {P.}~\bibnamefont {Virtanen}}, \bibinfo {author} {\bibfnamefont {R.}~\bibnamefont {Gommers}}, \bibinfo {author} {\bibfnamefont {T.~E.}\ \bibnamefont {Oliphant}}, \bibinfo {author} {\bibfnamefont {M.}~\bibnamefont {Haberland}}, \bibinfo {author} {\bibfnamefont {T.}~\bibnamefont {Reddy}}, \bibinfo {author} {\bibfnamefont {D.}~\bibnamefont {Cournapeau}}, \bibinfo {author} {\bibfnamefont {E.}~\bibnamefont {Burovski}}, \bibinfo {author} {\bibfnamefont {P.}~\bibnamefont {Peterson}}, \bibinfo {author} {\bibfnamefont {W.}~\bibnamefont {Weckesser}}, \bibinfo {author} {\bibfnamefont {J.}~\bibnamefont {Bright}}, \bibinfo {author} {\bibfnamefont {S.~J.}\ \bibnamefont {van~der Walt}}, \bibinfo {author} {\bibfnamefont {M.}~\bibnamefont {Brett}}, \bibinfo {author} {\bibfnamefont {J.}~\bibnamefont {Wilson}}, \bibinfo {author} {\bibfnamefont {K.~J.}\ \bibnamefont {Millman}}, \bibinfo {author} {\bibfnamefont {N.}~\bibnamefont {Mayorov}}, \bibinfo {author} {\bibfnamefont
  {A.~R.~J.}\ \bibnamefont {Nelson}}, \bibinfo {author} {\bibfnamefont {E.}~\bibnamefont {Jones}}, \bibinfo {author} {\bibfnamefont {R.}~\bibnamefont {Kern}}, \bibinfo {author} {\bibfnamefont {E.}~\bibnamefont {Larson}}, \bibinfo {author} {\bibfnamefont {C.~J.}\ \bibnamefont {Carey}}, \bibinfo {author} {\bibfnamefont {{\.I}.}~\bibnamefont {Polat}}, \bibinfo {author} {\bibfnamefont {Y.}~\bibnamefont {Feng}}, \bibinfo {author} {\bibfnamefont {E.~W.}\ \bibnamefont {Moore}}, \bibinfo {author} {\bibfnamefont {J.}~\bibnamefont {VanderPlas}}, \bibinfo {author} {\bibfnamefont {D.}~\bibnamefont {Laxalde}}, \bibinfo {author} {\bibfnamefont {J.}~\bibnamefont {Perktold}}, \bibinfo {author} {\bibfnamefont {R.}~\bibnamefont {Cimrman}}, \bibinfo {author} {\bibfnamefont {I.}~\bibnamefont {Henriksen}}, \bibinfo {author} {\bibfnamefont {E.~A.}\ \bibnamefont {Quintero}}, \bibinfo {author} {\bibfnamefont {C.~R.}\ \bibnamefont {Harris}}, \bibinfo {author} {\bibfnamefont {A.~M.}\ \bibnamefont {Archibald}}, \bibinfo {author}
  {\bibfnamefont {A.~H.}\ \bibnamefont {Ribeiro}}, \bibinfo {author} {\bibfnamefont {F.}~\bibnamefont {Pedregosa}}, \bibinfo {author} {\bibfnamefont {P.}~\bibnamefont {van Mulbregt}}, \ and\ \bibinfo {author} {\bibnamefont {{SciPy 1.0 Contributors}}},\ }\href@noop {} {\bibfield  {journal} {\bibinfo  {journal} {Nat. Methods}\ }\textbf {\bibinfo {volume} {17}},\ \bibinfo {pages} {261} (\bibinfo {year} {2020})}\BibitemShut {NoStop}%
\bibitem [{\citenamefont {Treinish}\ \emph {et~al.}(2023)\citenamefont {Treinish}, \citenamefont {Gambetta}, \citenamefont {Thomas}, \citenamefont {Nation}, \citenamefont {{qiskit-bot}}, \citenamefont {Kassebaum}, \citenamefont {Rodr{\'\i}guez}, \citenamefont {de~la Puente~Gonz{\'a}lez}, \citenamefont {Lishman}, \citenamefont {Hu}, \citenamefont {Bello}, \citenamefont {Garrison}, \citenamefont {Krsulich}, \citenamefont {Huang}, \citenamefont {Yu}, \citenamefont {Marques}, \citenamefont {Arellano}, \citenamefont {Gacon}, \citenamefont {McKay}, \citenamefont {Gomez}, \citenamefont {Capelluto}, \citenamefont {{Travis-S-IBM}}, \citenamefont {Mitchell}, \citenamefont {Panigrahi}, \citenamefont {{lerongil}}, \citenamefont {Rahman}, \citenamefont {Wood}, \citenamefont {Itoko}, \citenamefont {Pozas-Kerstjens},\ and\ \citenamefont {Wood}}]{Treinish2023-co}%
  \BibitemOpen
  \bibfield  {author} {\bibinfo {author} {\bibfnamefont {M.}~\bibnamefont {Treinish}}, \bibinfo {author} {\bibfnamefont {J.}~\bibnamefont {Gambetta}}, \bibinfo {author} {\bibfnamefont {S.}~\bibnamefont {Thomas}}, \bibinfo {author} {\bibfnamefont {P.}~\bibnamefont {Nation}}, \bibinfo {author} {\bibnamefont {{qiskit-bot}}}, \bibinfo {author} {\bibfnamefont {P.}~\bibnamefont {Kassebaum}}, \bibinfo {author} {\bibfnamefont {D.~M.}\ \bibnamefont {Rodr{\'\i}guez}}, \bibinfo {author} {\bibfnamefont {S.}~\bibnamefont {de~la Puente~Gonz{\'a}lez}}, \bibinfo {author} {\bibfnamefont {J.}~\bibnamefont {Lishman}}, \bibinfo {author} {\bibfnamefont {S.}~\bibnamefont {Hu}}, \bibinfo {author} {\bibfnamefont {L.}~\bibnamefont {Bello}}, \bibinfo {author} {\bibfnamefont {J.}~\bibnamefont {Garrison}}, \bibinfo {author} {\bibfnamefont {K.}~\bibnamefont {Krsulich}}, \bibinfo {author} {\bibfnamefont {J.}~\bibnamefont {Huang}}, \bibinfo {author} {\bibfnamefont {J.}~\bibnamefont {Yu}}, \bibinfo {author} {\bibfnamefont {M.}~\bibnamefont
  {Marques}}, \bibinfo {author} {\bibfnamefont {E.}~\bibnamefont {Arellano}}, \bibinfo {author} {\bibfnamefont {J.}~\bibnamefont {Gacon}}, \bibinfo {author} {\bibfnamefont {D.}~\bibnamefont {McKay}}, \bibinfo {author} {\bibfnamefont {J.}~\bibnamefont {Gomez}}, \bibinfo {author} {\bibfnamefont {L.}~\bibnamefont {Capelluto}}, \bibinfo {author} {\bibnamefont {{Travis-S-IBM}}}, \bibinfo {author} {\bibfnamefont {A.}~\bibnamefont {Mitchell}}, \bibinfo {author} {\bibfnamefont {A.}~\bibnamefont {Panigrahi}}, \bibinfo {author} {\bibnamefont {{lerongil}}}, \bibinfo {author} {\bibfnamefont {R.~I.}\ \bibnamefont {Rahman}}, \bibinfo {author} {\bibfnamefont {S.}~\bibnamefont {Wood}}, \bibinfo {author} {\bibfnamefont {T.}~\bibnamefont {Itoko}}, \bibinfo {author} {\bibfnamefont {A.}~\bibnamefont {Pozas-Kerstjens}}, \ and\ \bibinfo {author} {\bibfnamefont {C.~J.}\ \bibnamefont {Wood}},\ }\href@noop {} {\enquote {\bibinfo {title} {Qiskit/qiskit: Qiskit 0.42.1},}\ } (\bibinfo {year} {2023})\BibitemShut {NoStop}%
\bibitem [{\citenamefont {Suzuki}\ \emph {et~al.}(2021)\citenamefont {Suzuki}, \citenamefont {Kawase}, \citenamefont {Masumura}, \citenamefont {Hiraga}, \citenamefont {Nakadai}, \citenamefont {Chen}, \citenamefont {Nakanishi}, \citenamefont {Mitarai}, \citenamefont {Imai}, \citenamefont {Tamiya}, \citenamefont {Yamamoto}, \citenamefont {Yan}, \citenamefont {Kawakubo}, \citenamefont {Nakagawa}, \citenamefont {Ibe}, \citenamefont {Zhang}, \citenamefont {Yamashita}, \citenamefont {Yoshimura}, \citenamefont {Hayashi},\ and\ \citenamefont {Fujii}}]{Suzuki2021-lj}%
  \BibitemOpen
  \bibfield  {author} {\bibinfo {author} {\bibfnamefont {Y.}~\bibnamefont {Suzuki}}, \bibinfo {author} {\bibfnamefont {Y.}~\bibnamefont {Kawase}}, \bibinfo {author} {\bibfnamefont {Y.}~\bibnamefont {Masumura}}, \bibinfo {author} {\bibfnamefont {Y.}~\bibnamefont {Hiraga}}, \bibinfo {author} {\bibfnamefont {M.}~\bibnamefont {Nakadai}}, \bibinfo {author} {\bibfnamefont {J.}~\bibnamefont {Chen}}, \bibinfo {author} {\bibfnamefont {K.~M.}\ \bibnamefont {Nakanishi}}, \bibinfo {author} {\bibfnamefont {K.}~\bibnamefont {Mitarai}}, \bibinfo {author} {\bibfnamefont {R.}~\bibnamefont {Imai}}, \bibinfo {author} {\bibfnamefont {S.}~\bibnamefont {Tamiya}}, \bibinfo {author} {\bibfnamefont {T.}~\bibnamefont {Yamamoto}}, \bibinfo {author} {\bibfnamefont {T.}~\bibnamefont {Yan}}, \bibinfo {author} {\bibfnamefont {T.}~\bibnamefont {Kawakubo}}, \bibinfo {author} {\bibfnamefont {Y.~O.}\ \bibnamefont {Nakagawa}}, \bibinfo {author} {\bibfnamefont {Y.}~\bibnamefont {Ibe}}, \bibinfo {author} {\bibfnamefont {Y.}~\bibnamefont {Zhang}},
  \bibinfo {author} {\bibfnamefont {H.}~\bibnamefont {Yamashita}}, \bibinfo {author} {\bibfnamefont {H.}~\bibnamefont {Yoshimura}}, \bibinfo {author} {\bibfnamefont {A.}~\bibnamefont {Hayashi}}, \ and\ \bibinfo {author} {\bibfnamefont {K.}~\bibnamefont {Fujii}},\ }\href@noop {} {\bibfield  {journal} {\bibinfo  {journal} {Quantum}\ }\textbf {\bibinfo {volume} {5}},\ \bibinfo {pages} {559} (\bibinfo {year} {2021})}\BibitemShut {NoStop}%
\bibitem [{\citenamefont {Gocho}\ \emph {et~al.}(2023)\citenamefont {Gocho}, \citenamefont {Nakamura}, \citenamefont {Kanno}, \citenamefont {Gao}, \citenamefont {Kobayashi}, \citenamefont {Inagaki},\ and\ \citenamefont {Hatanaka}}]{Gocho2023-uy}%
  \BibitemOpen
  \bibfield  {author} {\bibinfo {author} {\bibfnamefont {S.}~\bibnamefont {Gocho}}, \bibinfo {author} {\bibfnamefont {H.}~\bibnamefont {Nakamura}}, \bibinfo {author} {\bibfnamefont {S.}~\bibnamefont {Kanno}}, \bibinfo {author} {\bibfnamefont {Q.}~\bibnamefont {Gao}}, \bibinfo {author} {\bibfnamefont {T.}~\bibnamefont {Kobayashi}}, \bibinfo {author} {\bibfnamefont {T.}~\bibnamefont {Inagaki}}, \ and\ \bibinfo {author} {\bibfnamefont {M.}~\bibnamefont {Hatanaka}},\ }\href@noop {} {\bibfield  {journal} {\bibinfo  {journal} {npj Computational Materials}\ }\textbf {\bibinfo {volume} {9}},\ \bibinfo {pages} {1} (\bibinfo {year} {2023})}\BibitemShut {NoStop}%
\bibitem [{\citenamefont {Aharonov}\ \emph {et~al.}(2009)\citenamefont {Aharonov}, \citenamefont {Jones},\ and\ \citenamefont {Landau}}]{Aharonov2009-aa}%
  \BibitemOpen
  \bibfield  {author} {\bibinfo {author} {\bibfnamefont {D.}~\bibnamefont {Aharonov}}, \bibinfo {author} {\bibfnamefont {V.}~\bibnamefont {Jones}}, \ and\ \bibinfo {author} {\bibfnamefont {Z.}~\bibnamefont {Landau}},\ }\href@noop {} {\bibfield  {journal} {\bibinfo  {journal} {Algorithmica}\ }\textbf {\bibinfo {volume} {55}},\ \bibinfo {pages} {395} (\bibinfo {year} {2009})}\BibitemShut {NoStop}%
\bibitem [{\citenamefont {Huggins}\ \emph {et~al.}(2020)\citenamefont {Huggins}, \citenamefont {Lee}, \citenamefont {Baek}, \citenamefont {O'Gorman},\ and\ \citenamefont {Birgitta~Whaley}}]{Huggins2020-fk}%
  \BibitemOpen
  \bibfield  {author} {\bibinfo {author} {\bibfnamefont {W.~J.}\ \bibnamefont {Huggins}}, \bibinfo {author} {\bibfnamefont {J.}~\bibnamefont {Lee}}, \bibinfo {author} {\bibfnamefont {U.}~\bibnamefont {Baek}}, \bibinfo {author} {\bibfnamefont {B.}~\bibnamefont {O'Gorman}}, \ and\ \bibinfo {author} {\bibfnamefont {K.}~\bibnamefont {Birgitta~Whaley}},\ }\href@noop {} {\bibfield  {journal} {\bibinfo  {journal} {New J. Phys.}\ }\textbf {\bibinfo {volume} {22}},\ \bibinfo {pages} {073009} (\bibinfo {year} {2020})}\BibitemShut {NoStop}%
\bibitem [{\citenamefont {Baek}\ \emph {et~al.}(2023)\citenamefont {Baek}, \citenamefont {Hait}, \citenamefont {Shee}, \citenamefont {Leimkuhler}, \citenamefont {Huggins}, \citenamefont {Stetina}, \citenamefont {Head-Gordon},\ and\ \citenamefont {Whaley}}]{Baek2023-xb}%
  \BibitemOpen
  \bibfield  {author} {\bibinfo {author} {\bibfnamefont {U.}~\bibnamefont {Baek}}, \bibinfo {author} {\bibfnamefont {D.}~\bibnamefont {Hait}}, \bibinfo {author} {\bibfnamefont {J.}~\bibnamefont {Shee}}, \bibinfo {author} {\bibfnamefont {O.}~\bibnamefont {Leimkuhler}}, \bibinfo {author} {\bibfnamefont {W.~J.}\ \bibnamefont {Huggins}}, \bibinfo {author} {\bibfnamefont {T.~F.}\ \bibnamefont {Stetina}}, \bibinfo {author} {\bibfnamefont {M.}~\bibnamefont {Head-Gordon}}, \ and\ \bibinfo {author} {\bibfnamefont {K.~B.}\ \bibnamefont {Whaley}},\ }\href@noop {} {\bibfield  {journal} {\bibinfo  {journal} {PRX Quantum}\ }\textbf {\bibinfo {volume} {4}},\ \bibinfo {pages} {030307} (\bibinfo {year} {2023})}\BibitemShut {NoStop}%
\bibitem [{\citenamefont {Lu}\ \emph {et~al.}(2021)\citenamefont {Lu}, \citenamefont {Ba{\~n}uls},\ and\ \citenamefont {Cirac}}]{Lu2021-on}%
  \BibitemOpen
  \bibfield  {author} {\bibinfo {author} {\bibfnamefont {S.}~\bibnamefont {Lu}}, \bibinfo {author} {\bibfnamefont {M.~C.}\ \bibnamefont {Ba{\~n}uls}}, \ and\ \bibinfo {author} {\bibfnamefont {J.~I.}\ \bibnamefont {Cirac}},\ }\href@noop {} {\bibfield  {journal} {\bibinfo  {journal} {PRX Quantum}\ }\textbf {\bibinfo {volume} {2}},\ \bibinfo {pages} {020321} (\bibinfo {year} {2021})}\BibitemShut {NoStop}%
\bibitem [{\citenamefont {Harada}\ \emph {et~al.}(2023)\citenamefont {Harada}, \citenamefont {Suzuki}, \citenamefont {Yang}, \citenamefont {Tokunaga},\ and\ \citenamefont {Endo}}]{Harada2023-il}%
  \BibitemOpen
  \bibfield  {author} {\bibinfo {author} {\bibfnamefont {H.}~\bibnamefont {Harada}}, \bibinfo {author} {\bibfnamefont {Y.}~\bibnamefont {Suzuki}}, \bibinfo {author} {\bibfnamefont {B.}~\bibnamefont {Yang}}, \bibinfo {author} {\bibfnamefont {Y.}~\bibnamefont {Tokunaga}}, \ and\ \bibinfo {author} {\bibfnamefont {S.}~\bibnamefont {Endo}},\ }\href@noop {} {\  (\bibinfo {year} {2023})},\ \Eprint {http://arxiv.org/abs/2309.15761} {arXiv:2309.15761 [quant-ph]} \BibitemShut {NoStop}%
\bibitem [{\citenamefont {Okada}\ \emph {et~al.}(2023)\citenamefont {Okada}, \citenamefont {Osaki}, \citenamefont {Mitarai},\ and\ \citenamefont {Fujii}}]{Okada2023-jm}%
  \BibitemOpen
  \bibfield  {author} {\bibinfo {author} {\bibfnamefont {K.~N.}\ \bibnamefont {Okada}}, \bibinfo {author} {\bibfnamefont {K.}~\bibnamefont {Osaki}}, \bibinfo {author} {\bibfnamefont {K.}~\bibnamefont {Mitarai}}, \ and\ \bibinfo {author} {\bibfnamefont {K.}~\bibnamefont {Fujii}},\ }\href@noop {} {\bibfield  {journal} {\bibinfo  {journal} {Phys. Rev. Res.}\ }\textbf {\bibinfo {volume} {5}},\ \bibinfo {pages} {043217} (\bibinfo {year} {2023})}\BibitemShut {NoStop}%
\bibitem [{\citenamefont {Zhang}\ \emph {et~al.}(2020)\citenamefont {Zhang}, \citenamefont {Kyaw}, \citenamefont {Kottmann}, \citenamefont {Degroote},\ and\ \citenamefont {Aspuru-Guzik}}]{Zhang2020-vv}%
  \BibitemOpen
  \bibfield  {author} {\bibinfo {author} {\bibfnamefont {Z.-J.}\ \bibnamefont {Zhang}}, \bibinfo {author} {\bibfnamefont {T.~H.}\ \bibnamefont {Kyaw}}, \bibinfo {author} {\bibfnamefont {J.~S.}\ \bibnamefont {Kottmann}}, \bibinfo {author} {\bibfnamefont {M.}~\bibnamefont {Degroote}}, \ and\ \bibinfo {author} {\bibfnamefont {A.}~\bibnamefont {Aspuru-Guzik}},\ }\href@noop {} {\  (\bibinfo {year} {2020})},\ \Eprint {http://arxiv.org/abs/2008.07553} {arXiv:2008.07553 [quant-ph]} \BibitemShut {NoStop}%
\bibitem [{\citenamefont {Zhang}\ \emph {et~al.}(2022{\natexlab{b}})\citenamefont {Zhang}, \citenamefont {Cincio}, \citenamefont {Negre}, \citenamefont {Czarnik}, \citenamefont {Coles}, \citenamefont {Anisimov}, \citenamefont {Mniszewski}, \citenamefont {Tretiak},\ and\ \citenamefont {Dub}}]{Zhang2022-mu}%
  \BibitemOpen
  \bibfield  {author} {\bibinfo {author} {\bibfnamefont {Y.}~\bibnamefont {Zhang}}, \bibinfo {author} {\bibfnamefont {L.}~\bibnamefont {Cincio}}, \bibinfo {author} {\bibfnamefont {C.~F.~A.}\ \bibnamefont {Negre}}, \bibinfo {author} {\bibfnamefont {P.}~\bibnamefont {Czarnik}}, \bibinfo {author} {\bibfnamefont {P.~J.}\ \bibnamefont {Coles}}, \bibinfo {author} {\bibfnamefont {P.~M.}\ \bibnamefont {Anisimov}}, \bibinfo {author} {\bibfnamefont {S.~M.}\ \bibnamefont {Mniszewski}}, \bibinfo {author} {\bibfnamefont {S.}~\bibnamefont {Tretiak}}, \ and\ \bibinfo {author} {\bibfnamefont {P.~A.}\ \bibnamefont {Dub}},\ }\href@noop {} {\bibfield  {journal} {\bibinfo  {journal} {npj Quantum Information}\ }\textbf {\bibinfo {volume} {8}},\ \bibinfo {pages} {1} (\bibinfo {year} {2022}{\natexlab{b}})}\BibitemShut {NoStop}%
\bibitem [{\citenamefont {Bergholm}\ \emph {et~al.}(2018)\citenamefont {Bergholm}, \citenamefont {Izaac}, \citenamefont {Schuld}, \citenamefont {Gogolin}, \citenamefont {Ahmed}, \citenamefont {Ajith}, \citenamefont {Sohaib~Alam}, \citenamefont {Alonso-Linaje}, \citenamefont {AkashNarayanan}, \citenamefont {Asadi}, \citenamefont {Arrazola}, \citenamefont {Azad}, \citenamefont {Banning}, \citenamefont {Blank}, \citenamefont {Bromley}, \citenamefont {Cordier}, \citenamefont {Ceroni}, \citenamefont {Delgado}, \citenamefont {Di~Matteo}, \citenamefont {Dusko}, \citenamefont {Garg}, \citenamefont {Guala}, \citenamefont {Hayes}, \citenamefont {Hill}, \citenamefont {Ijaz}, \citenamefont {Isacsson}, \citenamefont {Ittah}, \citenamefont {Jahangiri}, \citenamefont {Jain}, \citenamefont {Jiang}, \citenamefont {Khandelwal}, \citenamefont {Kottmann}, \citenamefont {Lang}, \citenamefont {Lee}, \citenamefont {Loke}, \citenamefont {Lowe}, \citenamefont {McKiernan}, \citenamefont {Meyer}, \citenamefont {Monta{\~n}ez-Barrera},
  \citenamefont {Moyard}, \citenamefont {Niu}, \citenamefont {O'Riordan}, \citenamefont {Oud}, \citenamefont {Panigrahi}, \citenamefont {Park}, \citenamefont {Polatajko}, \citenamefont {Quesada}, \citenamefont {Roberts}, \citenamefont {S{\'a}}, \citenamefont {Schoch}, \citenamefont {Shi}, \citenamefont {Shu}, \citenamefont {Sim}, \citenamefont {Singh}, \citenamefont {Strandberg}, \citenamefont {Soni}, \citenamefont {Sz{\'a}va}, \citenamefont {Thabet}, \citenamefont {Vargas-Hern{\'a}ndez}, \citenamefont {Vincent}, \citenamefont {Vitucci}, \citenamefont {Weber}, \citenamefont {Wierichs}, \citenamefont {Wiersema}, \citenamefont {Willmann}, \citenamefont {Wong}, \citenamefont {Zhang},\ and\ \citenamefont {Killoran}}]{Bergholm2018-cn}%
  \BibitemOpen
  \bibfield  {author} {\bibinfo {author} {\bibfnamefont {V.}~\bibnamefont {Bergholm}}, \bibinfo {author} {\bibfnamefont {J.}~\bibnamefont {Izaac}}, \bibinfo {author} {\bibfnamefont {M.}~\bibnamefont {Schuld}}, \bibinfo {author} {\bibfnamefont {C.}~\bibnamefont {Gogolin}}, \bibinfo {author} {\bibfnamefont {S.}~\bibnamefont {Ahmed}}, \bibinfo {author} {\bibfnamefont {V.}~\bibnamefont {Ajith}}, \bibinfo {author} {\bibfnamefont {M.}~\bibnamefont {Sohaib~Alam}}, \bibinfo {author} {\bibfnamefont {G.}~\bibnamefont {Alonso-Linaje}}, \bibinfo {author} {\bibfnamefont {B.}~\bibnamefont {AkashNarayanan}}, \bibinfo {author} {\bibfnamefont {A.}~\bibnamefont {Asadi}}, \bibinfo {author} {\bibfnamefont {J.~M.}\ \bibnamefont {Arrazola}}, \bibinfo {author} {\bibfnamefont {U.}~\bibnamefont {Azad}}, \bibinfo {author} {\bibfnamefont {S.}~\bibnamefont {Banning}}, \bibinfo {author} {\bibfnamefont {C.}~\bibnamefont {Blank}}, \bibinfo {author} {\bibfnamefont {T.~R.}\ \bibnamefont {Bromley}}, \bibinfo {author} {\bibfnamefont {B.~A.}\
  \bibnamefont {Cordier}}, \bibinfo {author} {\bibfnamefont {J.}~\bibnamefont {Ceroni}}, \bibinfo {author} {\bibfnamefont {A.}~\bibnamefont {Delgado}}, \bibinfo {author} {\bibfnamefont {O.}~\bibnamefont {Di~Matteo}}, \bibinfo {author} {\bibfnamefont {A.}~\bibnamefont {Dusko}}, \bibinfo {author} {\bibfnamefont {T.}~\bibnamefont {Garg}}, \bibinfo {author} {\bibfnamefont {D.}~\bibnamefont {Guala}}, \bibinfo {author} {\bibfnamefont {A.}~\bibnamefont {Hayes}}, \bibinfo {author} {\bibfnamefont {R.}~\bibnamefont {Hill}}, \bibinfo {author} {\bibfnamefont {A.}~\bibnamefont {Ijaz}}, \bibinfo {author} {\bibfnamefont {T.}~\bibnamefont {Isacsson}}, \bibinfo {author} {\bibfnamefont {D.}~\bibnamefont {Ittah}}, \bibinfo {author} {\bibfnamefont {S.}~\bibnamefont {Jahangiri}}, \bibinfo {author} {\bibfnamefont {P.}~\bibnamefont {Jain}}, \bibinfo {author} {\bibfnamefont {E.}~\bibnamefont {Jiang}}, \bibinfo {author} {\bibfnamefont {A.}~\bibnamefont {Khandelwal}}, \bibinfo {author} {\bibfnamefont {K.}~\bibnamefont {Kottmann}},
  \bibinfo {author} {\bibfnamefont {R.~A.}\ \bibnamefont {Lang}}, \bibinfo {author} {\bibfnamefont {C.}~\bibnamefont {Lee}}, \bibinfo {author} {\bibfnamefont {T.}~\bibnamefont {Loke}}, \bibinfo {author} {\bibfnamefont {A.}~\bibnamefont {Lowe}}, \bibinfo {author} {\bibfnamefont {K.}~\bibnamefont {McKiernan}}, \bibinfo {author} {\bibfnamefont {J.~J.}\ \bibnamefont {Meyer}}, \bibinfo {author} {\bibfnamefont {J.~A.}\ \bibnamefont {Monta{\~n}ez-Barrera}}, \bibinfo {author} {\bibfnamefont {R.}~\bibnamefont {Moyard}}, \bibinfo {author} {\bibfnamefont {Z.}~\bibnamefont {Niu}}, \bibinfo {author} {\bibfnamefont {L.~J.}\ \bibnamefont {O'Riordan}}, \bibinfo {author} {\bibfnamefont {S.}~\bibnamefont {Oud}}, \bibinfo {author} {\bibfnamefont {A.}~\bibnamefont {Panigrahi}}, \bibinfo {author} {\bibfnamefont {C.-Y.}\ \bibnamefont {Park}}, \bibinfo {author} {\bibfnamefont {D.}~\bibnamefont {Polatajko}}, \bibinfo {author} {\bibfnamefont {N.}~\bibnamefont {Quesada}}, \bibinfo {author} {\bibfnamefont {C.}~\bibnamefont {Roberts}},
  \bibinfo {author} {\bibfnamefont {N.}~\bibnamefont {S{\'a}}}, \bibinfo {author} {\bibfnamefont {I.}~\bibnamefont {Schoch}}, \bibinfo {author} {\bibfnamefont {B.}~\bibnamefont {Shi}}, \bibinfo {author} {\bibfnamefont {S.}~\bibnamefont {Shu}}, \bibinfo {author} {\bibfnamefont {S.}~\bibnamefont {Sim}}, \bibinfo {author} {\bibfnamefont {A.}~\bibnamefont {Singh}}, \bibinfo {author} {\bibfnamefont {I.}~\bibnamefont {Strandberg}}, \bibinfo {author} {\bibfnamefont {J.}~\bibnamefont {Soni}}, \bibinfo {author} {\bibfnamefont {A.}~\bibnamefont {Sz{\'a}va}}, \bibinfo {author} {\bibfnamefont {S.}~\bibnamefont {Thabet}}, \bibinfo {author} {\bibfnamefont {R.~A.}\ \bibnamefont {Vargas-Hern{\'a}ndez}}, \bibinfo {author} {\bibfnamefont {T.}~\bibnamefont {Vincent}}, \bibinfo {author} {\bibfnamefont {N.}~\bibnamefont {Vitucci}}, \bibinfo {author} {\bibfnamefont {M.}~\bibnamefont {Weber}}, \bibinfo {author} {\bibfnamefont {D.}~\bibnamefont {Wierichs}}, \bibinfo {author} {\bibfnamefont {R.}~\bibnamefont {Wiersema}}, \bibinfo
  {author} {\bibfnamefont {M.}~\bibnamefont {Willmann}}, \bibinfo {author} {\bibfnamefont {V.}~\bibnamefont {Wong}}, \bibinfo {author} {\bibfnamefont {S.}~\bibnamefont {Zhang}}, \ and\ \bibinfo {author} {\bibfnamefont {N.}~\bibnamefont {Killoran}},\ }\href@noop {} {\  (\bibinfo {year} {2018})},\ \Eprint {http://arxiv.org/abs/1811.04968} {arXiv:1811.04968 [quant-ph]} \BibitemShut {NoStop}%
\end{thebibliography}%


\begin{thebibliography}{0}%
\makeatletter
\providecommand \@ifxundefined [1]{%
 \@ifx{#1\undefined}
}%
\providecommand \@ifnum [1]{%
 \ifnum #1\expandafter \@firstoftwo
 \else \expandafter \@secondoftwo
 \fi
}%
\providecommand \@ifx [1]{%
 \ifx #1\expandafter \@firstoftwo
 \else \expandafter \@secondoftwo
 \fi
}%
\providecommand \natexlab [1]{#1}%
\providecommand \enquote  [1]{``#1''}%
\providecommand \bibnamefont  [1]{#1}%
\providecommand \bibfnamefont [1]{#1}%
\providecommand \citenamefont [1]{#1}%
\providecommand \href@noop [0]{\@secondoftwo}%
\providecommand \href [0]{\begingroup \@sanitize@url \@href}%
\providecommand \@href[1]{\@@startlink{#1}\@@href}%
\providecommand \@@href[1]{\endgroup#1\@@endlink}%
\providecommand \@sanitize@url [0]{\catcode `\\12\catcode `\$12\catcode `\&12\catcode `\#12\catcode `\^12\catcode `\_12\catcode `\%12\relax}%
\providecommand \@@startlink[1]{}%
\providecommand \@@endlink[0]{}%
\providecommand \url  [0]{\begingroup\@sanitize@url \@url }%
\providecommand \@url [1]{\endgroup\@href {#1}{\urlprefix }}%
\providecommand \urlprefix  [0]{URL }%
\providecommand \Eprint [0]{\href }%
\providecommand \doibase [0]{https://doi.org/}%
\providecommand \selectlanguage [0]{\@gobble}%
\providecommand \bibinfo  [0]{\@secondoftwo}%
\providecommand \bibfield  [0]{\@secondoftwo}%
\providecommand \translation [1]{[#1]}%
\providecommand \BibitemOpen [0]{}%
\providecommand \bibitemStop [0]{}%
\providecommand \bibitemNoStop [0]{.\EOS\space}%
\providecommand \EOS [0]{\spacefactor3000\relax}%
\providecommand \BibitemShut  [1]{\csname bibitem#1\endcsname}%
\let\auto@bib@innerbib\@empty
\end{thebibliography}%
\end{document}